\documentclass[preprint, unsortedaddress, showpacs, nofootinbib, eqsecnum]{revtex4}
\usepackage{amsfonts, amsbsy, amssymb, amsmath, graphicx,  float, color} 
\usepackage{subfigure}
\usepackage{color}
\begin{document}

\title{Nonstatistical dynamics on the caldera}

\author{Peter Collins}
\affiliation{School of Mathematics  \\
University of Bristol\\Bristol BS8 1TW\\United Kingdom}

\author{Zeb C. Kramer}
\email[]{zck3@cornell.edu}
\affiliation{Department of Chemistry and Chemical Biology\\
Baker Laboratory\\
Cornell University\\
Ithaca, NY 14853\\USA}

\author{Barry K. Carpenter}
\email[]{CarpenterB1@cardiff.ac.uk}
\affiliation{School of Chemistry\\
Cardiff University\\
Cardiff\\
CF10 3AT\\
United Kingdom
}

\author{Gregory S. Ezra}
\email[]{gse1@cornell.edu}
\affiliation{Department of Chemistry and Chemical Biology\\
Baker Laboratory\\
Cornell University\\
Ithaca, NY 14853\\USA}

\author{Stephen Wiggins}
\email[]{stephen.wiggins@mac.com}
\affiliation{School of Mathematics \\
University of Bristol\\Bristol BS8 1TW\\United Kingdom}

\date{\today}

\begin{abstract}

We explore both classical and quantum dynamics of a model potential
exhibiting a caldera: that is, a shallow potential well with two pairs of symmetry 
related index one saddles associated with entrance/exit channels.
Classical trajectory simulations at several different energies 
confirm the existence of the `dynamical matching' phenomenon 
originally proposed by Carpenter, where the momentum direction
associated with an incoming trajectory initiated at a high energy saddle point
determines to a considerable extent the outcome of the 
reaction (passage through the diametrically opposing exit channel).  By studying
a `stretched' version of the caldera model, we have uncovered a generalized
dynamical matching: bundles of trajectories can reflect off a hard potential wall
so as to end up exiting predominantly through the transition state opposite the 
reflection point.  We also investigate the effects of dissipation on the classical dynamics.
In addition to classical trajectory studies, we examine
the dynamics of quantum wave packets on the caldera potential (stretched and unstretched).
These computations reveal a quantum mechanical analogue of the `dynamical matching'
phenomenon, where the initial expectation value of the momentum 
direction for the wave packet determines the exit channel through which most
of the probability density passes to product.  

\end{abstract}


\maketitle

\newpage

\section{Introduction}
\label{sec:intro}

\begin{figure}[htbp!]
\begin{center}
\subfigure[]{\includegraphics[width=2.85in]{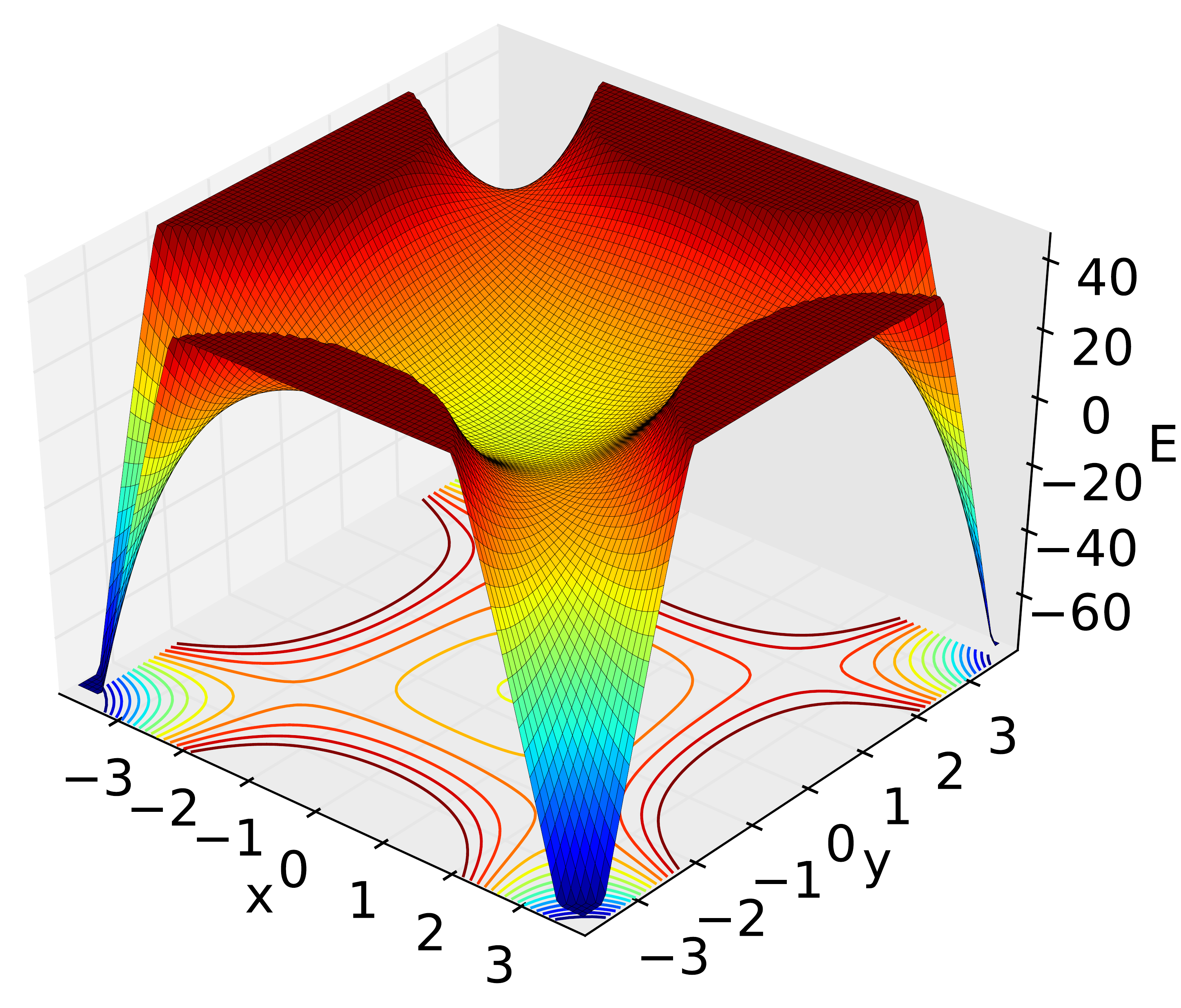}}
\hskip 0.5cm
\subfigure[]{\includegraphics[width=2.5in]{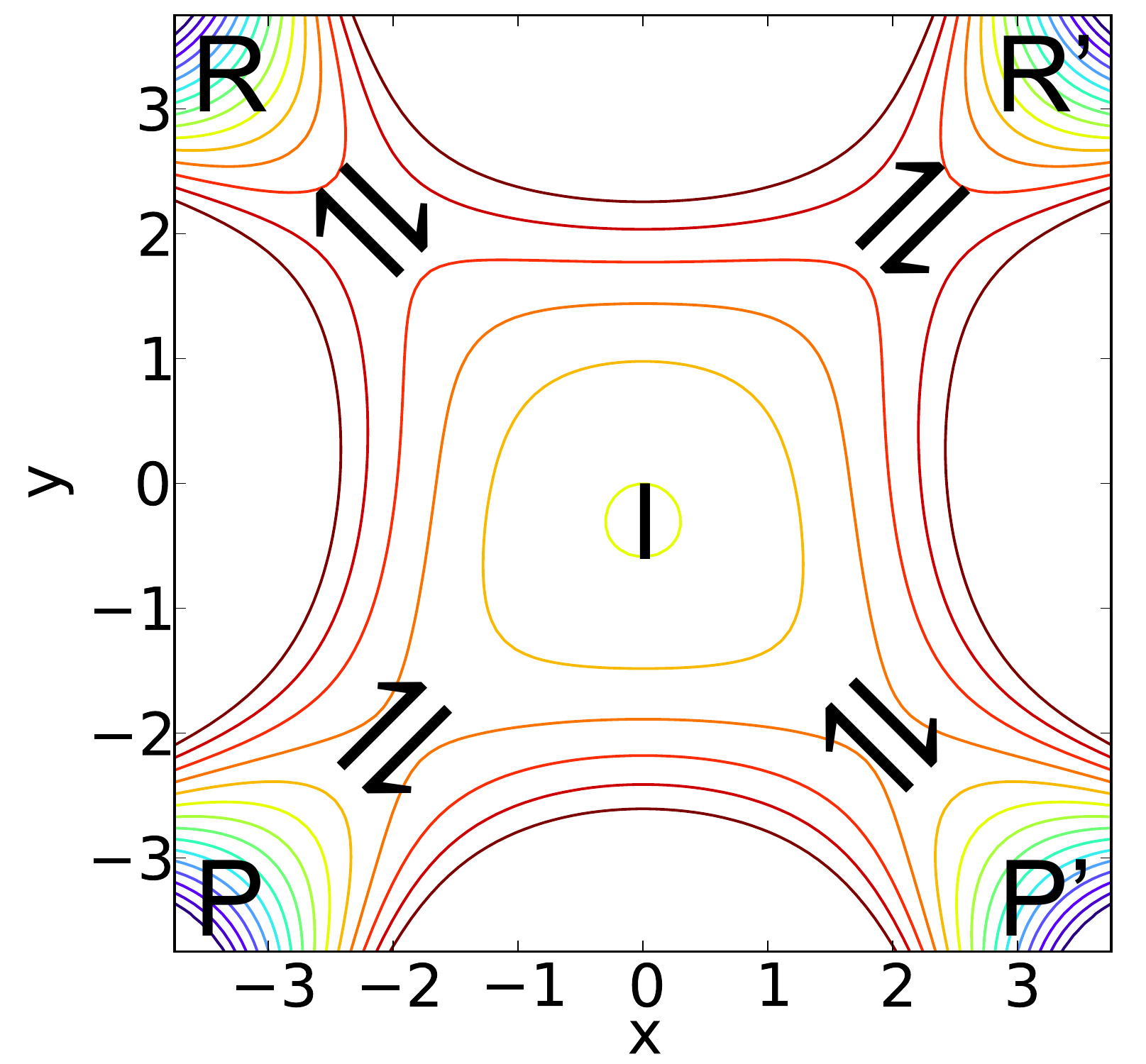}}
\caption{In (a), an example two\--dimensional caldera feature, $E\left(x,y\right)$, looking through a low energy index\--1 saddle toward the high energy index\--1 saddle.  In (b), the contour of the surface shown in (a) with a superimposed reaction scheme.  The energy and distance units are the same
used in the classical trajectory calculations and the potential is of parameter $\lambda=1$, see Section \ref{sec:pes}.}
\label{fig:scheme1}
\end{center}
\end{figure}

The `caldera', a term used by von Doering and coworkers \cite{Doering02,Doering04} due to its similarity in shape 
to the collapsed region within an erupted volcano, is an important multidimensional structure found embedded in 
`reaction\--determining' coordinates of the full dimensional potential energy surfaces (PESs) in 
a variety of organic chemical reactions.  In particular, calderas often arise in reactions involving transient singlet\--state biradicals 
which are able to undergo facile large amplitude motions.  The caldera is primarily characterized by the flat region or shallow minimum at its center 
surrounded by potential walls and multiple symmetry\--related first\--order saddle points that allow entrance and exit 
from this intermediate region.  These saddle points are usually quite low in energy with respect to the caldera center 
so that the salient feature of the caldera is its flatness.  Calderas have also been called ``twixtyls'' \cite{Hoffmann70}, 
``continuous diradicals''\cite{Doering74}, and ``mesas'' \cite{Carpenter13}, sometimes depending on 
the depth of the center, if any.   In this work, we adopt the
``caldera" terminology and the calderas we study here have a depressed center.  Nevertheless, the 
conclusions in this work are such that immediate implications follow for the more general potential features listed above.  
An example caldera, one which we shall study in detail below, 
is shown in Figure \ref{fig:scheme1}(a) where all the features, e.g. the shallow minimum and the saddle points,
described above can be easily identified.   Looking at the 
reaction scheme imposed on the potential contour map in Figure \ref{fig:scheme1}(b), one might naively 
suppose that the caldera is a ``decision point" or ``crossroads" in the reaction where selectivity between P, P', R, and R' 
is determined.  The notion of reaction path so commonly employed in understanding reaction mechanisms and in 
applications of statistical rate theory is no longer necessarily dynamically relevant due to the relatively flat topography and
the ``crossroads" nature of the caldera.
The subject of this work is to explore what governs reactions at the caldera crossroads and what mechanistic 
conclusions can be drawn from the presence of an embedded caldera feature in the PES topography.  The multidimensionality
of the caldera and breakdown of the reaction path picture at this ``crossroads" suggest the result of the caldera
intermediate on the reaction depends on the details of the passage dynamics.

Much recent experimental and theoretical work has focused on recognizing and understanding 
the manifestations of nonstatistical dynamics in thermal reactions of organic molecules 
(for reviews, see refs 
\onlinecite{Carpenter92,Carpenter98,Hase98,Gruebele04,Carpenter03,Carpenter05,Ess08,Yamataka10,Rehbein11}; 
see also the representative refs 
\onlinecite{Biswas14,Bogle12,Quijano11,Thomas08,Ussing06,Goldman11,Litovitz08,Reyes02,Nummela02,Debbert02,Black12,Xu10,Xu11,Doubleday02,
Doubleday06,Akimoto12,Yamamoto11,Ammal03,Siebert11,Siebert12,Hong12}).
Such research has convincingly demonstrated that, for an ever-growing number of cases, 
standard transition state theory (TST) and Rice\--Ramsperger\--Kassel\--Marcus (RRKM) approaches \cite{Wigner38,Eyring35,Baer96,Rice27,Robinson72}
for prediction of rates, 
product ratios, stereospecificity and isotope effects can fail completely. 
This work is changing the basic textbook paradigms of physical organic chemistry 
(cf. ref. \onlinecite{Bachrach07}, Ch. 7). 

A fundamental dynamical assumption underlying conventional statistical theories of reaction rates 
and selectivities is the existence of intramolecular vibrational energy redistribution (IVR) 
that is rapid compared to the rate of reaction/isomerization 
\cite{Rice81,Brumer81,Dumont86,Brumer88,Noid81}.  
Such rapid IVR leads to a `loss of memory' of particular initial conditions \cite{Brass93}. 
Although the TST and RRKM models were developed as methods to describe the kinetics 
of elementary (single-step) unimolecular reactions, the assumption of rapid 
IVR would appear to justify their application to multi-step reactions. 
Following the rate determining step, the subsequent fate of reactants 
depends solely (according to these theories) on available
energy, distributed according to the statistical
partitioning at the transition state.
Under such circumstances, standard computations based on features (usually critical points, 
such as minima and saddle points) of the potential energy surface (PES) should allow 
predictions for relative rates associated with competing reactive channels, as well as 
temperature dependencies of branching ratios \cite{Carpenter84,Wales03}.
Specifically addressing the caldera, a statistical description of 
reaction through the caldera assumes that IVR is much faster than the caldera passage time,
so that a statistical ensemble of reactants entering the caldera
essentially ``loses its way", i.e. promptly spreads out over the accessible phase space of the caldera intermediate, 
during the passage and a quasi\--equilibrium population of intermediate forms in the caldera 
bowl.  Decay of this intermediate is then determined solely by the transition states of exit that are
nominally associated with first\--order saddle points in elementary statistical rate theory. 

However, the work cited above has shown that the assumption of rapid IVR 
for thermally generated reactive intermediates is not universally valid.  
The essential underlying reason is the `failure of ergodicity', a property which is 
notoriously difficult either to predict or to diagnose.  
In brief, the finding is that the physical mechanisms promoting IVR 
and those that lead to chemical reaction share enough common features that it is 
not feasible for them to occur on very different timescales, as the TST and RRKM models assume
\cite{Bunker62,Bunker64,Hase76,Hase98,Grebenshchikov03,Bach06,Lourderaj09,
Schofield94,Schofield95,Schofield95a,Leitner96,Leitner97a,Hase94,Gruebele04,Leitner06,Leitner08,Leitner11}. 
Hence, if an intermediate is energized in a nonstatistical fashion by the first step 
of a reaction, as will often be the case, 
then the nature of its excitation can have an influence on its subsequent chemistry.  
In other words, it is not just the total amount of energy available to the intermediate, 
but also the energy flow within it that can influence the subsequent reactions.  
The manifestations of this nonstatistical behavior include branching ratios and/or stereochemistries 
that differ significantly from statistical predictions, lack of temperature dependence of 
product ratios, and unusual intramolecular kinetic isotope effects
\cite{Carpenter03,Carpenter05,Rehbein11}.  

The range of thermal organic reactions now believed to manifest some kind of nonstatistical behavior 
is extraordinarily diverse (see, for example, refs. \onlinecite{Carpenter05} and \onlinecite{Rehbein11}, 
and references therein). 
A general characteristic shared by the systems for which the standard 
statistical theories fail is that the associated PES corresponds poorly if 
at all to the standard textbook picture of a one dimensional (1$d$) reaction coordinate 
passing over high barriers connecting deep wells (intermediates or reactants/products) 
\cite{Carpenter92,Carpenter05,Rehbein11}. 
More specifically, the reaction coordinate (understood in the broadest sense 
\cite{Heidrich95}) is inherently multidimensional, 
as are corresponding relevant phase space structures \cite{Waalkens08}.
In terms of the caldera, the multidimensional flatness of the caldera and the ``crossroads'' nature of its
intermediate suggest that any definition of reaction path through the caldera based on the 
potential topography has little connection to the dynamics of passage.

Other systems for which the standard 1$d$ reaction coordinate picture is not valid
include the growing class of so-called non-MEP (minimum energy path) reactions
\cite{Mann02,Sun02,Debbert02,Ammal03,Carpenter04,Lopez07,Lourderaj08,Lourderaj09} and ``roaming'' mechanisms
\cite{Townsend04,Bowman06,Shepler07,Shepler08,Suits08,Heazlewood08,Bowman11,Bowman11a,Mauguiere14,Mauguiere14a}; 
the dynamics of these reactions is not mediated by
a single conventional transition state associated with an index 1 saddle.

It is commonly the case
in reactions with caldera intermediates that the caldera is entered from the reactant via a higher energy transition
state and exits to the products from a lower energy transition state \cite{Hrovat97,Carpenter95}.  The caldera potentials in this
work contain reflectional symmetry, with a set of two higher energy saddle points and a set of two lower energy saddle
points equivalent under the reflection, see Figure \ref{fig:scheme1}.  In these cases, a statistical theory would predict that
any reaction flux entering either of the upper entrance transition states, or for that matter any flux entering via the lower energy
transition states, will result in a 1:1 ratio of product formation due to the reflectional equivalence of the product saddles.
However, in reality chemical reactions which possess caldera regions on their PES almost
never exhibit the expected symmetry in their product ratios \cite{Hrovat97,Carpenter95,Reyes00,Baldwin94}.  
For decades mechanistic organic chemists would explain away these discrepancies 
by postulating the existence of competing channels on the PES (commonly so-called concerted reactions) 
which would allow some molecules to react by pathways that avoided the 
caldera altogether (e.g., ref.\ \onlinecite{Gajewski96}). 
However, the advent of high-level electronic 
structure calculations has enabled testing of these hypotheses, 
and in many cases they have been found to be at odds with the computational results 
(e.g., ref.\ \onlinecite{Davidson97}).  
It has subsequently been recognized that additional features on the PES are unnecessary 
to explain the results.  Instead, one must abandon the rapid-IVR approximation of the 
statistical theories, and take account of the fact that the intermediate 
in the caldera region carries a dynamical ``memory'' of its origins 
-- a memory that can persist for a time comparable to the time required 
for product formation \cite{Carpenter85}.  
The ``memory'' is encoded in the tendency for momentum direction to be conserved, 
for trajectories traversing flat potential regions of the caldera .  
As we shall show below, in the simplest cases on the caldera, the aforementioned ``memory" is just a manifestation of the First Law of Motion, where the
inertia of the motion in a particular ``reaction" direction resists deflection from its course, and in more general caldera features 
it is possible that the shape of the caldera gives rise to deflections within the caldera
bowl such that the dynamical memory is maintained.
In other words, the reaction must be understood in a 
phase space rather than just a configuration space perspective.

There have 
been significant recent theoretical and computational advances in the 
application of dynamical systems theory  \cite{MacKay87,Lichtenberg92,Wiggins92,Arnold06} 
to study reaction dynamics and phase space structure in multimode models of 
molecular systems and to probe the
dynamical origins of nonstatistical behavior 
\cite{Wiggins90,wwju,ujpyw,WaalkensBurbanksWiggins04,WaalkensWiggins04,WaalkensBurbanksWigginsb04,
WaalkensBurbanksWiggins05,WaalkensBurbanksWiggins05c,SchubertWaalkensWiggins06,Waalkens08,Ezra09,Ezra09a,Collins11}
(see also refs
\onlinecite{Komatsuzaki00,Komatsuzaki02,Toda02,Komatsuzaki05,Wiesenfeld03,Wiesenfeld04,Wiesenfeld04a,Toda05,Gabern05,Gabern06,Shojiguchi08}).
A phase space approach is essential to obtain a rigorous 
dynamical definition of the transition state (TS) in multimode systems,
this being the Normally Hyperbolic Invariant Manifold (NHIM) \cite{Waalkens08}.
The NHIM generalizes the concept of the
periodic orbit (PO) dividing surface \cite{Pollak78,Pechukas81,Pechukas82}
to $N \geq 3$ mode systems.  The Poincar\'{e}\--Birkoff normalization theory has been implemented
as an efficient computational tool for realizing such phase space structures as the NHIM and the TS
dividing surface.  This method provides a rigorously defined phase space TS
that allows us to sample the caldera entrance channel on a surface with a local non\--recrossing
property and ensures that all possible caldera\--entry states are uniformly represented.  Therefore, to reiterate, all possible
momenta of entry into the caldera region will be represented in the sampling of 
initial conditions, a fact that is critical for correctly describing the
caldera dynamics and the ``dynamical matching" phenomenon we observe below. We note that a configuration space defined dividing 
surface is not guaranteed to have such desirable properties.
The constructions of NHIMs and associated phase space dividing surfaces
have been used in several recent contributions to the field of nonstatistical reaction dynamics.
In particular, a reappraisal of the \emph{gap time} formalism for 
unimolecular rates \cite{Ezra09a} has led to novel 
diagnostics for nonstatistical behavior (`nonexponential decay') in
isomerization processes, leading to a necessary condition for ergodicity, also see references \onlinecite{Quapp10} and \onlinecite{Aquilanti10}.

In the present paper, we study dynamics on model caldera potential energy surfaces. 
These potentials have reflection symmetry, and exit from the caldera 
region occurs via four index\--1 saddles appearing in symmetry-related pairs 
at two different energies. 
A computed normal form (NF) is used to sample the dividing surface at
fixed total energy, at one of the two saddle points on the left of the caldera.
The `incoming' TS is located at a point of high, or low, potential energy
depending upon the saddle point chosen.  
(For previous discussion of sampling using normal forms, see ref.\ 
\onlinecite{Collins11}.)
Bundles of trajectories so defined
are then followed into the caldera and the subsequent dynamics studied. 
We also study quantum dynamics on the caldera by examining the fate of wave packets
initiated at the saddle points.

The structure of this paper is as follows: in Sec.\ \ref{sec:model} we introduce the model 
system to be studied.  We describe the 
potential energy function used in our study and 
formulate the equations of motion used to calculate reaction
dynamics on our model surface and discuss the 
specification of initial conditions on the dividing surface.  
We also discuss the methodology used to calculate
the time evolution of quantum mechanical wave packets on the model potential.
Classical trajectory results are presented in Sec.\ \ref{sec:classical_results}:
we discuss the dynamics of trajectory bundles and product ratios at fixed energies.
The variation in the dynamics with the `stretching' of the potential (i.e., scaling one coordinate axis) is also
examined, and the classical results section concludes with a discussion of dissipation effects. 
Section \ref{sec:quantum_results} describes our calculations on
the dynamics of quantum wave packets on the caldera potential (stretched and unstretched).
Our computations reveal a quantum mechanical analogue of the `dynamical matching'
phenomenon, where the initial expectation value of the momentum 
direction for the wave packet determines the exit channel through which most
of the probability density passes to product.  
Sec.\ \ref{sec:summary} concludes.

\newpage

\section{Theoretical model and methods}
\label{sec:model}
In this section, we describe the model Hamiltonian used in our studies of 
dynamics on the caldera.
We introduce the potential energy function and discuss 
the sampling of initial conditions and the incorporation of dissipation.
We also discuss the methodology used to 
investigate the quantum dynamics of wave packets initiated 
in the vicinity of saddle points on the caldera.

\subsection{Model potential energy surface}
\label{sec:pes}
The model potential used in the present study describes 
a radially symmetric barrier surrounding a
central minimum, modulated by a term with a 4-fold symmetric angular dependence
leading to four symmetric index 1 saddles. Addition of a  linear
term $\propto y$ then yields two high energy saddles for $y > 0$ and two lower
energy saddles with $y < 0$. 
The potential function is:
\begin{subequations}
\label{eq:pot1}
\begin{align}
V(x, y) &=  c_1 r^2 + c_2 y - c_3 r^4 \cos[ 4 \theta] \\
& = c_1 (y^2 + x^2)  + c_2 y -  c_3 (x^4 + y^4 - 6 x^2 y^2)
 \end{align}
 \end{subequations}
where $r^2 = x^2 + y^2$, $\cos[\theta] = x/r$.
Potential parameters are taken to be $c_1 = 5$, $c_2 = 3$, $c_3 = -3/10$.

The potential \eqref{eq:pot1} is symmetric with respect to the reflection
operation $x \to -x$, so that critical points with $x \neq 0$ appear in
symmetrically related pairs.
Stationary points for potential \eqref{eq:pot1} are listed in Table \ref{tab:equi_pts}.

We also study a deformed or `stretched' version of the potential \eqref{eq:pot1},
where one of the coordinates ($x$) is scaled by a parameter $0 < \lambda \le 1$,
$x \to \lambda x$:
\begin{equation}
\label{eq:pot2}
V(x, y; \lambda) =  c_1 (y^2 + x^2 \lambda^2) + c_2 y - 
 c_3 (y^4 - 6 x^2 y^2 \lambda^2 + x^4 \lambda^4).
\end{equation}

\begin{figure}[H]
\begin{center}
\includegraphics[width=3in]{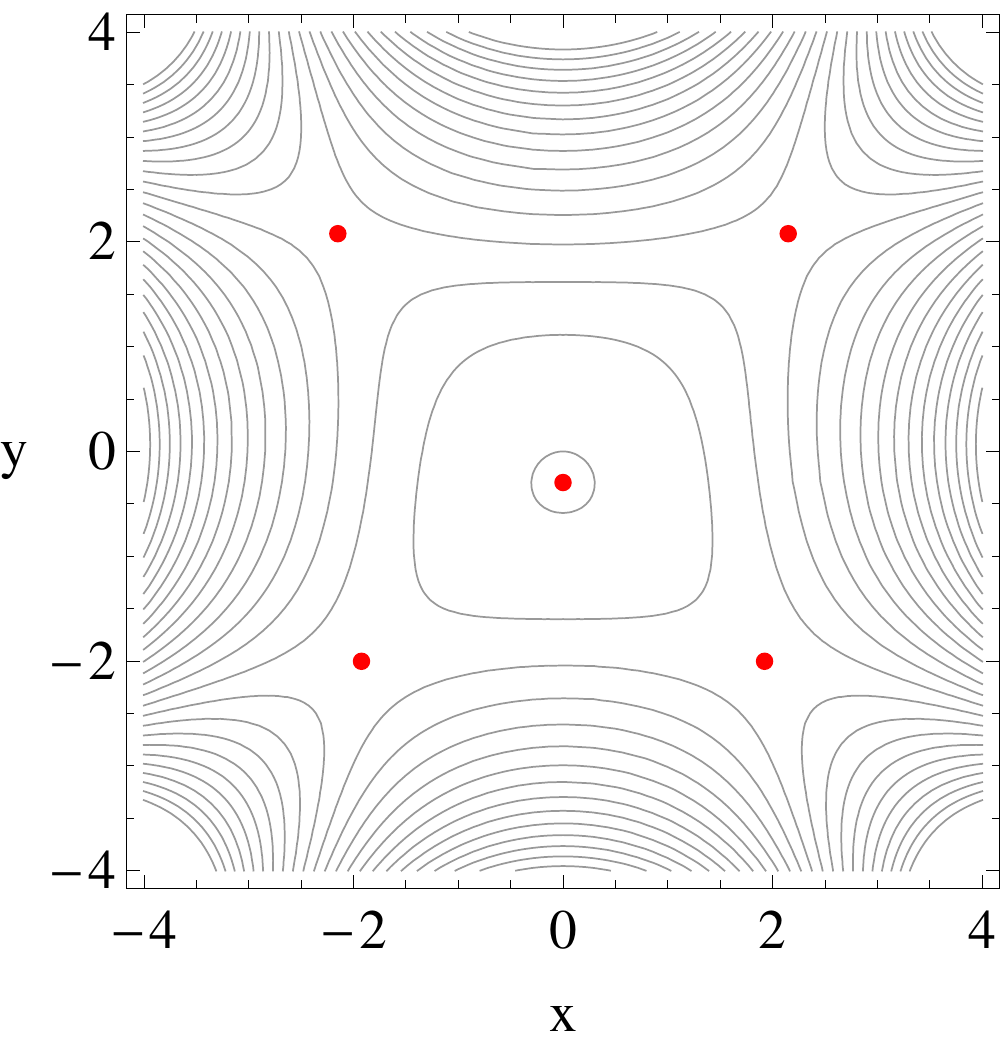}
\includegraphics[width=3in]{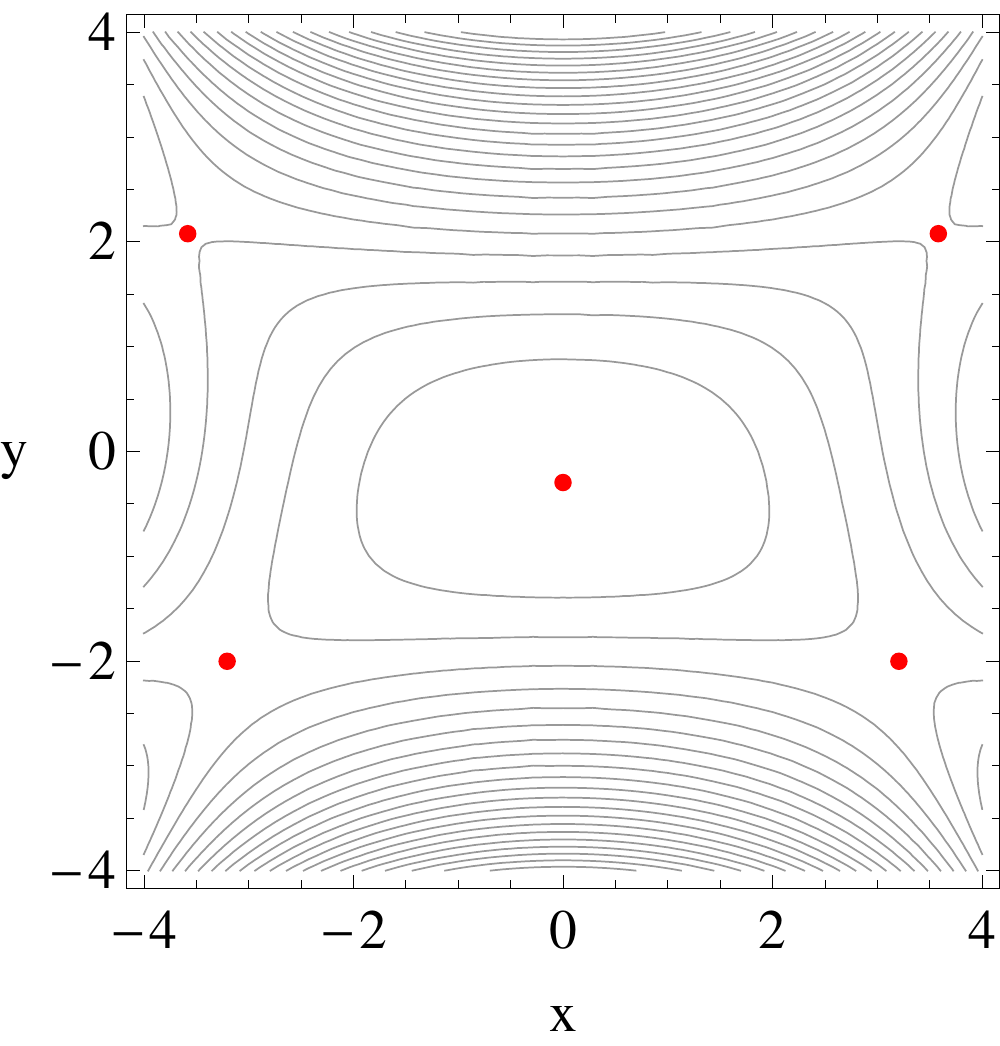}\\
\includegraphics[width=3in]{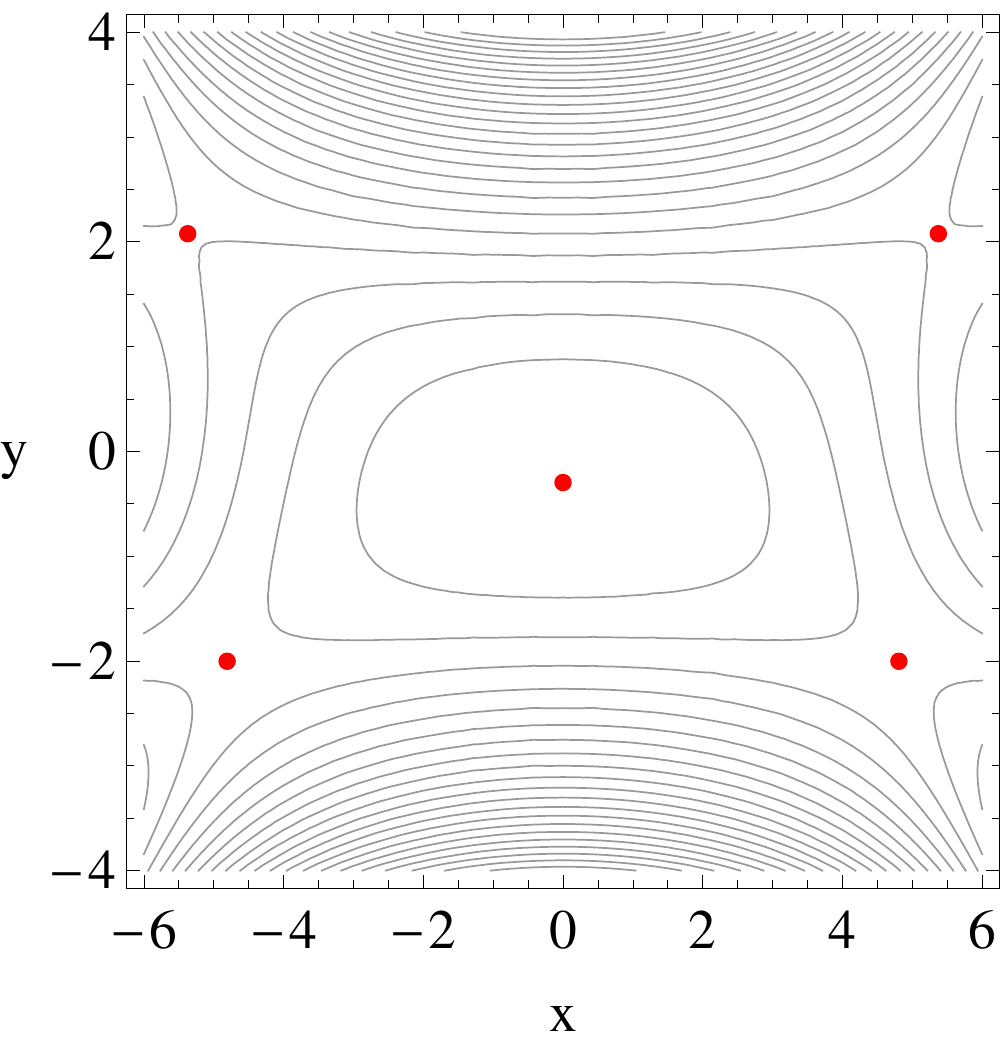}
\caption{Contours of potential energy surface $V(x, y)$, eq.\ \eqref{eq:pot2}.
(a)  Scaling parameter $\lambda = 1$. (b) Scaling parameter $\lambda = 0.6$.  (c) Scaling parameter $\lambda = 0.4$.}
\label{fig:pot1}
\end{center}
\end{figure}

The undistorted potential is recovered when $\lambda \to 1$.  Figure \ref{fig:pot1} shows contour plots of the potential:
$\lambda = 1$ (Fig.\ \ref{fig:pot1}a), $\lambda = 0.6$ (Fig.\ \ref{fig:pot1}b)
and $\lambda = 0.4$ (Fig.\ \ref{fig:pot1}c), respectively.

\subsection{Equations of motion}
The model Hamiltonian has the form:
\begin{equation}
\label{eq:ham1}
H(x, y, p_x, p_y) = \frac{p_x^2}{2 m} + \frac{p_y^2}{2 m} + V(x, y),
\end{equation}
with potential $V(x,y)$ given by eq.\ \eqref{eq:pot2}, and $m=1$.  
(For the physical values of mass and other physical parameters 
used in our quantum computations, see Sec.\ \ref{subsubsec:parameters}.)
Hamilton's equations of motion are:
\begin{subequations}
\label{eq:hameq1}
\begin{align}
\dot{x} & = \frac{p_x}{m},  \\ 
\dot{y} & = \frac{p_y}{m},  \\
\dot{p}_x & =  -\frac{\partial V}{\partial x} (x, y), \\
\dot{p}_y & =  -\frac{\partial V}{\partial y } (x, y).
\end{align}
\end{subequations}
The effects of dissipation are modelled by adding a simple damping term to the equations 
of motion \eqref{eq:hameq1} as follows:
\begin{subequations}
\label{eq:hameq_diss1}
\begin{align}
\dot{x} & =  \frac{p_x}{m}, \\ 
\dot{y} & =  \frac{p_y}{m}, \\
\dot{p}_x & =  -\frac{\partial V}{\partial x} (x, y)- \gamma_x p_x, \\
\dot{p}_y & =  -\frac{\partial V}{\partial y } (x, y) - \gamma_y p_y.
\end{align}
\end{subequations}
for some $\gamma_x, \, \gamma_y >0$, so that the
kinetic energy monotonically decreases along the trajectory.
We set  $\gamma_x= \gamma_y \equiv \gamma$
and study the effects of dissipation for a range of $\gamma$ values $0 \leq \gamma \leq 1$.
In the present calculations, random thermal fluctuations (e.g., Langevin dynamics \cite{Zwanzig01}) are not
considered.

\subsection{Trajectory ensembles and dividing surface sampling}
As in previous work \cite{Collins11,Collins13}, we calculate the NF 
at a given saddle and sample trajectory initial conditions 
on the associated dividing surface (with no local recrossing) as computed
using the NF, adjusting for inherent NF inaccuracies.
Specifically, we sample on a grid
in the $\{Q_2, P_2\}$ plane, where $Q_2$ and $P_2$ are
NF coordinates associated with the NF bath mode.
We then use the inverse transformation provided by the NF computation
to transform these points into the physical coordinates $\{x, y, p_x, p_y\}$.

\subsection{Wave packet dynamics in the caldera}
In order to augment the classical trajectory 
analysis presented in this work with a heuristic understanding of the 
quantum effects inherent in the caldera dynamics,
we have propagated quantum wave packets on the caldera potentials.

\subsubsection{Potential parameters and units}
\label{subsubsec:parameters}
To present the wave packet dynamics on a chemically relevant  
model system, the potentials used
in the classical analysis have been scaled 
so that the unit of energy corresponds to 0.20 kcal mol$^{-1}$.  
The scaling sets the energies of the upper and lower saddle 
points to 5.49 kcal mol$^{-1}$ and 3.04 kcal mol$^{-1}$, respectively, 
relative to the caldera minimum.  
In order to test
how the wave packet dynamics depended on the scaling factor, some trial wave packet calculations were 
performed on potentials with reduced 
scaling, i.e. further flattening the caldera. It was found that the observed dynamics did not contradict 
the results we present here and our choice of scaling seems qualitatively representative over
a range of possible energy scaling factors.
Our choice places the potential 
in an energy regime of chemical interest based on comparison 
to previous caldera features that have been found to be 
embedded in the 
potentials of a number of organic reactions \cite{Carpenter13}. 

The general form of the Hamiltonian of the caldera system is given 
in eq.\ \eqref{eq:ham1}.
For the quantum computations the mass of the one-particle system 
is set equal to an atomic mass, 
and wave packets corresponding to particles both of the mass of $^{12}$C 
and of the mass of $^{1}$H are considered in order to illustrate mass effects.  
The units of distance are taken to be atomic units (bohr), so that 
the locations of the potential features remain unchanged 
from those given in Table \ref{tab:equi_pts}.  
The time units used in the classical simulations can be related to the quantum simulations
given the scaling factor and the particle mass:  specifically, each time unit of the classical
simulations corresponds to 200 fs (58 fs) in the $^{12}$C ($^1$H) particle quantum wave 
packet propagations.

Since the wave packet is represented on a finite grid in configuration space
subject to periodic boundary conditions, any wave packet amplitude that passes 
out through any of the transition states to the regions of decreasing potential,
i.e. passes out the caldera and ``down the mountain", will reappear on the opposite side of the grid.  
That is, rather than escaping the caldera, the wave packet will be subject 
to an unphysical periodicity.  
In order to avoid this effect, a (linear) negative imaginary 
potential (NIP) \cite{Sathyamurthy,Baer} of the form 
\begin{equation}
\label{eq:z_wp_7}
u\left(x,y\right) = \left\{
   \begin{array}{ll}
   0 & \quad \left(x,y\right) \in \mathbb{S}_{{\textnormal{Caldera}}} \\
   -i U_0 \left( \frac{R_{\perp}\left(x,y\right)}{R_{\perp_{max}}}\right) & 
   \quad \left(x,y\right) \notin \mathbb{S}_{{\textnormal{Caldera}}} \\
   \end{array}
   \right.
\end{equation}
is used to absorb the outgoing components of the wave packet. 
The NIP is non-vanishing in the regions of the potential outside the caldera 
and past the nominal transition state dividing surfaces defined by the saddle 
points.
The caldera region is denoted in eq.\ \eqref{eq:z_wp_7} by 
$\mathbb{S}_{{\textnormal{Caldera}}}$, 
and the function $R_{\perp}\left(x,y\right)$ represents the shortest, 
equivalently perpendicular, 
distance from the point $\left(x,y\right)$ to the inner boundary, 
i.e. the boundary not defined by the edges of the 2$d$ grid, of the NIP.  
The constant value $R_{\perp_{max}}$ is the perpendicular distance 
from the corner of the 2$d$ grid to the inner boundary of the NIP.  
In this study, only simple linear boundaries are used to define $\mathbb{S}_{{\textnormal{Caldera}}}$.  
The real, positive factor $U_0$ determines 
the magnitude of the NIP. 

\subsubsection{Initial wave packet construction \& representation}
The wave packets are initiated at the upper saddle point in the 2$^{{\textnormal{nd}}}$ 
quadrant of the $xy$\--plane.  To construct the wave packet, 
a normal mode analysis (2$^{\textnormal{\tiny{nd}}}$ order normal form) is performed on the system at this saddle point, 
resulting in one normal mode corresponding to an imaginary frequency,
the ``reactive" mode, defining coordinate $Q_r$, and another with a positive real frequency, the ``bath" mode, 
defining coordinate $Q_b$.  
The saddle point is located at $(Q_r = 0, Q_b = 0)$.  

The initial wave packet is constructed in normal mode coordinates in product form,
\begin{equation}
\label{eq:z_wp_1}
\Psi^{t=0}_{{\textnormal{nm}}}(Q_r,Q_b) = \phi_r \left(Q_r\right) \phi_b\left(Q_b\right)
\end{equation}
The first factor on the right hand side (rhs) of eq.\ \eqref{eq:z_wp_1} is of gaussian form,
\begin{equation}
\label{eq:z_wp_2}
\phi_r(Q_r) = \frac{1}{\sqrt{ \Delta_r \sqrt{2 \pi}}}  e^{-\frac{Q_r^2}{4 \Delta_r^2}+i \frac{P_r}{\hbar}}
\end{equation}
The wave packet factor has two parameters: $\Delta_r$ defines the width of the gaussian 
in the reactive mode direction and $P_r$ defines an expected momentum.  
In order to more intuitively construct the reactive component of the wave packet, 
the parameters $\delta_r$ and $p_r$, the width and expected momentum 
in the reactive mode direction in cartesian coordinate system ($x$,$y$), respectively, 
can be used to define $\Delta_r$ and $P_r$,
\begin{subequations}
\begin{align}
\label{eq:z_wp_3}
\Delta_r & = \delta_r \sqrt{m_x {(\nu^{\eta_x}_r)}^2 + m_y {(\nu^{\eta_y}_r)}^2}\\
\label{eq:z_wp_4}
P_r & = p_r  \left( {(\nu^{\eta_x}_r)}^2 \sqrt{  \frac{ m_y / m_x }{m_x {(\nu^{\eta_y}_r)}^2 + m_y {(\nu^{\eta_x}_r)}^2 }  }  + {(\nu^{\eta_y}_r)}^2 \sqrt{  \frac{ m_x / m_y }{m_x {(\nu^{\eta_y}_r)}^2 + m_y {(\nu^{\eta_x}_r)}^2 } }  \right)
\end{align}
\end{subequations}
where $m_x$ and $m_y$ are the masses in the $x$ and $y$ directions 
and $\left[\nu^{\eta_x}_r, \nu^{\eta_y}_r\right]$ ( $\left[\nu^{\eta_x}_b, \nu^{\eta_y}_b\right]$) 
is the normalized reactive (bath) normal mode eigenvector in the 
mass-weighted cartesian coordinate system ($\eta_x=\sqrt{m_x}x$,$\eta_y=\sqrt{m_y}y$).  
Note that, as in the cases treated here, when $m_x=m_y=m$ the form of the above equations 
can be significantly simplified to
\begin{subequations}
\begin{align}
\Delta_r & = \sqrt{m} \delta_r \\
P_r & = p_r / \sqrt{m} .
\end{align}
\end{subequations}

The $\phi_b\left(Q_b\right)$ factor on the rhs of Equation \ref{eq:z_wp_1} is 
chosen to be an eigenstate of a ``local'' one-dimensional ($1d$) Hamiltonian  $\hat{h}_b $ of 
the bath normal mode coordinate, 
\begin{equation}
\label{eq:z_wp_5}
\hat{h}_b  \psi_v \left( Q_b \right)  = 
\left[ \tfrac{1}{2}\hat{P_b}^2 + \hat{V}\left(Q_r=0,Q_b \right) \right] \psi_v \left( Q_b \right)
= \epsilon_v \psi_v \left( Q_b \right)
\end{equation}

At distances far from the saddle point ($Q_b$ large), the potential $V\left(Q_r=0,Q_b \right)$ 
might ``turn over''.  Indeed, values of the potential below the saddle point may lie 
along the $Q_b$ contour.  
However, only the portion of $V\left(Q_r=0,Q_b \right)$ in a neighborhood of the 
saddle point where the potential is increasing as a 
function of distance from the saddle is considered in the construction of $\hat{h}_b$, 
i.e. the $1d$ Hamiltonian is constructed locally in the vicinity of the saddle point in the 
Fourier grid representation \cite{Tannor}.  
For the system parameterizations used in this work, 
the region of the potential $V\left(Q_r=0,Q_b \right)$ 
is able to support several, quite localized, bound states significantly 
below the values of the potential at the endpoints of the locally defined region.  
The $\psi_v \left( Q_b \right)$ can be taken as an approximation 
to the state of the bath degree of freedom that 
would result if the motion along the reaction coordinate to the 
saddle point region proceeded adiabatically and tunneling 
through any barrier along the $Q_b$ contour is unimportant.  
As will be seen in the results that follow, tunneling through
potential barriers along the $Q_b$ contour on the potentials considered 
is not observed in the initial dynamics of the wave packets which proceed directly into the caldera region.

Since the normal mode coordinates are rectilinear with 
respect to the mass\--weighted cartesian coordinates, the initial wave packet is
readily expressed in cartesian coordinates:
\begin{multline}
\label{eq:z_wp_6}
\Psi^{t=0}_{{\textnormal{xy}}}(x,y) =  \Psi^{t=0}_{{\textnormal{nm}}}
(\sqrt{m_x}\left(x-x_{s}\right)\nu^{\eta_x}_r+\sqrt{m_y}\left(y-y_s\right)\nu^{\eta_x}_r,\\ 
\sqrt{m_x}\left(x-x_{s}\right)\nu^{\eta_x}_b+\sqrt{m_y}\left(y-y_s\right)\nu^{\eta_x}_b) 
\left( m_x m_y \right)^{\frac{1}{4}} 
\end{multline}
where $\left(x_s,y_s\right)$ is the position of the upper saddle point.  
The 2$d$ wave packets therefore depend on the three parameters $\{ \delta_r, p_r, v\}$.  
For all wave packets considered, $p_r$ is chosen so that the wave 
packet has an initial momentum directed from the saddle point into 
the caldera region and $v$ is set to zero, the local ground state of the bath degree of freedom.

\subsubsection{Propagation algorithm}
The wave packets are propagated using an implementation of the split operator (SO) method accurate to third order in the time step \cite{Bandrauk,Sathyamurthy}.  In this method, the time evolution operator is decomposed into an approximate sequence of kinetic and potential time propagators,
\begin{equation}
\label{eq:z_wp_8}
e^{-i\frac{\hat{H} \Delta t}{\hbar}} \approx  e^{-i \Gamma \frac{\hat{T} \Delta t}{2 \hbar}} e^{-i \Gamma \frac{\hat{V} \Delta t}{\hbar}} e^{-i \left(1-\Gamma \right) \frac{\hat{T} \Delta t}{2 \hbar}} e^{-i \left(1-2\Gamma\right) \frac{\hat{V} \Delta t}{\hbar}} e^{-i \left(1-\Gamma \right) \frac{\hat{T} \Delta t}{2 \hbar}} e^{-i \Gamma \frac{\hat{V} \Delta t}{\hbar}} e^{-i \Gamma \frac{\hat{T} \Delta t}{2 \hbar}} 
\end{equation}
where $\Gamma=1/(2-2^{\frac{1}{3}})$.  
The potential evolution operators are diagonal in the $\left(x,y\right)$ 
representation and can, therefore, be calculated on the $\left(x,y\right)$ grid.  
The action of the potential evolution operators can then be directly applied to the 
wave packet also expressed in the coordinate representation on the grid.  
While the kinetic evolution operators are non-local in the coordinate 
representation, the ``kinetic plus potential" form of the system Hamiltonian results 
in these operators being diagonal in the momentum, $\left(p_x,p_y\right)$, representation.  
Therefore, application of the kinetic evolution operators is efficiently performed by 
first transforming the wave packet on the coordinate grid to the momentum, 
or $k$\--space, representation using a discrete Fourier transform (DFT) and 
then applying a diagonal kinetic evolution operator in the momentum, or the $k$, representation.  
The evolved wave packet can then be expressed on the coordinate grid using an inverse DFT.  
The time step, $\Delta t$, used to propagate the wave packets 
was chosen to be 4 atomic units of time ($\sim 0.10$ fs), 
which was found sufficient to yield well-converged results.

\newpage

\section{Results: classical mechanics}
\label{sec:classical_results}
In this section we investigate, using classical trajectories,
reaction dynamics on a model potential surface exhibiting 
a caldera.  Entry into and escape from the caldera occurs via 
four index 1 saddles, which occur as symmetry-related pairs at two
different energies.

For the 2-fold symmetric potential under study,  `statistical' behavior
of the ensemble would imply that an ensemble of trajectories initiated on a given dividing surface
should lead to equal numbers of left/right symmetry related `products'.
However, as discovered by Carpenter, reaction dynamics on the caldera
tends to be highly nonstatistical \cite{Carpenter85,Carpenter95,Carpenter05}.  
In fact it is found that many trajectories exhibit `direct'
dynamics where the associated product
is determined by the momentum  \emph{direction} 
at the dividing surface entering the caldera \cite{Carpenter95}. 

For the first part of our study of classical dynamics on the caldera, 
trajectories are initiated on the 
dividing surface associated with one of the pair of
symmetry equivalent higher energy saddles, the upper lefthand (LH) and upper
righthand (RH) saddles.  
Any of four products (exit channels) can in principle be accessed.
For the second  part of the study,  trajectories are initiated 
on the dividing surface associated with the lower energy saddle, and 
some products are classically inaccessible due to
energy constraints. 
In the third part, we consider dynamics on the `stretched' potential,
where the $x$ coordinate is scaled by a distortion factor $\lambda$, $0 < \lambda <1$. 
Finally, we examine the effects of dissipation on the dynamics.
For consistency, we shall always choose our entrance TS to be on the lefthand side.

\subsection{Higher energy entrance transition state}
\label{ssec:hes}
\begin{figure}[htbp!]
\begin{center}
\hskip 0.25cm
\subfigure[]{\includegraphics[width=0.335\linewidth]{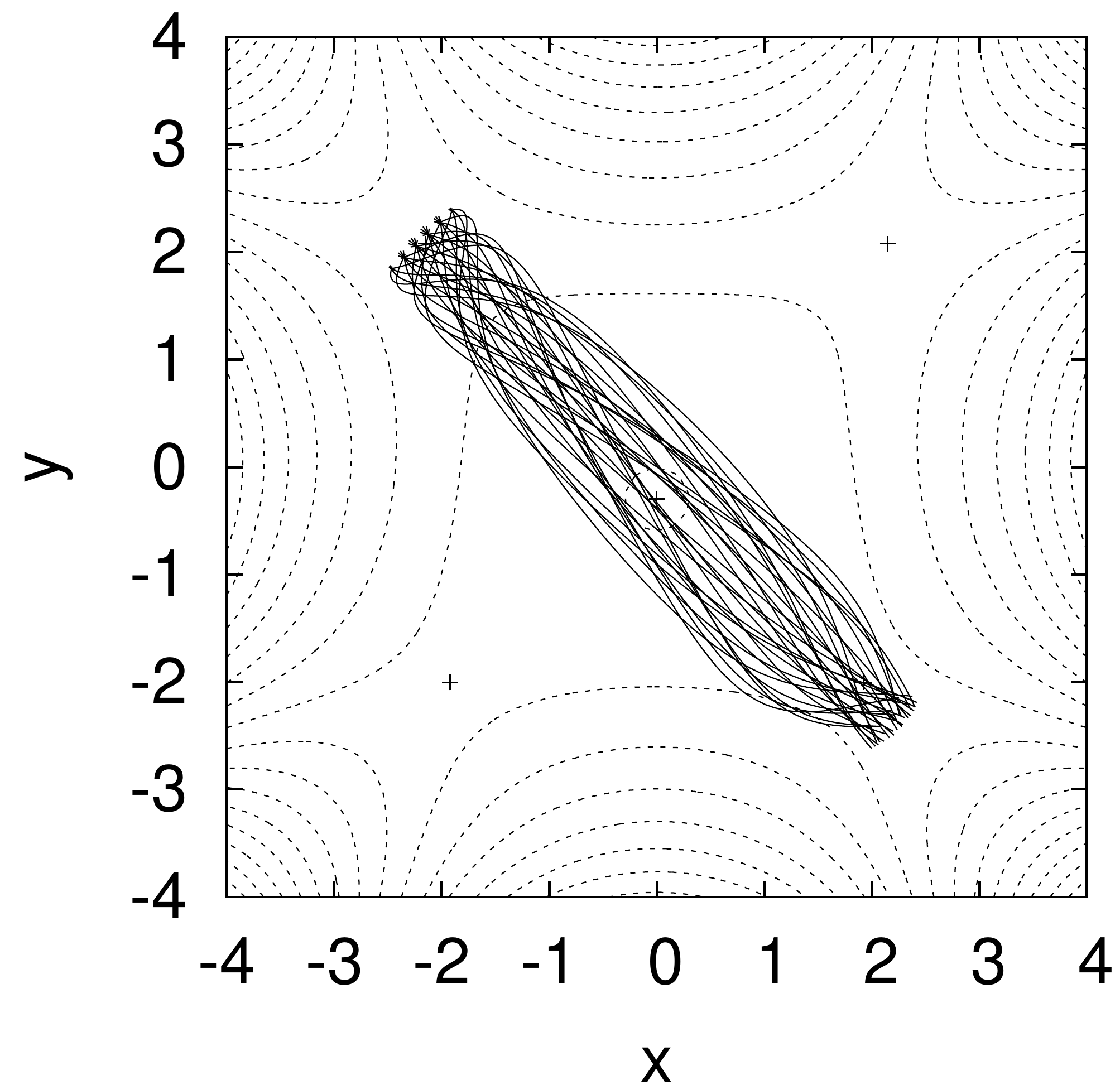}}
\hskip 0.25cm
\subfigure[]{\includegraphics[width=0.35\linewidth]{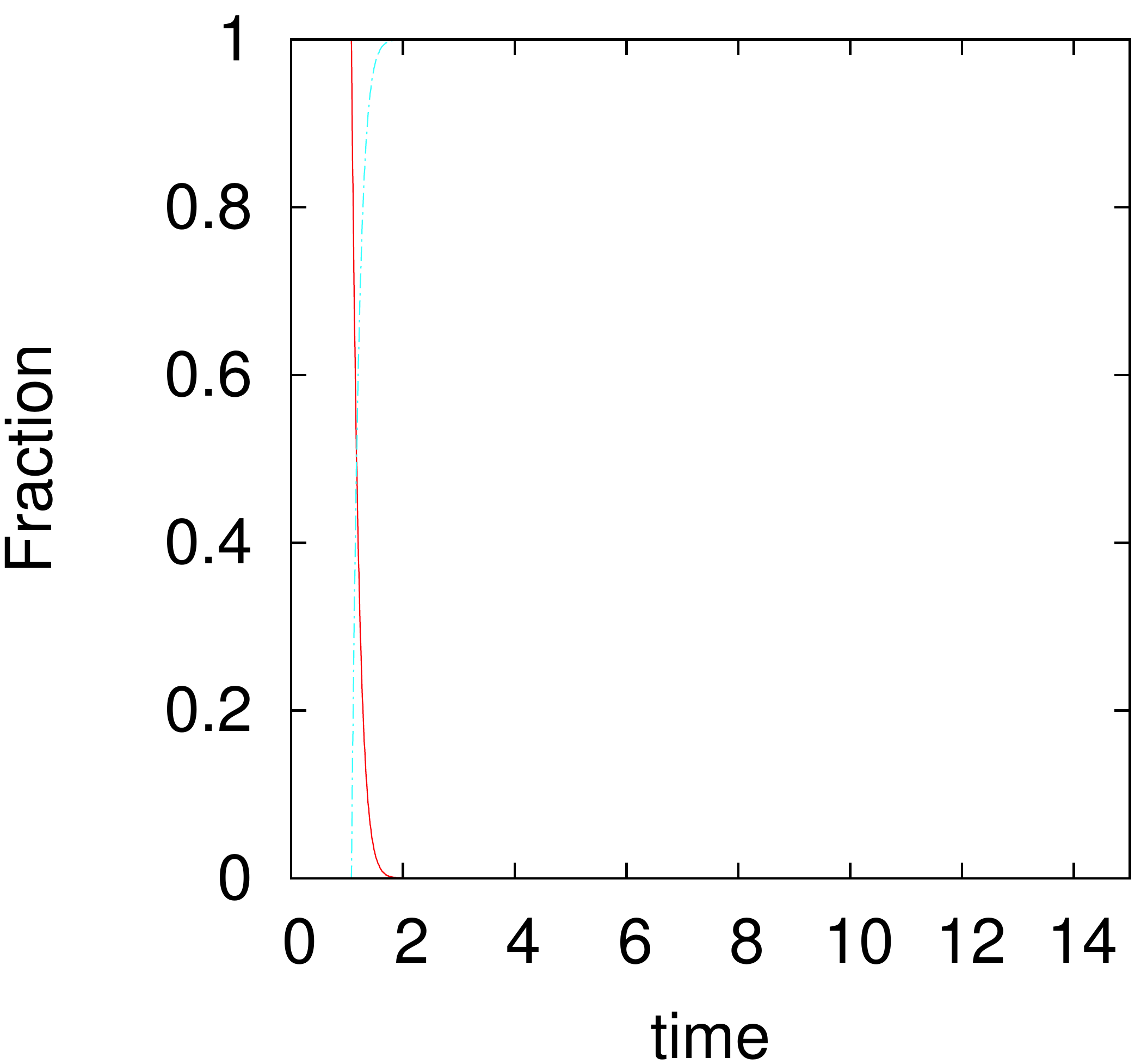}}
\hskip 4.5cm\\
\hskip 0.25cm
\subfigure[]{\includegraphics[width=0.35\linewidth]{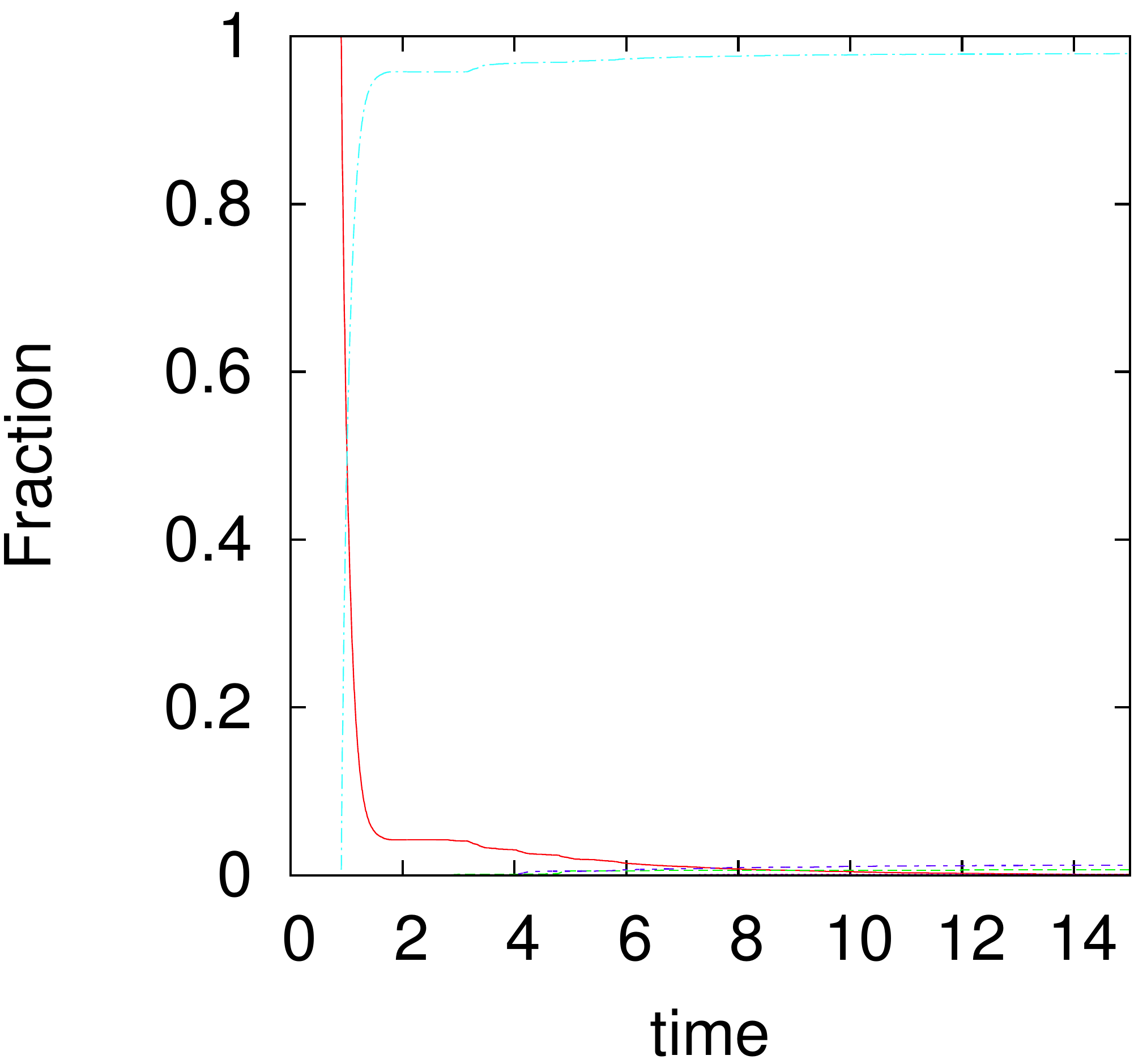}}
\hskip 0.25cm
\subfigure[]{\includegraphics[width=0.35\linewidth]{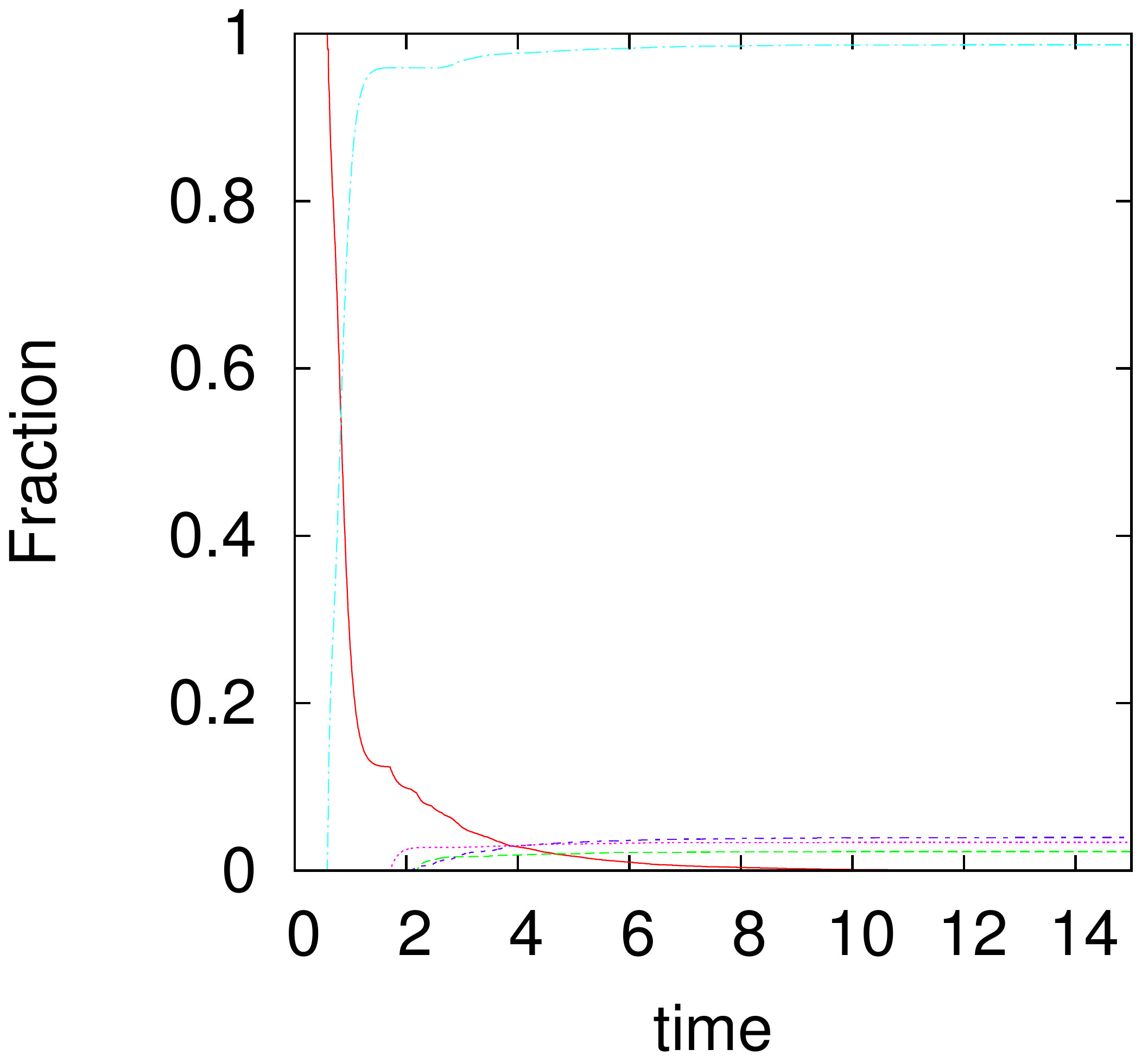}}
\caption{All panels are for trajectories initiated on dividing surface at upper LH saddle ($x < 0$, $y > 0$) at energy $E$ above the saddle.
In (a), initial low excitation energy, $E$=5, trajectories traced through configuration space.  Dynamical matching is evident.
In (b), (c), and (d), cumulative fractions of products versus propagation time are shown for $E=5$, $E=15$, and $E=30$, respectively.
Color key: fraction of trajectories remaining in well (red); fraction exiting lower LH saddle (blue);
 fraction exiting lower RH saddle (cyan); fraction exiting upper RH saddle (magenta);
 fraction exiting top LH saddle (green).}
\label{fig:r1}
\end{center}
\end{figure}

An ensemble of trajectories is initiated at fixed energy $E$ 
above the energy of the upper lefthand (LH) ($x<0$, $y >0$) saddle.
A few of the
sampled trajectories for $E=5$ are plotted in Fig.\ \ref{fig:r1}(a),
together with a plot of the time dependence of the
cumulative fractions of trajectories having exited 
the central well by crossing one of the saddles (Fig.\ \ref{fig:r1} (b)).
It can be seen that, as found originally by Carpenter \cite{Carpenter85,Carpenter95},
all trajectories in the sampled ensemble go straight across the
caldera and exit via the lower righthand (RH) saddle. This result is a manifestation of ``dynamical matching" \cite{Carpenter95};  the 
direction of the momentum at the dividing surface essentially determines
the outcome (product channel) for the reaction. There must be a small
fraction of trajectories having initial conditions close to the NHIM, which
do not show dynamical matching but these do not appear in our samples until
the energy is higher.

\begin{figure}[htbp!]
\begin{center}
\subfigure[]{\includegraphics[width=0.35\linewidth]{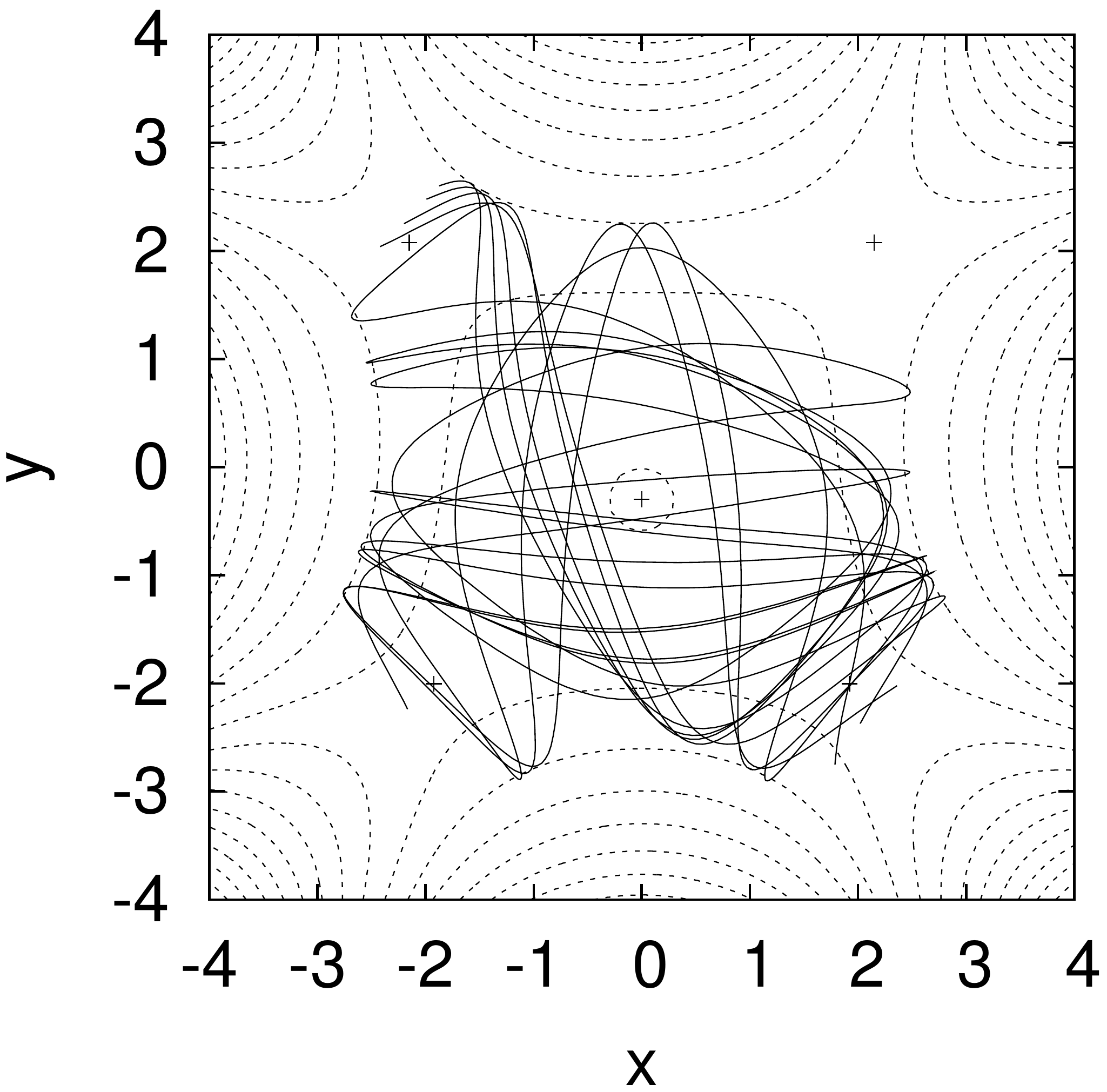}}
\hskip 0.25cm
\subfigure[]{\includegraphics[width=0.35\linewidth]{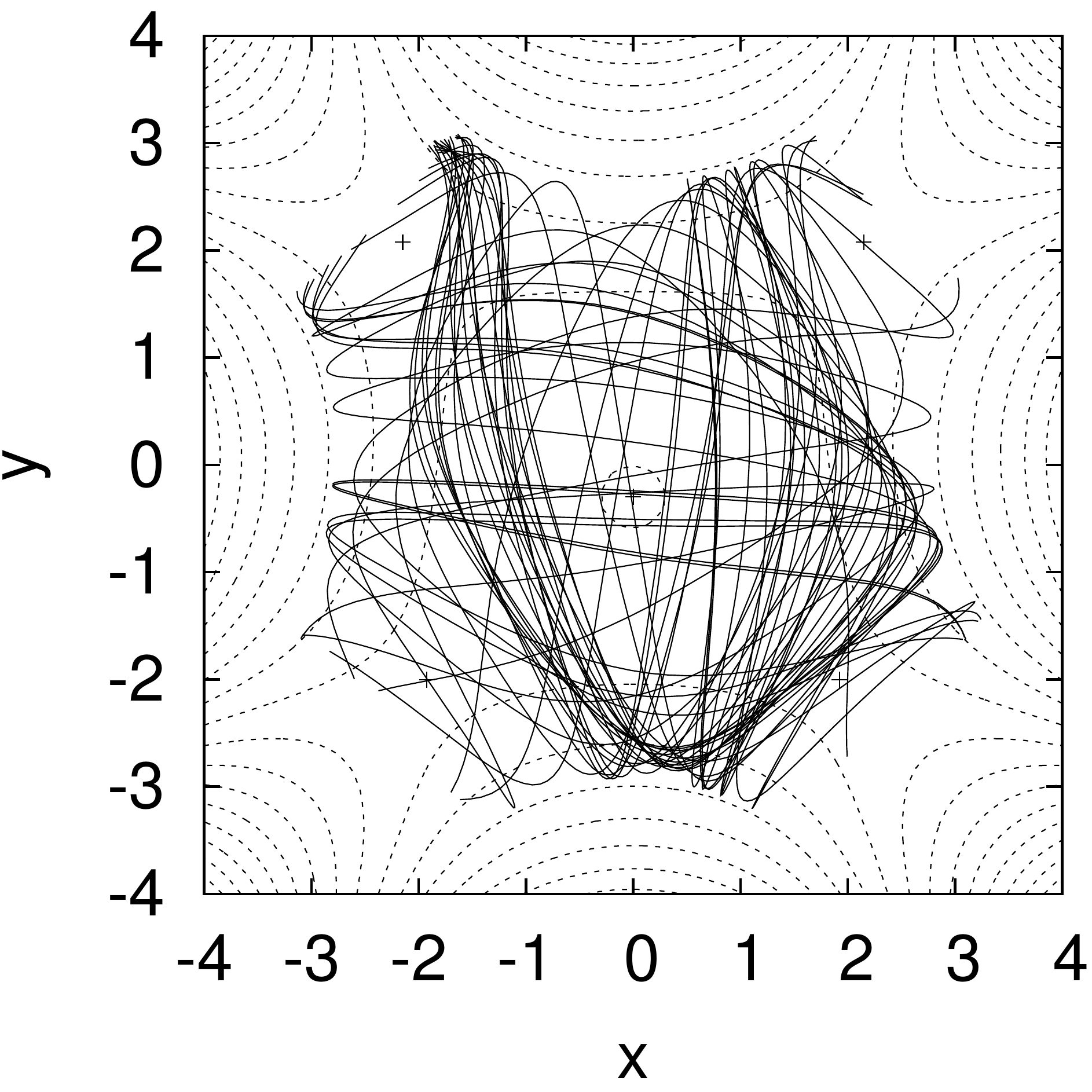}}
\caption{All panels are for trajectories initiated on dividing surface at upper LH saddle ($x < 0$, $y > 0$) at energy $E$ above the saddle.
In (a), for initial excitation energy $E=15$, the few trajectories that do not experience dynamical matching are shown.  Panel (b)
is the same as (a) except $E=30$.}
\label{fig:r1_2}
\end{center}
\end{figure}

Cumulative product fractions over time are shown for $E=15$ and $E=30$ initiated above the top LH saddle,
in panels (c) and (d) of Figure \ref{fig:r1}, respectively.  At  these higher energies
we start to see trajectories (those initiated 
near the `edge' (NHIM) of the dividing surface and thus have only a small momentum directed along the reaction coordinate, i.e.
the unstable saddle degree of freedom) 
which do not traverse
the caldera and exit immediately but instead are reflected back into the potential well.
Examples of such trajectories which do not exibit dynamical matching are shown
in Fig. \ref{fig:r1_2}, Note that such trajectories are a
small fraction of the total, see Fig. \ref{fig:r1} in panels (b), (c), and (d).
These trajectories exhibit `indirect' dynamics, and
a few trajectories exit via saddles other than the lower RH saddle at times
longer than the 2 time units which is roughly the limit of the transit time
for direct trajectories.
Nevertheless, branching ratios, as determined  from the long\--time limits of
the cumulative product yields, are still by no means statistical, as seen in Figs. \ref{fig:r1}(b), (c), and (d).

To conclude this section, the degree of dynamical matching in our model system decreases as the initial excitation
energy of the trajectory ensemble increases.  This effect can be explained by noting that at low $E$, the momentum imparted by the forces from the 
caldera potential on the
trajectory after it crosses the saddle dwarf the momentum in the bath degree of freedom, essentially eliminating any chance of deflection of the trajectory
during the passage through the caldera.  In other words, at lower $E$ not enough energy is available relative to what is imparted on the fall
into the caldera to alter the course of the trajectory.  At higher excitation energies, excess energy is available for  
the ``bath" degree of freedom, that parallel to the configuration space projection of the dividing surface, so that trajectories can enter the caldera with a momentum direction that does not
align with the diametrically opposed, dynamically\--matched, transition state.  Such trajectories are like those shown in Fig. \ref{fig:r1_2}(a) and (b).
The relative population of such `indirect' trajectories is anticipated to grow as the caldera entrance energy is raised further, however, since caldera intermediates tend to be of high potential energy, for most thermal reactions
entrance into the caldera is likely to occur with relatively low kinetic energy due to the bias of the Boltzmann factor.  In 
this sense, the results we present above are particularly relevant.

\subsection{Lower energy entrance transition state}
\label{ssec:lee}
\begin{figure}[htbp!]
\begin{center}
\hskip -2.0cm
\subfigure{\includegraphics[width=0.5\linewidth]{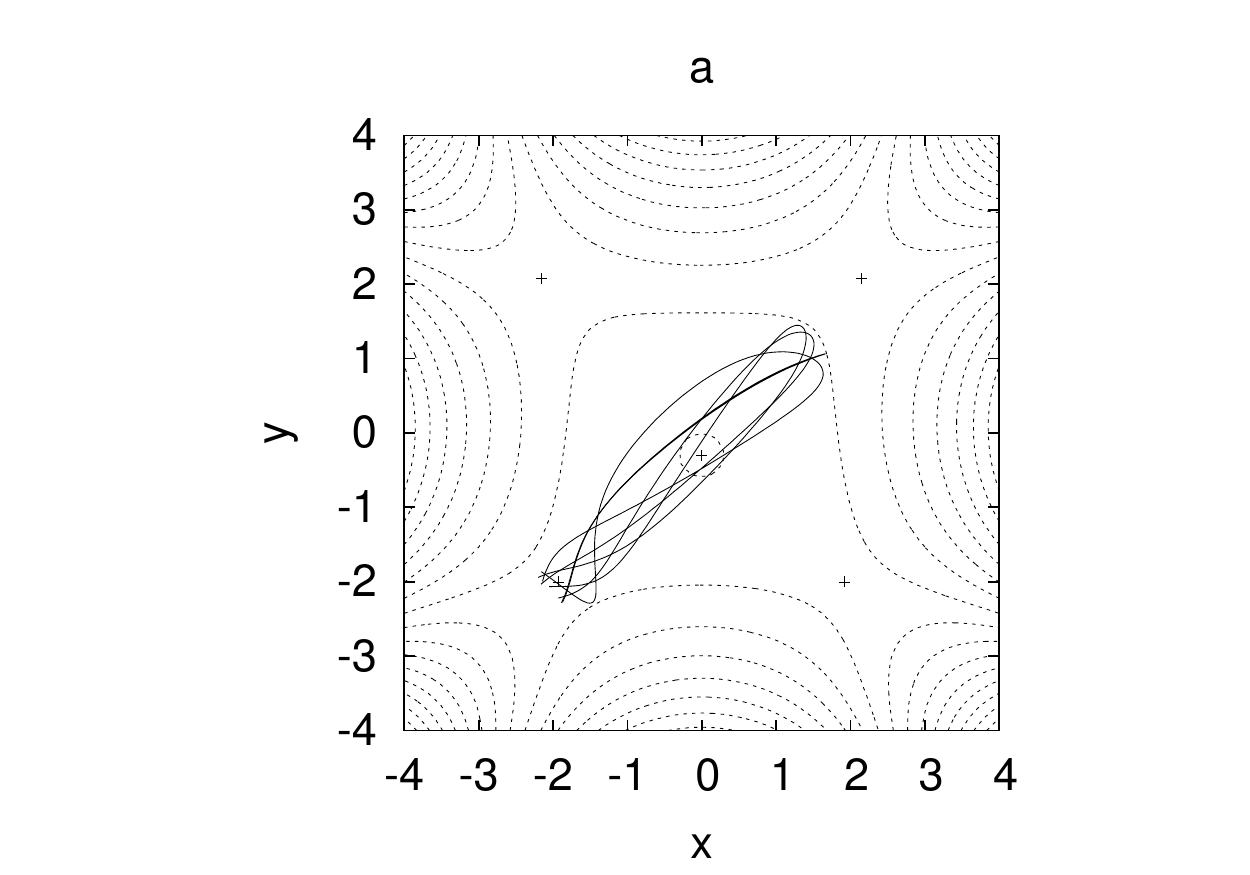}}
\hskip -2.0cm
\subfigure{\includegraphics[width=0.5\linewidth]{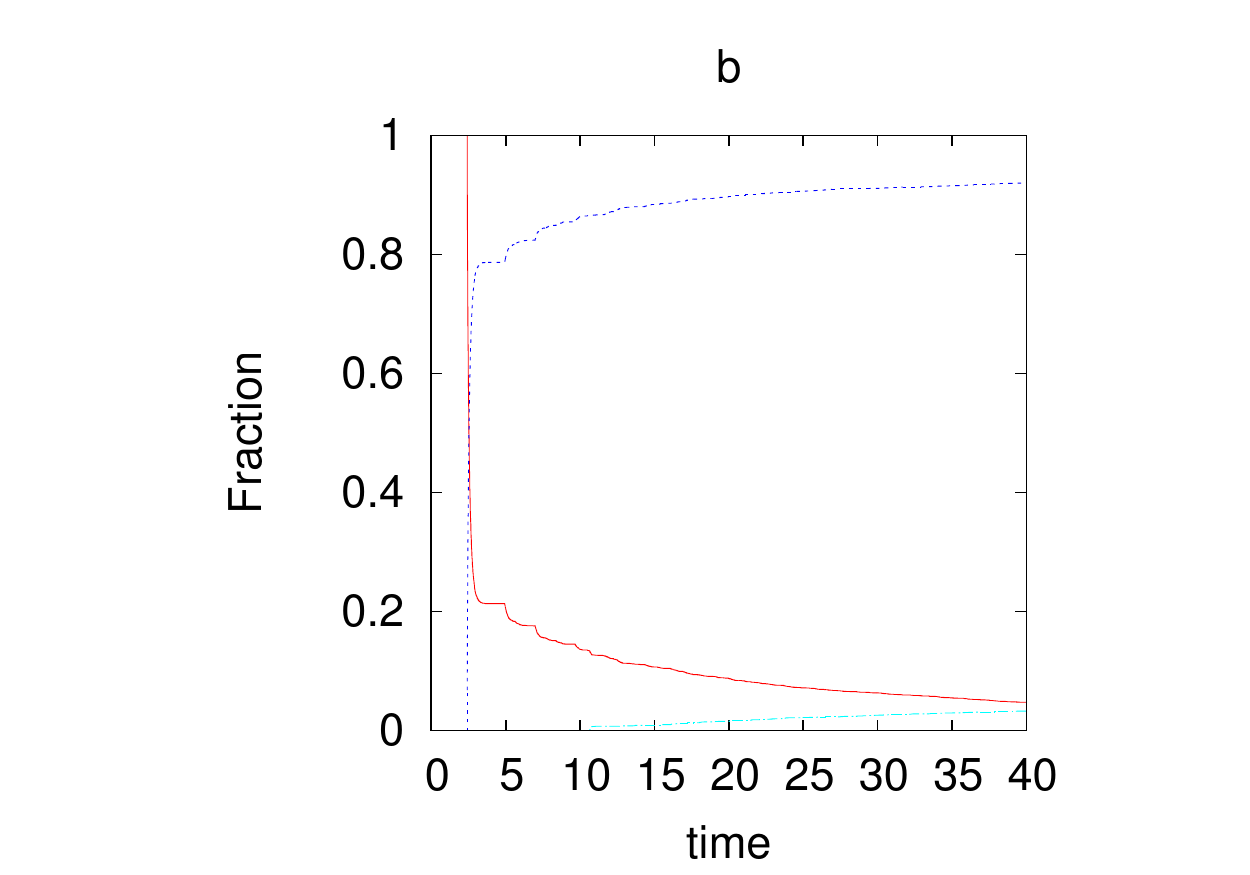}}
\hskip -5.5cm\\
\hskip -2.0cm
\subfigure{\includegraphics[width=0.5\linewidth]{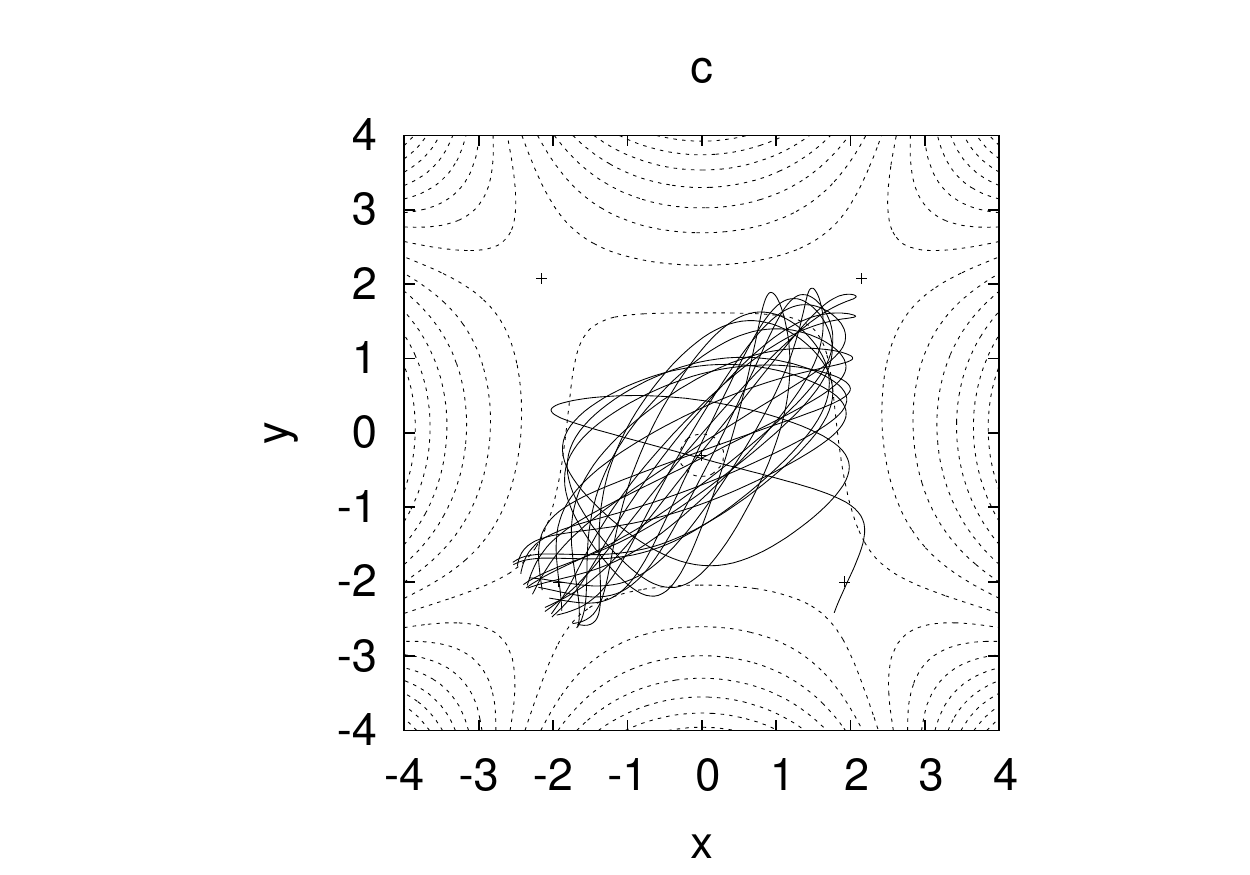}}
\hskip -2.0cm
\subfigure{\includegraphics[width=0.5\linewidth]{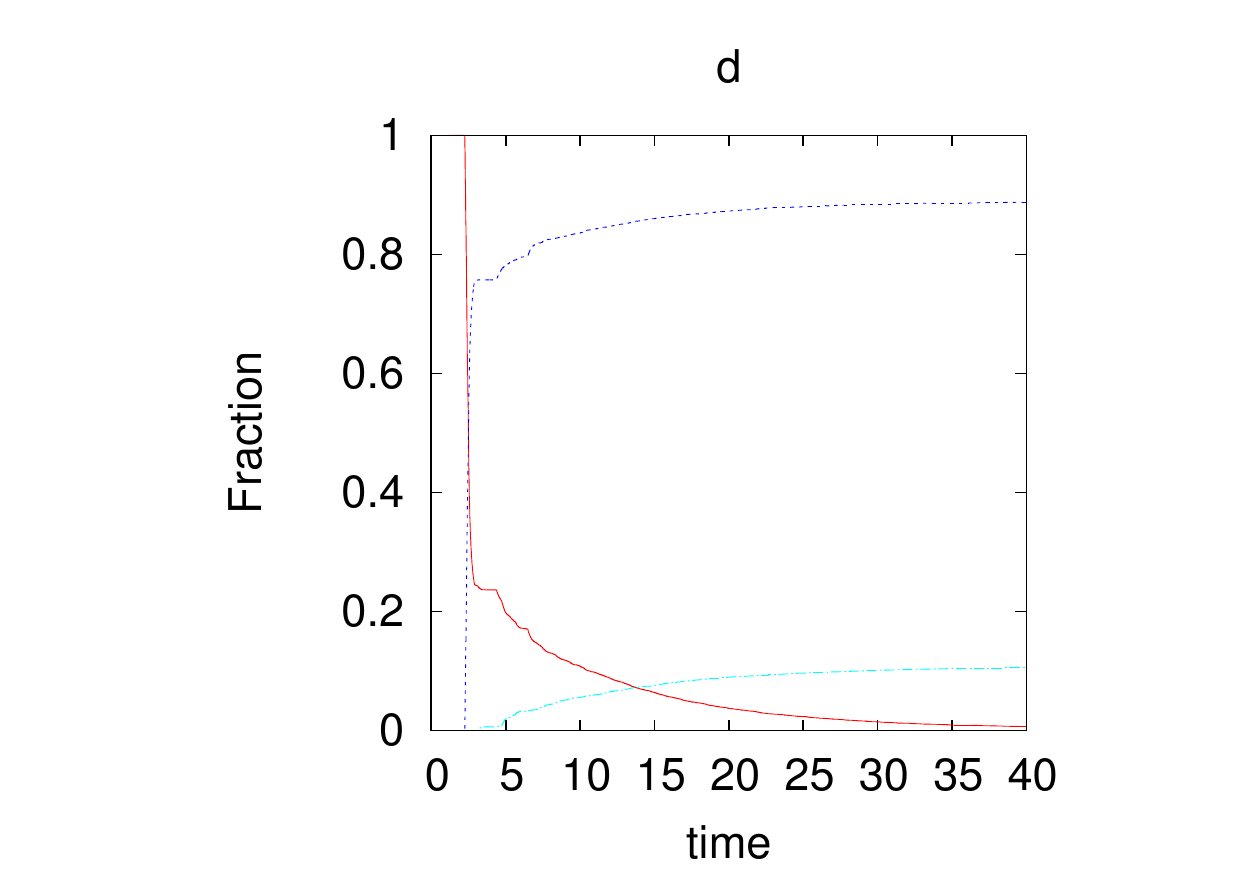}}
\hskip -5.5cm\\
\hskip -2.0cm
\subfigure{\includegraphics[width=0.5\linewidth]{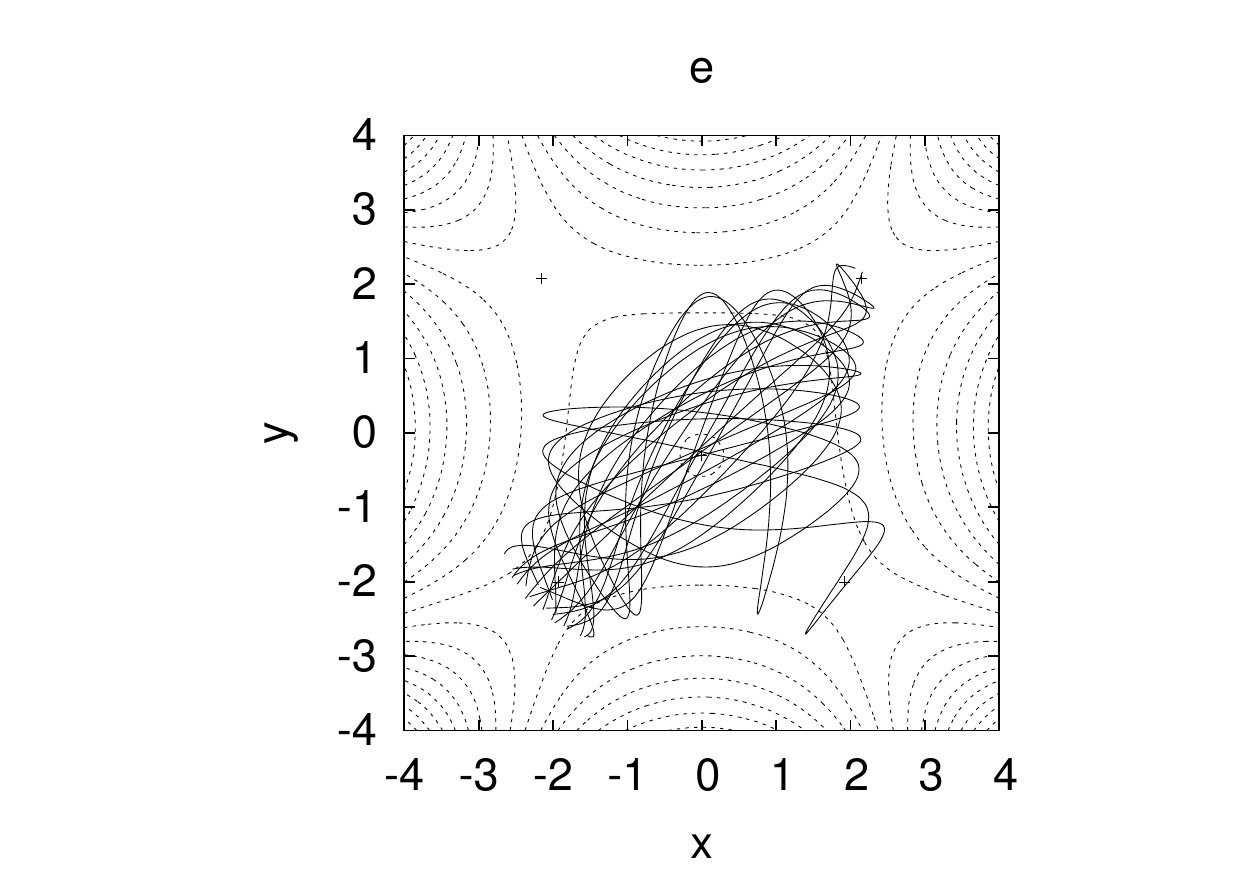}}
\hskip -2.0cm
\subfigure{\includegraphics[width=0.5\linewidth]{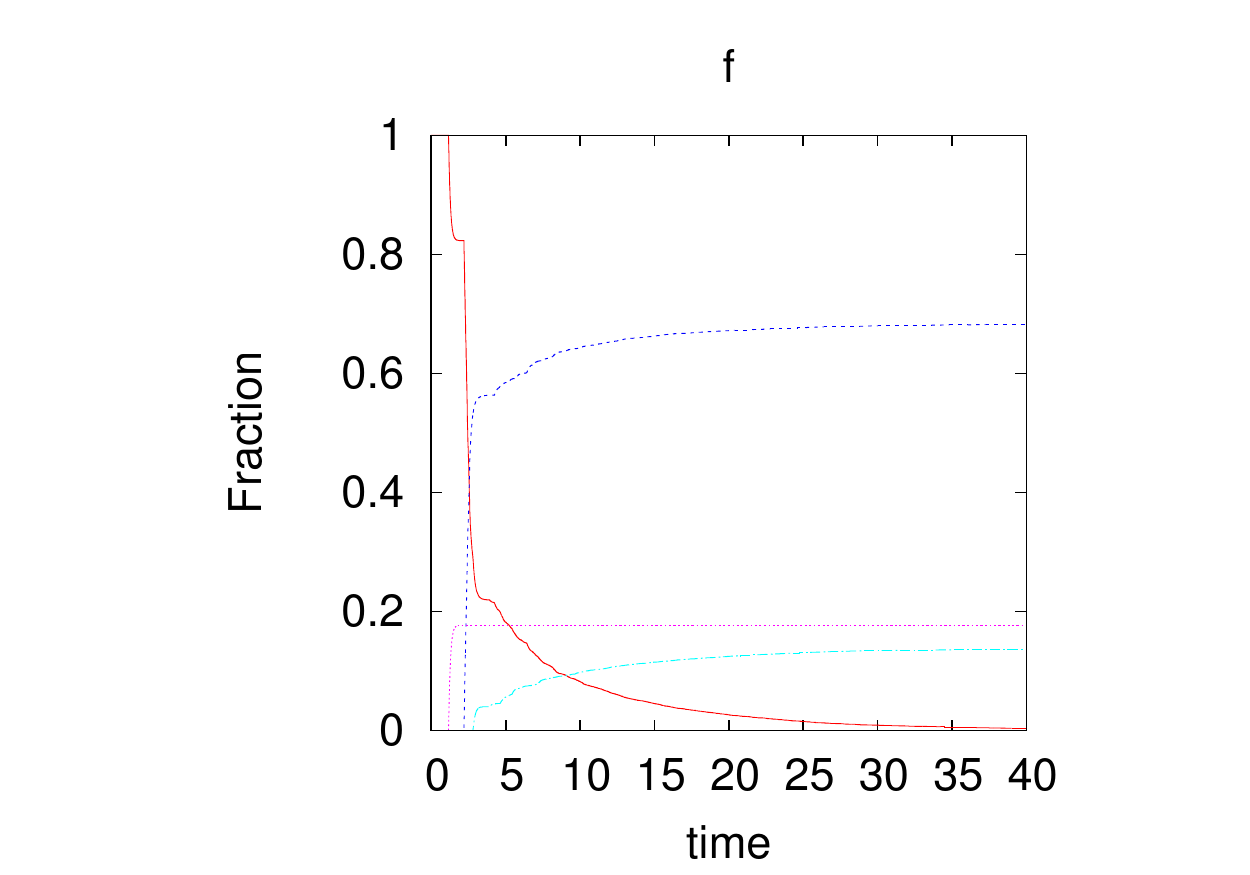}}
\hskip -5.5cm\\
\caption{Trajectories initiated on dividing surface at lower LH saddle ($x < 0$, $y < 0$) at energy $E$ above the saddle:  
(a)  $E = 5$; (c) $E = 12$; (e) $E = 15$.
 Cumulative fractions (b)  $E = 5$; (d) $E = 12$; (f) $E = 15$.
 Color key:  fraction of trajectories remaining in well (red); fraction exiting lower LH saddle (blue);
 fraction exiting lower RH saddle (cyan); fraction exiting upper RH saddle (magenta);
 fraction exiting top LH saddle (green).}
\label{fig:r2}
\end{center}
\end{figure}

We now consider trajectory ensembles initiated at the dividing surface
associated with the lower energy LH saddle ($x<0$, $y<0$).
At an energy of $E = 5$ above the saddle almost 
all trajectories pass directly through the central caldera well, 
bounce off the potential wall in the vincinity of the upper RH (higher energy) 
saddle, and then pass back out of the lower LH saddle
(Figs. \ref{fig:r2}a,b).
For the ensemble used in our calculations, 
only one trajectory ($<$ 0.05\% ) exits out of the lower RH saddle.
Note that $ \sim 5\%$ of the trajectory ensemble is still 
trapped in the central well at $t=20$, where, again for context, the passage time back and 
forth between the diametrically opposing saddles is very roughly 2 time units so that $t=20$ is approximately
10 times this caldera ``transit time".

At an energy of $E=12$ above the lower LH saddle, $\sim 0.23$ below the energy of
the upper saddles, again
almost all trajectories cross the well, are reflected and pass 
back out of the lower LH saddle (Figs \ref{fig:r2}c,d).
Only $\sim 3\%$ of trajectories go out of the bottom RH saddle.
Note that $\sim 1\%$ of the ensemble still in the central well at $t=20$.

At an energy of $E=15$ above the bottom saddle, $ \sim 2.77$ above the top saddle, 
$\sim 65\%$ goes in and back out of the bottom LH saddle, while
$\sim 15\%$ of the ensemble goes straight across and out of the top RH saddle
(Figs \ref{fig:r2}e,f). 
At this energy $\sim 15\%$ of trajectories go out of the lower RH saddle:
many trajectories almost get over the top right saddle but 
subsequently fall back into the well while moving in apparently
`unpredictable' directions.
At $t=20$, $\sim 1\%$ of the initial ensemble is still in the central well.

At an energy of $E=20$ above the bottom saddle, $\sim 7.77$ above the top saddle,
a larger fraction $ \sim 35\%$ of the ensemble 
goes straight across and out of the top RH saddle, while
$ \sim 45\%$ goes in and back out of the bottom LH saddle
(trajectories not shown).
Again $\sim 15\%$ of trajectories go out of the bottom RH saddle.
Note that $\sim 1\%$ of the ensemble 
is still in the central well at $t=20$;
these are very long-lived trajectories.

To summarize, a significant dynamical matching effect occurs for the dynamics entering through the 
lower energy saddle point vicinity into the caldera region.  When the total energy is insufficient to pass through the 
diametrically opposing higher energy TS, the dynamical matching effect is slightly different than in Section \ref{ssec:hes},
where it could be understood in terms of the First Law of Motion and the inertia of the ``reactive" caldera\--crossing motion.  
Here the dynamical matching involves a turning point in the dynamics where upon reaching the diametrically 
opposed upper saddle region, the trajectories are reflected essentially straight back and pass out of the caldera through
the lower saddle they entered.  The entrance TS is essentially ``self" matched.  
In this case, the dynamical matching is due not to inertia in the reactive motion, but rather, and
more generally, to poor coupling between the reactive motion and the degree of freedom orthogonal to it.  That is, it is a manifestation of the 
failure of ergodicity mentioned in Section \ref{sec:intro}.  When the total energy just above the lower barrier entrance TS exceeds the 
energy of the upper saddle, one might expect that the reaction mechanism would be a mixture of the passing through the 
diametrically opposing upper TS and back reflection passing out the lower energy entrance TS.  The dynamics are somewhat
more complicated in this case.  As seen in Section \ref{ssec:hes}, when the ``bath" degree of freedom is significantly excited,
trajectories entering the caldera no longer need adhere to the straight crossing path and significant reflection occurs within the
caldera region.  In this case, the mechanism is split between 
the the reflective dynamical matching out the initial entrance channel and the wandering out either of the energetically and entropically favored lower saddles, but passage out the opposing upper TS is insignificant.  When the initial excitation energy is 
raised even higher, a further split in mechanism is observed between the reflective dynamical matching,
the direct crossing dynamical matching of proceeding out the high energy TS, and the wandering mechanism out the lower saddles.  We reiterate,
however, that at all the energies we have studied, the dominant pathway is the ``self\--matched" back reflection out the entrance channel in contrast the to 
Section \ref{ssec:hes} where the dominant pathway was the direct crossing from the high energy TS out through the diametrically opposing lower energy TS.
Also, in both this section and Section \ref{ssec:hes} we note the tendency for the increase in energy, due to excitations in the ``bath" motion, to 
diversify the dynamics.  Whether higher energies give rise to separate statistical and dynamical mechanistic components to the caldera mechanism,
or whether our observations above simply indicate another, more subtle, `indirect' dynamical path to reaction is an interesting question.

\subsection{Dynamics on the stretched potential}
\label{ssec:stretched}
As seen above, the phenomenon of dynamical matching is apparent when the entrance and exit TSs are essentially aligned.  Whether this matching continues to occur when the caldera potential is distorted and the 
opposing TSs are no longer diametrically opposed is the subject of this section.

We now consider dynamics on the stretched potential, 
eq.\ \eqref{eq:pot2}.  Recall that the distortion parameter $\lambda$ is taken to be in the 
range $0 < \lambda < 1$.
For any $\lambda$, the stationary points are at the same energies as for
the unstretched case with $\lambda = 1$, eq.\ \eqref{eq:pot1}
(cf.\ Table \ref{tab:equi_pts}). 
The $y$ coordinates of the critical points are unchanged 
from the unscaled case while $x$ values are 
simply scaled by the factor $\lambda^{-1}$.
While the stretched potential retains the reflection symmetry of the
unscaled case ($\lambda = 1$), in allowing the stretching distortion of the potential we attempt to study a more 
general, and therefore possibly more realistic and chemically interesting, caldera surface.  It also provides a means of testing 
whether the dynamical matching effects observed on the undistorted potential are simply a product of the potential topography 
or a more general effect.

We consider first the stretching parameter values $\lambda = 0.6$ and $\lambda = 0.4$.
These values correspond to the quantum calculations reported in 
Section \ref{sec:quantum_results}.  Trajectory initial conditions are sampled on 
the NF dividing surface associated with the upper LH saddle, i.e. the trajectories enter
the caldera via the high energy upper LH TS, as in Section \ref{ssec:hes}.

\begin{figure}[htbp!]
\begin{center}
\hskip 0.50cm
\subfigure[]{\includegraphics[width=0.35\linewidth]{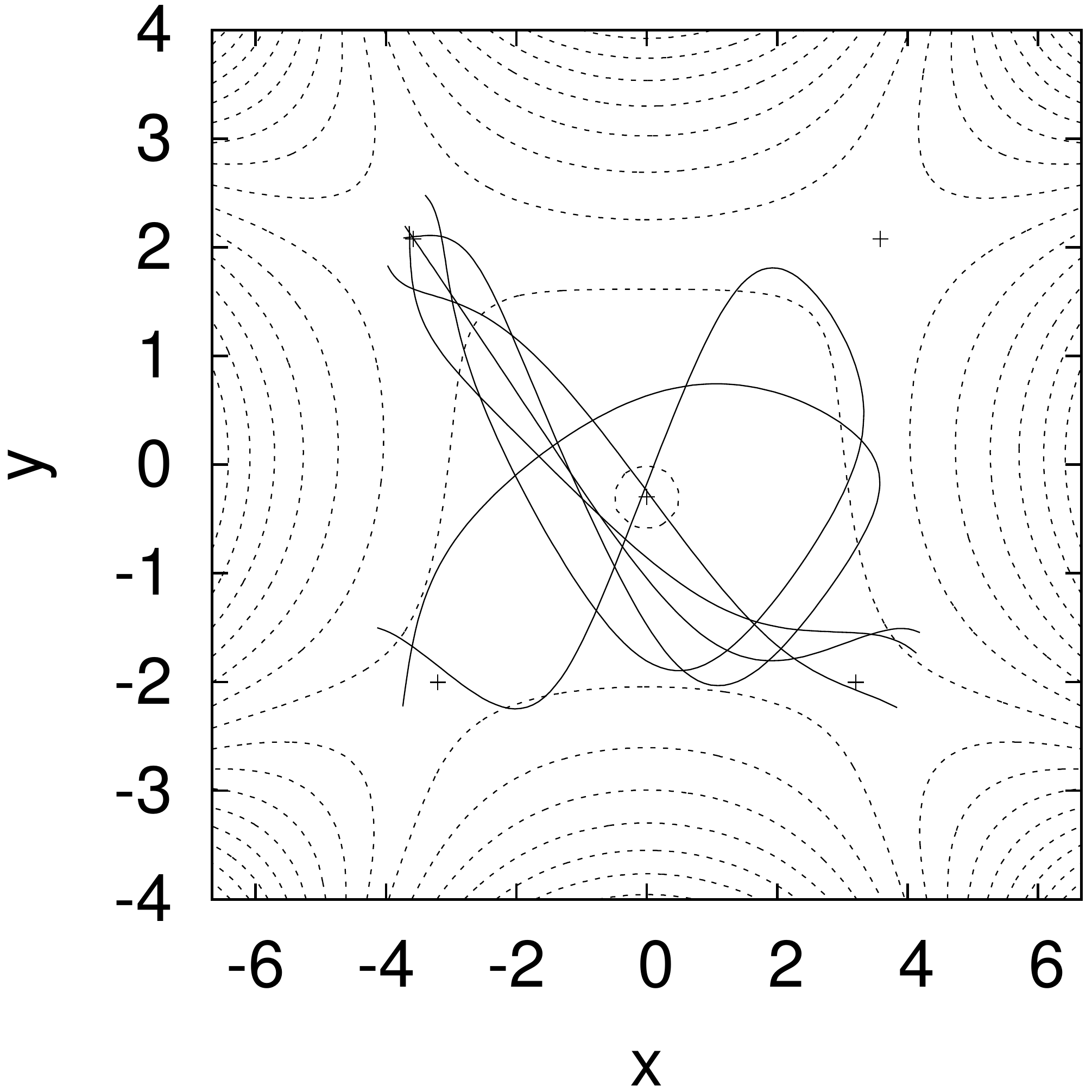}}
\hskip 0.50cm
\subfigure[]{\includegraphics[width=0.387\linewidth]{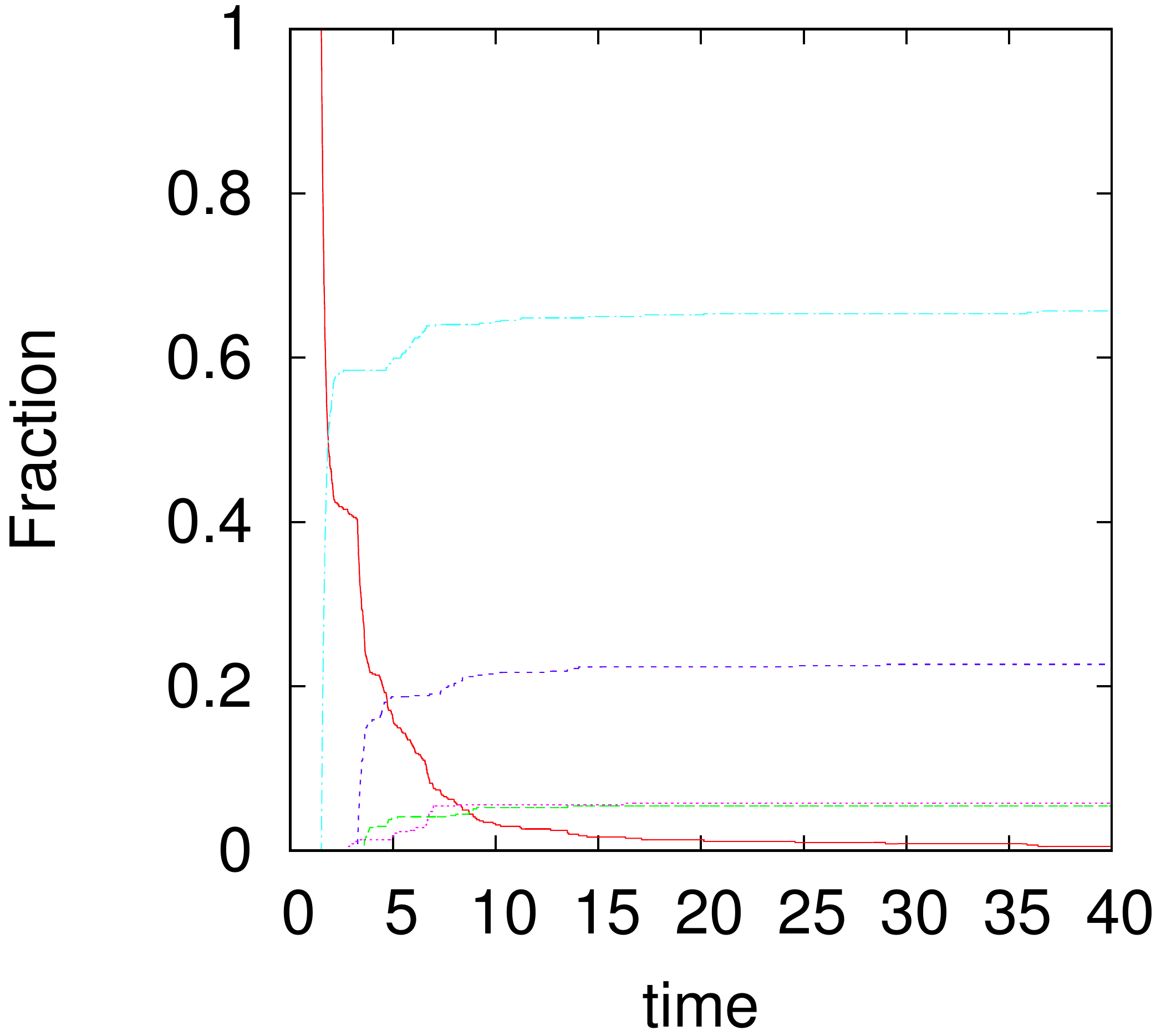}}
\\
\hskip 0.50cm
\subfigure[]{\includegraphics[width=0.35\linewidth]{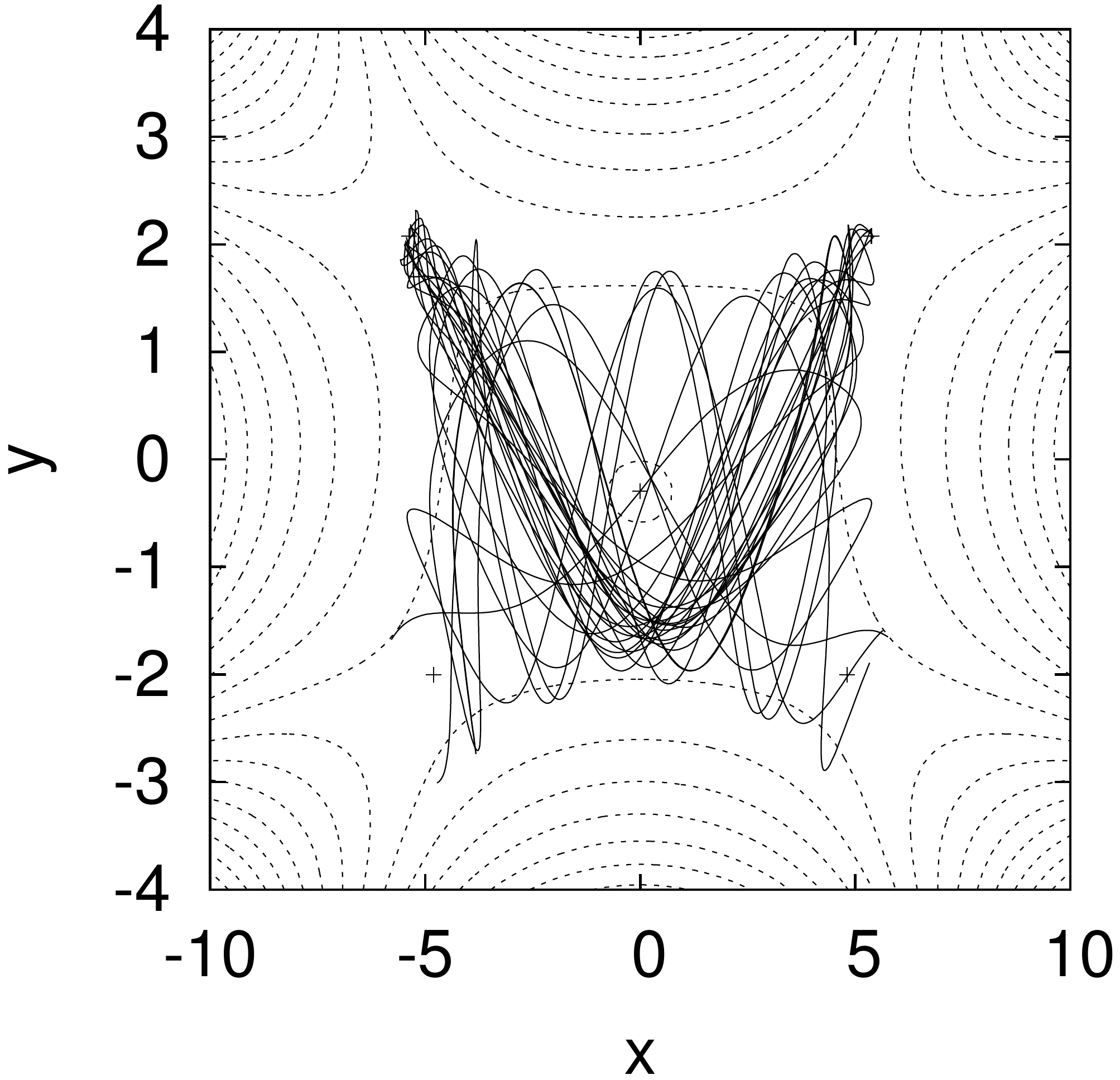}}
\hskip 0.50cm
\subfigure[]{\includegraphics[width=0.38\linewidth]{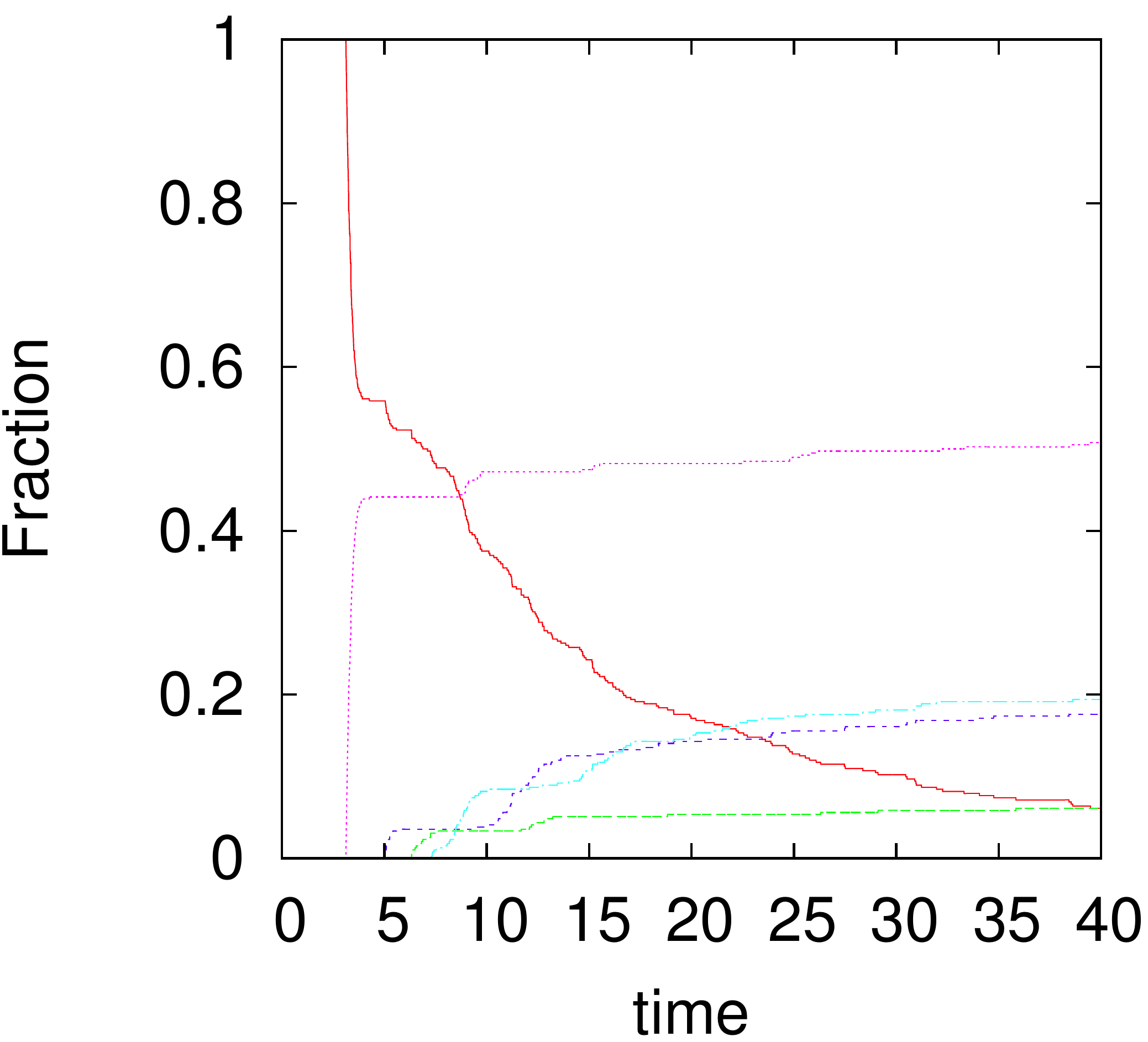}}
\caption{Trajectories shown for energies $E=5$  (top panels) and $E=1$ (bottom panels) above the LH saddle of the potential. Distortion parameter $\lambda = 0.6$ (top panels) and $0.4$ (bottom panels).
Color key: fraction of trajectories remaining in well (red); fraction exiting lower LH saddle (blue);
 fraction exiting lower RH saddle (cyan); fraction exiting upper RH saddle (magenta);
 fraction exiting top LH saddle (green).}
\label{fig:r7}
\end{center}
\end{figure}

In general, as $\lambda$ decreases, the potential becomes more elongated in the $x$\--direction, and
the bundle of trajectories initiated at the dividing surface has its first encounter with the potential wall further
and further away from the vicinity of the lower LH saddle.  
For $\lambda=0.6$ and $E=5$, a significant contribution of the direct dynamically matched reaction mechanism
is still observed, see Figure \ref{fig:r7}b and the fast initial rise of the lower RH product.  The effect, however, is not 
as pronounced as on the undistorted potential of Section \ref{ssec:hes}, which is obvious on comparing Figure \ref{fig:r7}b 
to the corresponding panels in Figure \ref{fig:r1}.  The trajectories in the $\lambda=0.6$ potential have a tendency to 
collide with some portion of the bottom potential wall, see the example trajectories in Figure \ref{fig:r7}a, and are 
subsequently reflected back into the caldera or are  ``shuttled'' over into the dynamically matched lower RH TS.  The effect of the reflection is quite 
apparent in Figure \ref{fig:r7}b.  Some trajectories undergo two reflections and proceed out the lower LH TS.  Other
trajectories wander about the caldera before finding their way out via `indirect' dynamics.  The key point is that the 
stretching of the potential and the removal of the diametrical opposition of the TSs results in more diverse dynamics,
but a dominant dynamical matching mechanism persists.

Further decreasing $\lambda$ to $0.4$, the observed dynamics changes dramatically.  The dynamical matching mechanism,
the direct uninterrupted crossing of the caldera, discussed above is virtually eliminated, but instead there is a matching between the 
entering upper LH TS and the upper RH TS.  This ``new" dynamical matching occurs when the trajectories enter the caldera and collide with the
lower potential wall.  The trajectories are then reflected toward the upper RH TS.  Essentially, the two higher energy saddles are matched via a `bank shot' 
mechanism.  Example trajectories for $E=1$ are shown in Figure \ref{fig:r7}c and the fast production of the upper RH product is clear from the cumulative
product fraction in Figure \ref{fig:r7}b.  The reflective dynamical matching is recognized in Figure \ref{fig:r7}c by the curving `U' shape formed
by a collection of trajectories in the ensemble.  Increasing the total energy is expected to have a similar effect we have discussed for 
undistorted potential, that is a greater variation in the dynamics, but this new direct ``reflective" dynamical matching mechanism is expected to 
significantly contribute.

\begin{figure}[htbp!]
\begin{center}
\hskip -2.0cm
\subfigure{\includegraphics[width=0.55\linewidth]{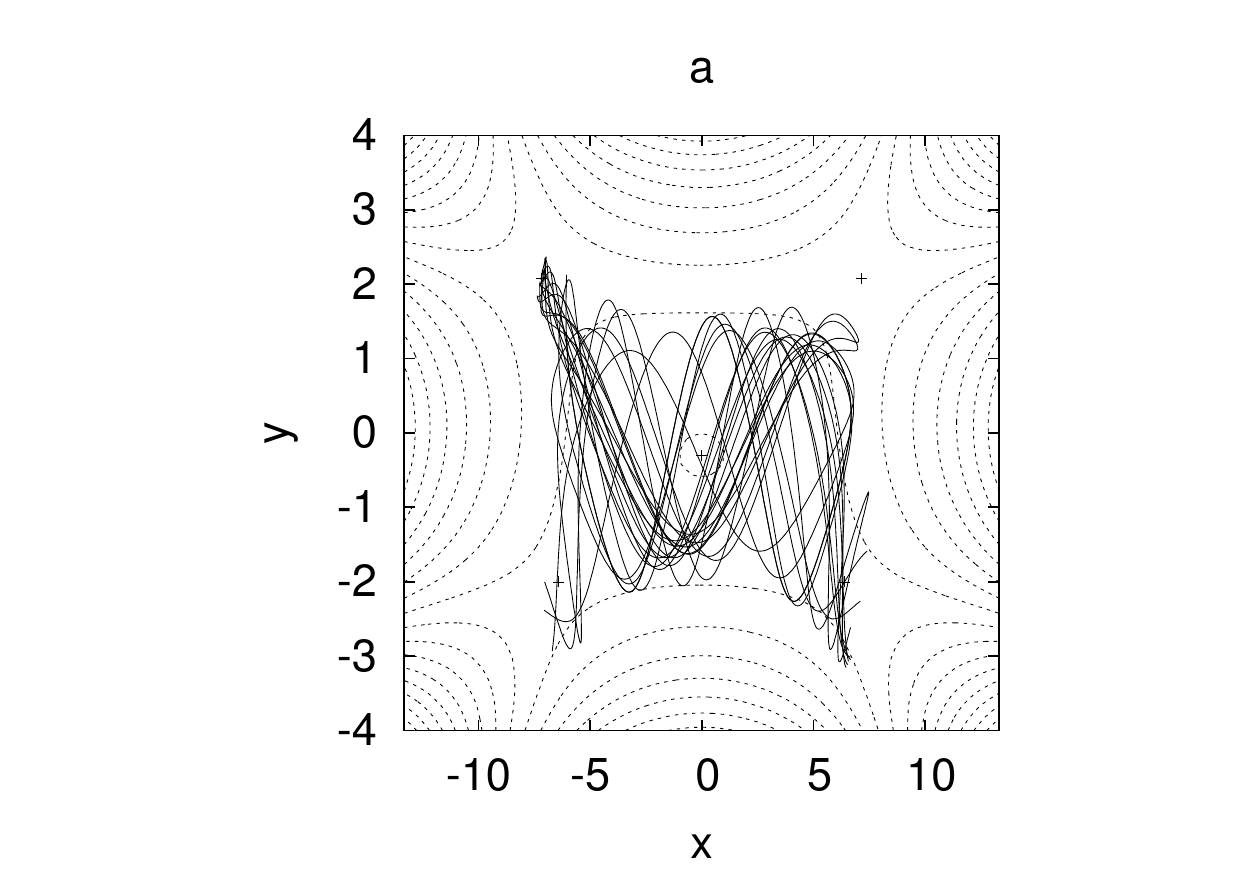}}
\hskip -2.0cm
\subfigure{\includegraphics[width=0.55\linewidth]{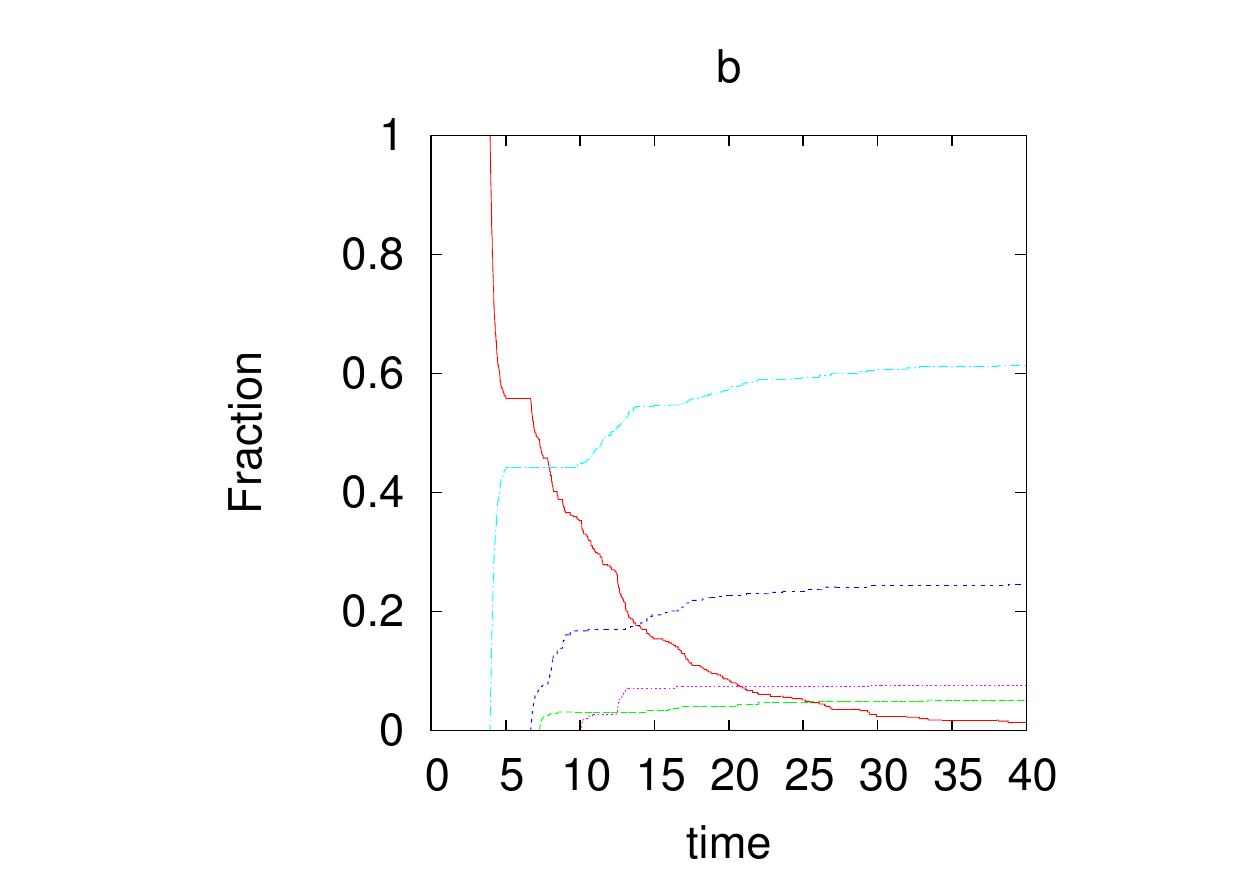}}
\\
\hskip -2.0cm
\subfigure{\includegraphics[width=0.55\linewidth]{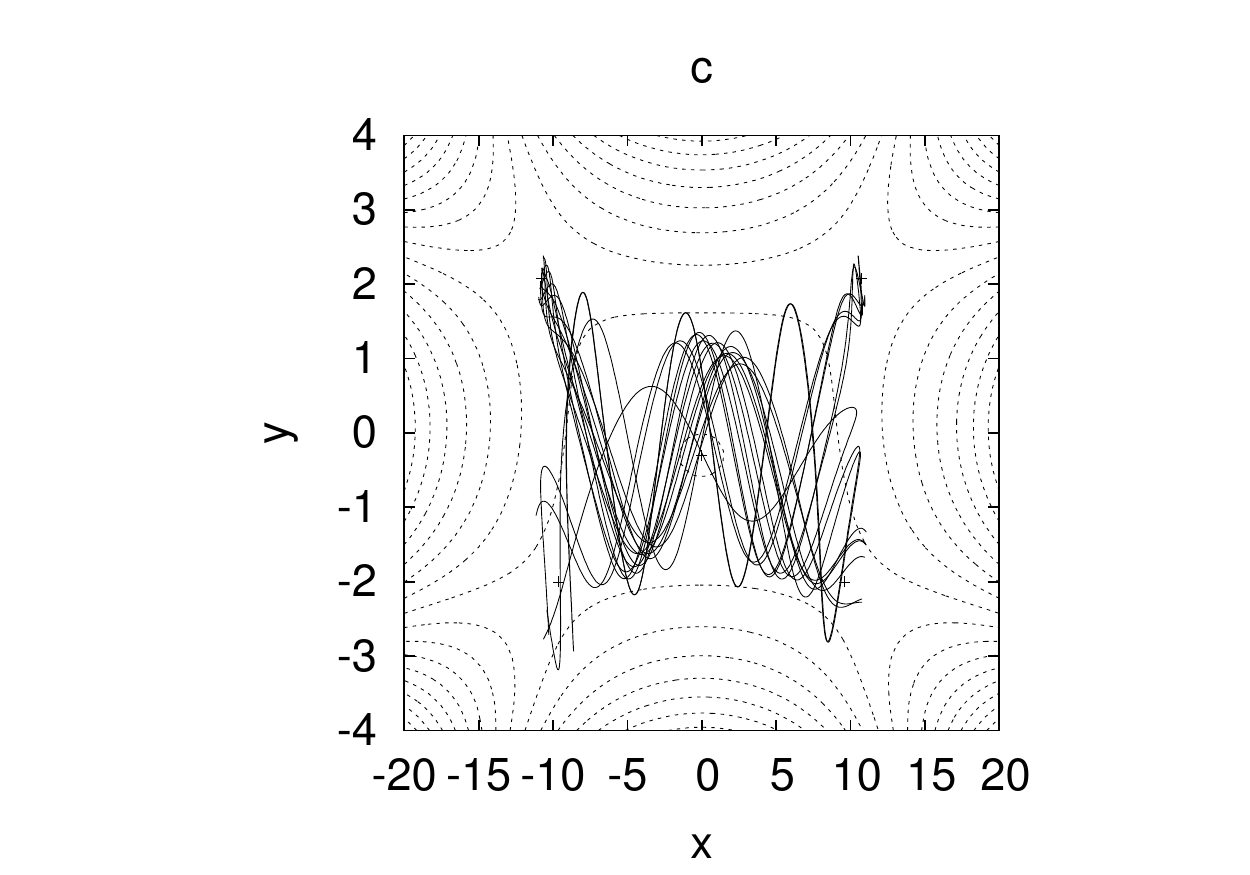}}
\hskip -2.0cm
\subfigure{\includegraphics[width=0.55\linewidth]{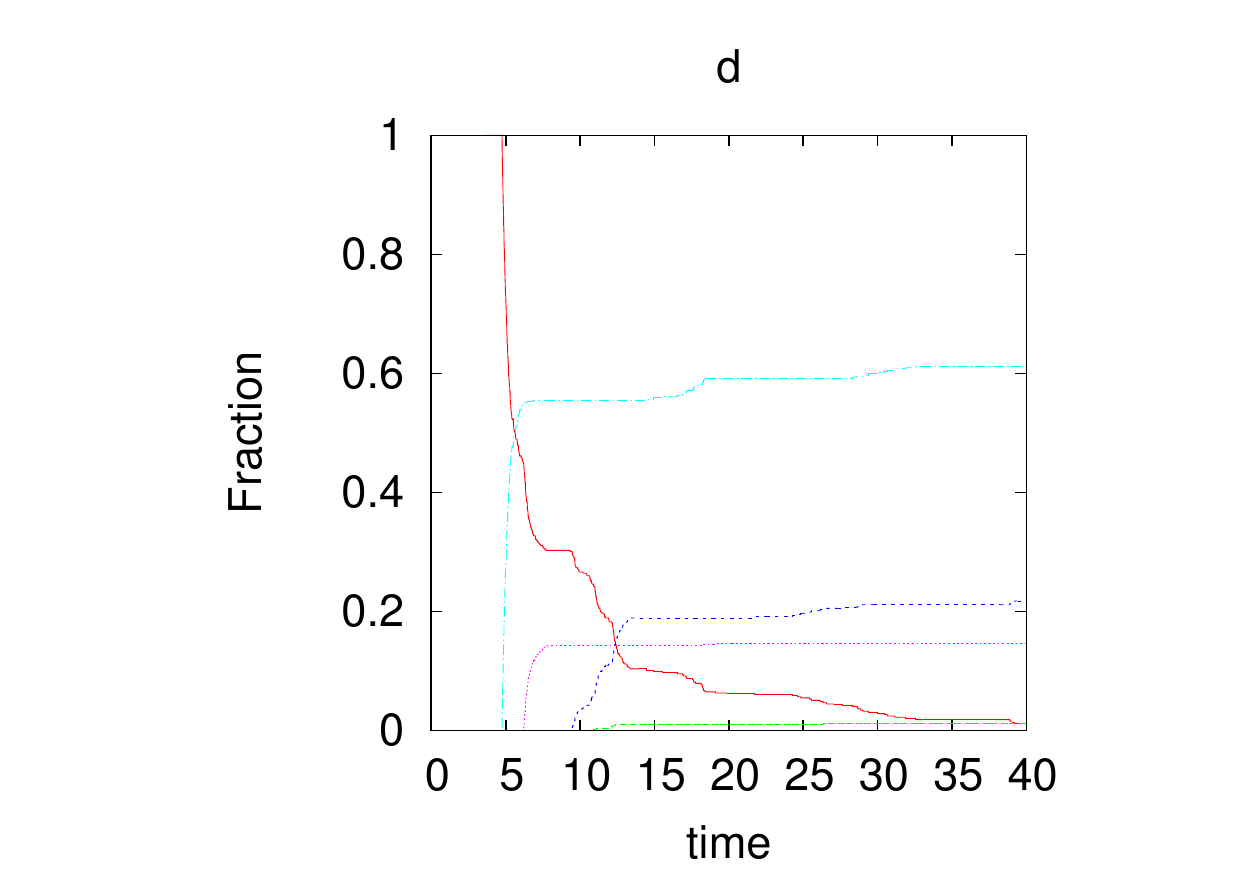}}
\\
\hskip -2.0cm
\subfigure{\includegraphics[width=0.55\linewidth]{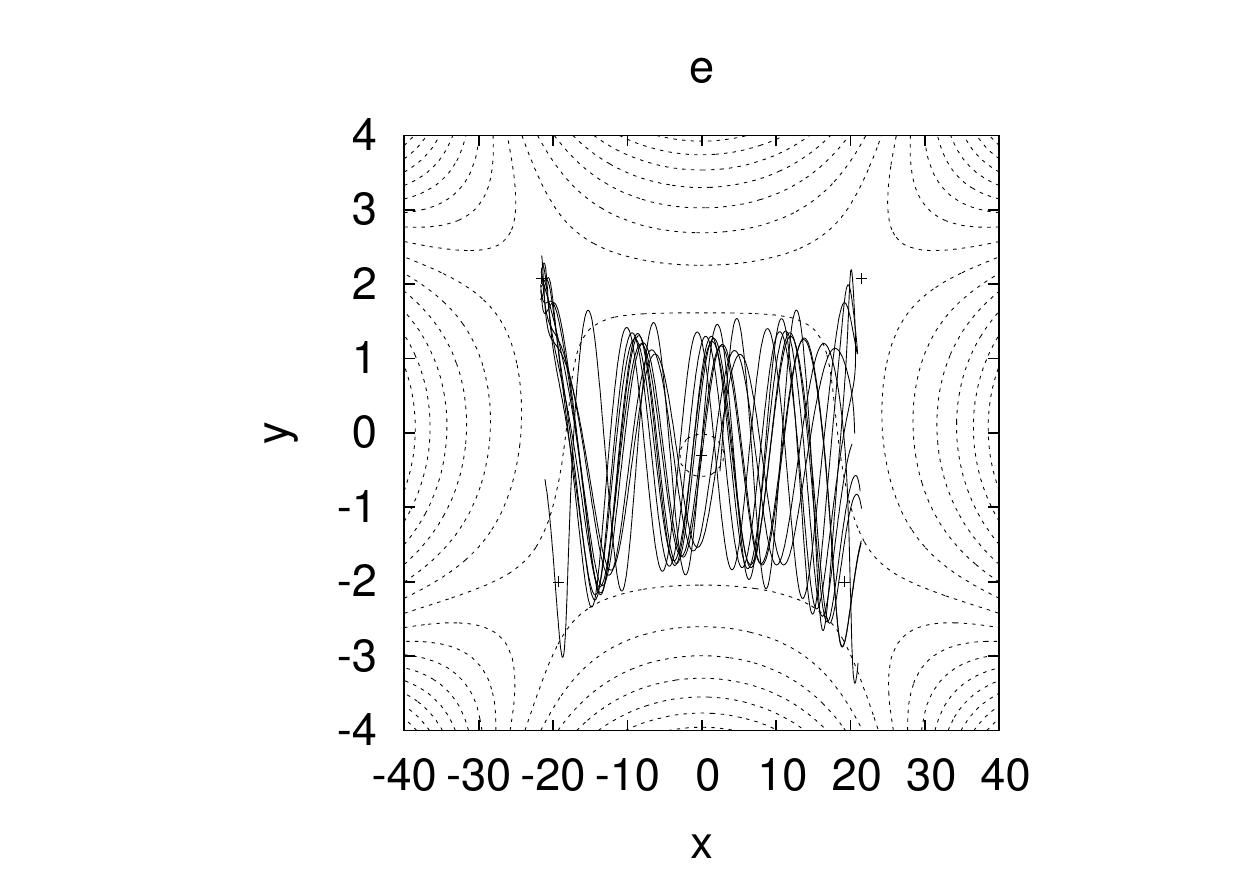}}
\hskip -2.0cm
\subfigure{\includegraphics[width=0.55\linewidth]{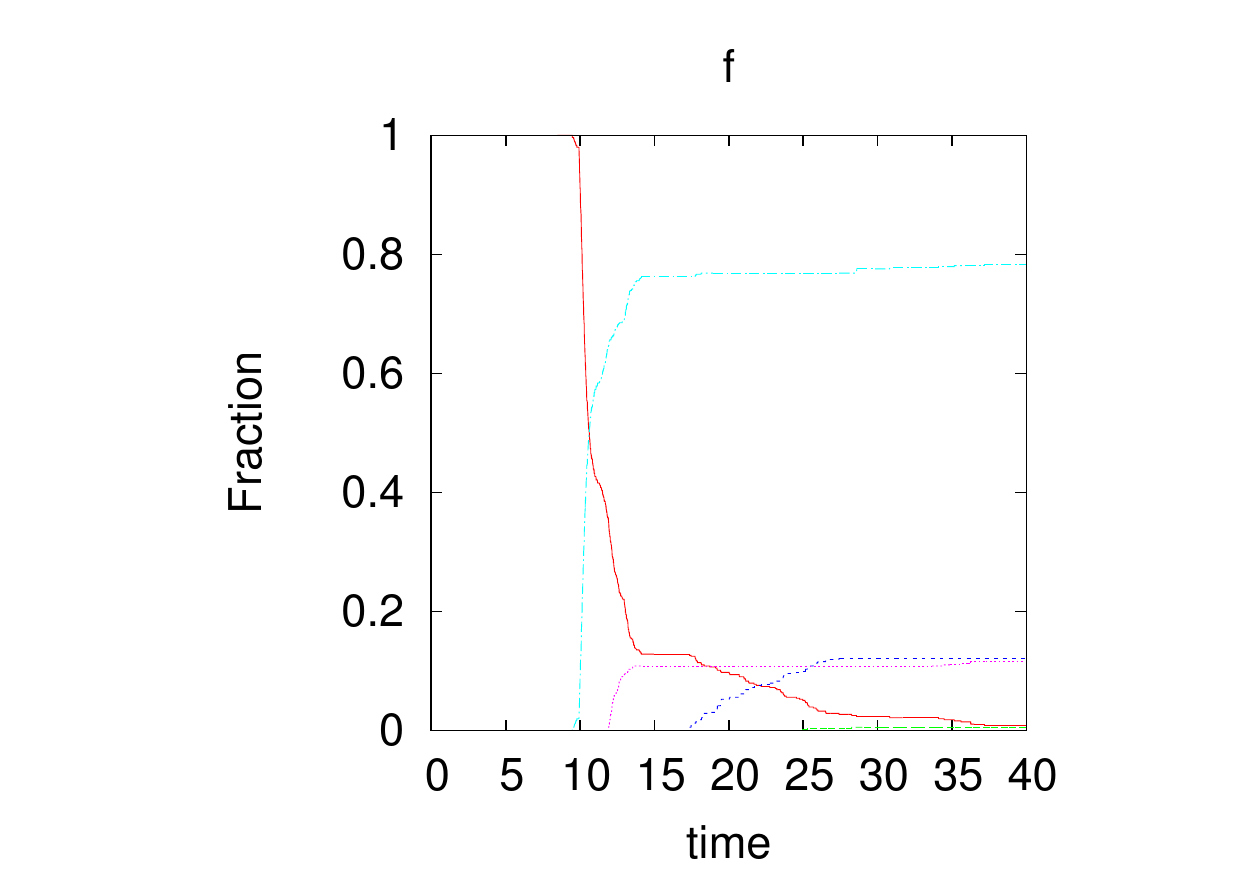}}
\caption{Trajectories shown for excitation energy $E=1$ above the LH saddle.  Potential distortion parameter $\lambda =0.3$ (top panels), 
$0.2$ (center panels), $0.1$ (lower panels).
Color key: fraction of trajectories remaining in well (red); fraction exiting lower LH saddle (blue);
 fraction exiting lower RH saddle (cyan); fraction exiting upper RH saddle (magenta);
 fraction exiting top LH saddle (green).}
\label{fig:r9}
\end{center}
\end{figure}

The results of even further distorting the potential, $\lambda=0.3$, $0.2$, and $0.1$ were also considered.  
Example trajectories and cumulative product probabilities are shown in Figure \ref{fig:r9} for $E=1$.  Several features are shared by 
these trajectories.  First, the residence time in the caldera for the trajectory ensemble is much longer as the time needed to 
traverse the extended caldera is much longer.  Second in the trajectory plots, Figures \ref{fig:r9}a, c, and d, the trajectory ensemble maintains a highly coherent structure as it enters and proceeds through the caldera.  All three cumulative product fraction plots, Figures \ref{fig:r9}b, d, and e, show that the 
major product in all cases is the lower RH product.  Note though, that the product fraction is highly dependent upon the distorted length of the caldera.  
For $\lambda=0.1$, the lower RH product makes up almost 80\% of the product yield, Figures \ref{fig:r9}c and f, 
suggesting that this caldera length matches the 
upper LH entry TS to the the lower RH product.  For $\lambda=0.3$, the product yield is relatively more split between the 
lower RH and the lower LH products, and that at this value of $\lambda$ the direct dynamical matching 
between the upper LH TS entry and the lower RH TS exit is reduced.  For the extended caldera lengths, the effect of raising the excitation energy on 
the contributions of direct dynamical reaction mechanisms was not explored but is expected to have the effect of increasing the variety of products 
contributing to the yield.  In short, our studies of significantly extended calderas have shown that the manifestations of reflective dynamical matching exhibits dependence upon the caldera shape.

\subsection{Dissipative dynamics}
\label{ssec:dissipation}
In the work above, we have studied the dynamics inherent to the key ``reaction" degrees of freedom of a caldera system but have not considered how such dynamics is modified by the presence of a chemical environment composed of either, or both, ``bath" modes of the reagent or the external environment, e.g. a solvent.  Caldera\--containing chemical reactions are high dimensional systems and the possibility of energy exchange between the key reaction degrees of freedom and the chemical environment, via IVR or collision, exists.  Similar to our work on a potential with a valley ridge inflection point (VRI) point in ref. \onlinecite{Collins13}, we have adopted a model of energy exchange from the reaction degrees of freedom to the environment as a simple irreversible dissipation, included by a velocity damping term in the equations of motion in Equation \ref{eq:hameq_diss1}.  This treatment of energy exchange is admittedly crude, but, as we outline below, contains the essence of the environmental effects.  We present the results for dissipative trajectories on the undistorted, $\lambda=1$, caldera potential, where, as before, the initial conditions for entry into the caldera are sampled on the NF dividing surface associated with the higher energy upper LH saddle point at an excitation energy of $E=5$.

The dissipation in the model is controlled by the parameter $\gamma$ in Equation \ref{eq:hameq_diss1}.  When $\gamma$ is large, energy dissipation from the reaction degrees of freedom occurs quickly while the energy\--conserving dynamics studied in Sections \ref{ssec:hes}, \ref{ssec:lee}, and \ref{ssec:stretched} is recovered when $\gamma\rightarrow0$.  Therefore, high values of $\gamma$ are analogous to systems where IVR or energy exchange with the external environment, e.g.  in high pressure systems, readily occurs on short timescales commensurate with, or faster than, reaction timescales.  At very high dissipation, $\gamma=0.7$, all the trajectories in the studied ensemble that entered the caldera quickly lost their energy and fell into the caldera well.  In the strict context of the model, these trajectories remain trapped in the caldera, however, they are representative of a mechanistic limit in a more realistic chemical system.  In this high dissipation regime, energy is quickly transferred out of the caldera crossing degrees of freedom before exit from the intermediate region can occur, and a long\--lived intermediate of the reaction is formed in the caldera.  Assuming the caldera potential bowl is sufficiently deep, that is the energy required to pass out one of the exiting TSs is $\gg k_{\textnormal{\tiny{B}}} T$, subsequent reaction of the intermediate depends upon reactivation, via some form of collisional energy transfer or energy redistribution back into the reaction coordinates.  Prior to reactivation to a reactive state, and in the absence of some subtle and unexpected nonstatistical IVR mechanism of energy transfer, the caldera intermediate is free to equilibrate and subsequent reaction is expected to proceed statistically.  Moreover, even relaxing the assumption above and assuming the caldera bowl can be quite shallow, such fast dissipation is compatible with at least quasi\--equilibration in the caldera region.  Simply put, in the presence of very fast statistical IVR or environmental dissipation from the reaction coordinates, the statistical picture of the reaction mechanism, with the formation of a caldera reaction intermediate and a product yield dictated solely by the nature of the exiting TSs, is recovered. 

When the dissipation is low, $\gamma=0.2$, the observed dynamics of the trajectories is similar to what is observed in the non dissipative $E=5$ ensemble.  A large majority of the trajectories cross the caldera region from the high energy entrance LH TS to the dynamically matching lower energy exit RH TS.  Such trajectories represent a system where energy transfer from the reaction degrees of freedom is not sufficiently fast to remove energy on the reaction timescale and is, therefore, of little consequence to the reaction mechanism.

Varying the dissipation parameter within the medial dissipation range, $0.2 < \gamma < 0.7$, a smooth transition between the high dissipation, where all trajectories become trapped in the caldera intermediate region, and the low dissipation, where the dynamics show little dependence on dissipation, situations is observed.  That is, in this medial dissipation regime some portion of the trajectory ensemble quickly passes out of the caldera region while another portion becomes trapped, where the ratio of reactive to trapped trajectories varies continuously with $\gamma$.  The medial dissipation regime corresponds to a reaction system where the reaction and the energy transfer out of the reaction degrees of freedom occur on comparable timescales.  In our simulation, those trajectories that were reactive were those that experienced the dynamical matching between the diametrically opposed TSs and, particularly for larger $\gamma$, were those trajectories on the sampled dividing surface that entered the caldera with a majority of the momentum localized in the direction of the diametric crossing.  The trajectories that tended to be trapped were those closer to the fringe of the sampled entry dividing surface with a substantial component of momentum directed orthogonal to the diametric crossing, including those few trajectories mentioned in the sections above that did not undergo dynamical matching but rather reflected several times in the caldera prior to reaction, i.e. those with `indirect' paths to reaction.  At higher dissipation, even the dynamically matched trajectories that did not promptly traverse the caldera region became trapped.  This suggests that in the medial dissipation regime, it is possible that the reaction mechanism may be split into two contributions: a `direct' mechanism where fast reaction occurs, represented by trajectories quickly traversing the caldera between dynamically matched entry and exit TSs, and a `statistical' mechanism, where energy transfer out of the reaction coordinates creates a long\--lived, equilibrated, caldera intermediate, represented by trajectories that either take an  `indirect' path in the caldera or by those dynamically matching trajectories that were simply too slow in traversing the caldera region to compete with the dissipation timescale.  These results are consistent with, for example, the experimental work of Reyes and Carpenter \cite{ReyesCarpenter} who studied the mechanism of the thermal deazetization of 2,3\--diazabicyclo[2.2.1]hept\--2\--ene (DBH), a reaction involving a cyclopentane\--1,3\--diyl biradical caldera intermediate, in supercritical propane and supercritical CO$_2$.  They found at lower pressures, presumably corresponding to lower dissipation in our model, the product yield of the reaction, the products were two stereoisomeric ring conformers that in the statistical limit was expected to give a 1:1 product ratio, was highly non\--statistical.  A dynamical matching argument was suggested in that work to explain the results.  However, when the pressure of the solvent fluid was increased, the product ratio approached the expected 1:1 statistical result.

In closing this section, we remark that while the above discussion provides a useful heuristic rationale of caldera mechanisms in the presence of a chemical environment, such mechanisms are still expected to be highly dependent upon the nature of the system and the environment.  Returning to                     the work of Reyes and Carpenter \cite{ReyesCarpenter}, when carbon dioxide supercritical fluid was substituted for propanol as the solvent, the pressure dependence of the product ratio was highly nonlinear.  This unusual response of the product yield to pressure was attributed to the ability of CO$_2$ molecules to undergo complexation with the biradical intermediate.  

\newpage

\section{Results: quantum mechanics}
\label{sec:quantum_results}
\subsection{Wave packet dynamics in the caldera}
\subsubsection{$\lambda=1$}
\begin{figure}[htbp!]
\centerline{\includegraphics[height=150mm]{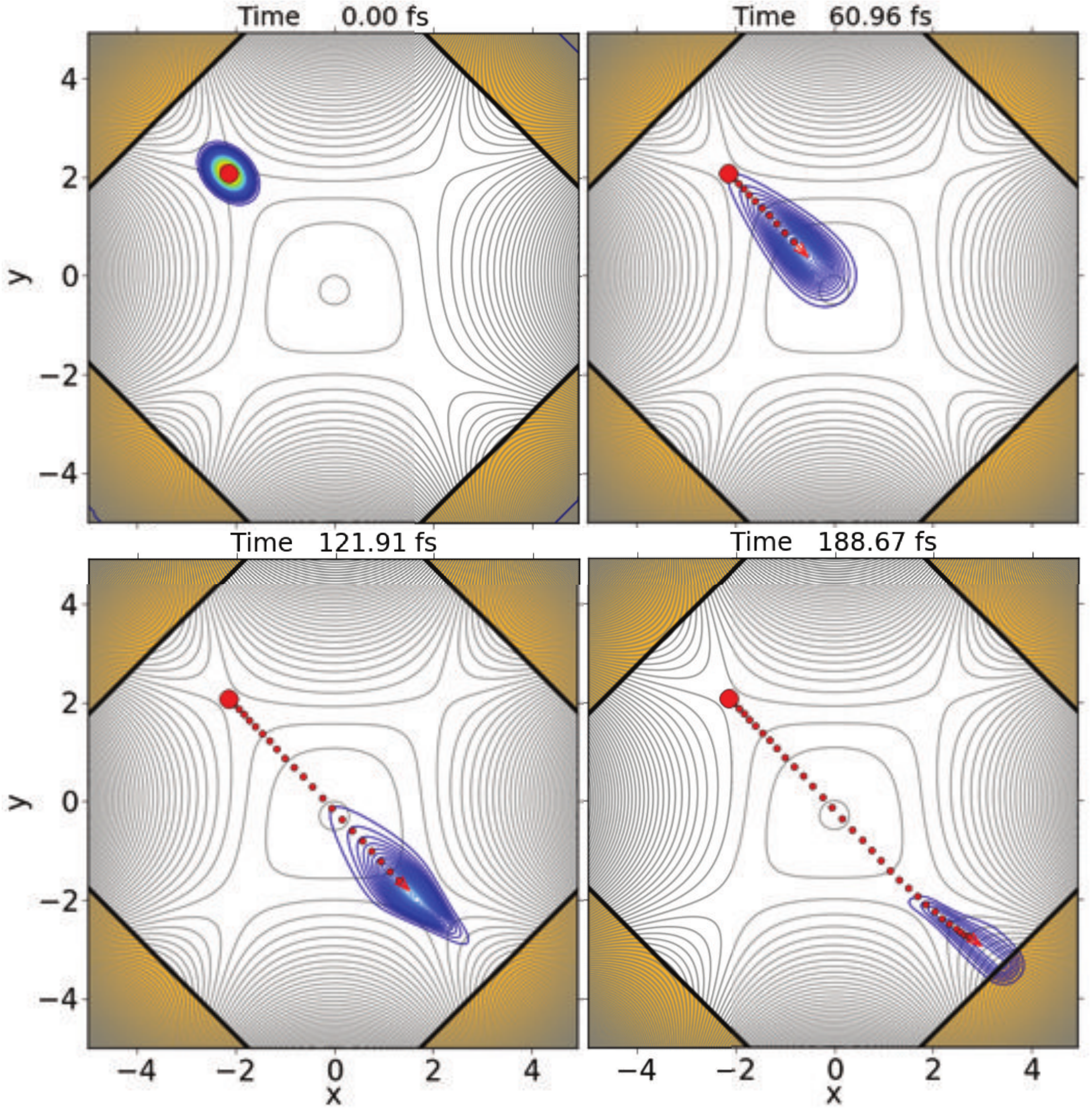}}
\caption{Snapshots of the wave packet (plotted as the probability density function 
$\Psi^{\star}_t \Psi_t \left(x,y\right)$) at various times.  
Here the wave packet is for a carbon particle with $\langle E \rangle_{t=0}=1$ kcal mol$^{-1}$
on the $\lambda=1$ PES, which is shown as grey contours in the background.  
The non-vanishing regions of the NIP are indicated by orange shading and thick black boundary lines.  
Points corresponding to expectation value pairs $\left(\langle x \rangle_{t'},\langle y \rangle_{t'}\right)$ 
for a discrete set of $t'$ values, $t' \in \left[0,t\right]$ are plotted as red circles and  
trace an approximate path of the portion of 
the packet in the caldera region.  These results show quantum mechanical ``dynamical matching" where 
the wave packet passes directly from the upper saddle through the lower TS through the caldera region.}
\label{fig:z_wp_1}
\end{figure}

On the unstretched PES ($\lambda=1$), the direction of motion associated with 
the unstable normal mode (`reaction coordinate')
at the higher energy saddle  
where the wave packet enters the caldera region is approximately aligned with the diametrically 
opposing lower energy saddle.  Therefore, a wave packet that enters the caldera through the
upper saddle is naturally steered by the PES (and any additional momentum in the reaction coordinate direction) 
toward the region of the lower energy TS.  
In this case, the wave packet passes directly 
over the caldera and out the opposing lower energy TS without significant reflection between 
the diametrically opposing and ``dynamically matching" \cite{Carpenter95} 
upper and lower transition states.  In Figure \ref{fig:z_wp_1}, various time snapshots 
of the  $^{12}$C particle on the $\lambda=1$ PES are shown and the dynamical matching between 
the two TSs is apparent.  While there is a slight mismatch between the initial 
average momentum vector of the wave packet and the direction connecting the two saddle points, 
the difference is small and energy of the opposing saddle region is low enough that 
the wave packet can completely pass over the lower energy TS while hugging the region 
just to the lower left of the lower energy saddle.  The $^{12}$C particle 
remains localized during its motion across the caldera.  
The behavior of the  $^{12}$C particle wave packet is the same for all the different 
values of $\langle E^{\star}\rangle_{t=0}$ we have considered.
The passage time of the $^{12}$C wave packet between the dynamically matched 
upper LH entry and lower RH exit TSs is $\sim150$ fs for $\langle E^{\star}\rangle_{t=0}=1$ kcal mol$^{-1}$
and is slightly less than 100 fs for $\langle E^{\star}\rangle_{t=0}=3$ and $6$ kcal mol$^{-1}$.  
In the classical simulations, the direct, dynamically matched, passage time over the undistorted ($\lambda=1$) 
caldera can be intuited from Figures \ref{fig:r1}(b), (c), and (d).  Specifically, for the lowest energy the classical passage time
is approximately 200 fs (1 time unit) and slightly less for the higher energies.  The classical and quantum $^{12}$C direct caldera transit times compare well, and we further note that this caldera passage timescale is consistent with the findings of Carpenter and Reyes in their work on the thermal deazetization of DBH \cite{ReyesCarpenter} ($\sim120$ fs) and in theoretical \cite{Hrovat97,Doubleday97,Doubleday98} and experimental \cite{Zewail94} studies
of the trimethylene biradical ($\sim140$ fs).

\begin{figure}[htbp!]
\centerline{\includegraphics[height=120mm]{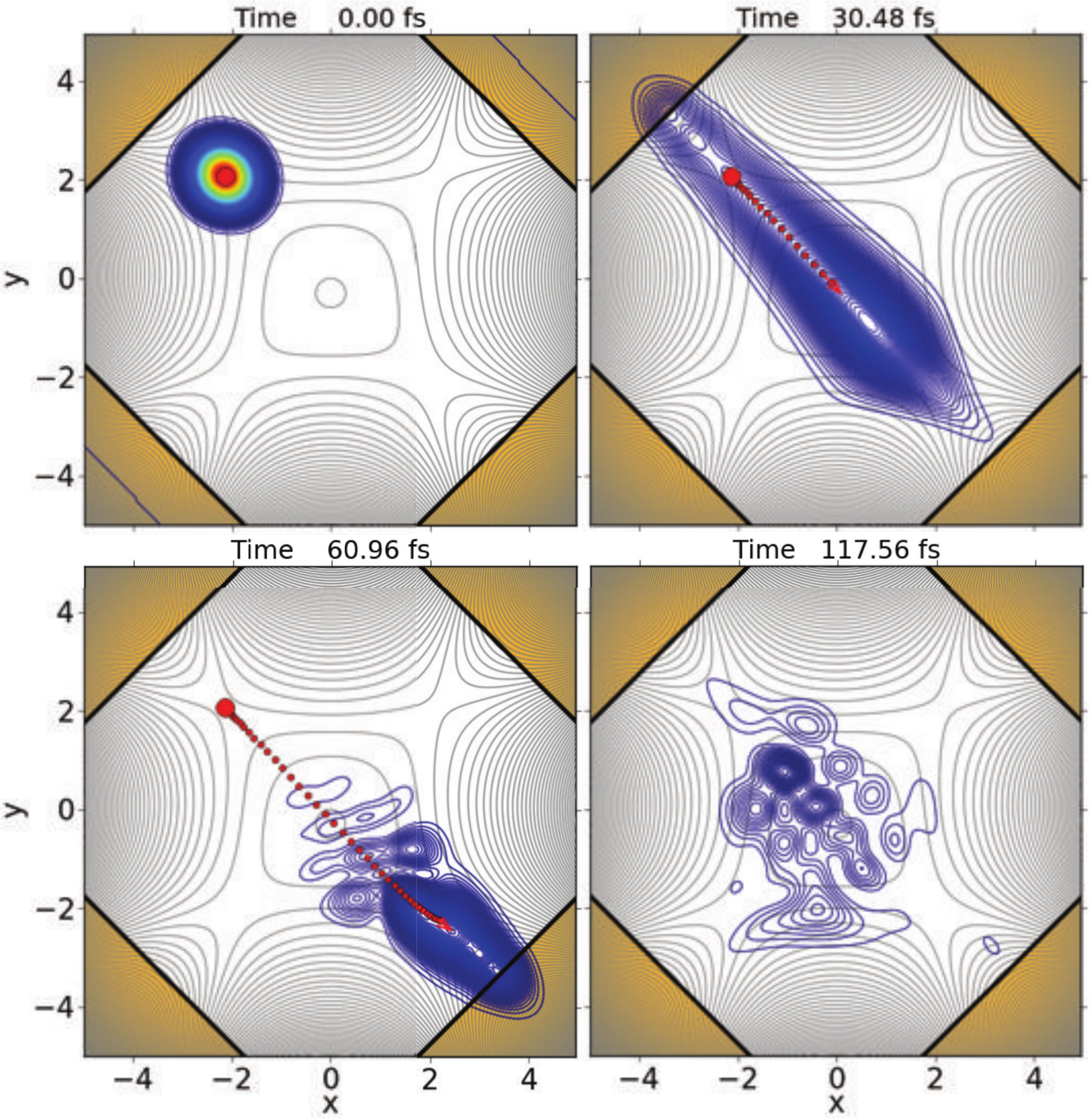}}
\caption{The wave packet on the $\lambda=1$ PES for a hydrogen wave packet with 
$\langle E \rangle_{t=0}=1$ kcal mol$^{-1}$.  A significant portion ($\sim20\%$) of the 
packet is lost due to spreading of the initial wave packet outside the caldera region.  
The portion of the packet that enters the caldera region initially mostly directly 
exits through the 
opposing lower TS.  However, a small delocalized remnant of the packet persists in the caldera 
region at longer times. The direct character of the reaction increases at higher energies (not shown).}
\label{fig:z_wp_5}
\end{figure}

The evolution of the $^{1}$H particle wave packet, shown in 
Figure \ref{fig:z_wp_5} for $\langle E^{\star}\rangle_{t=0}=1$ kcal mol$^{-1}$, is 
qualitatively similar to the $^{12}$C, on the $\lambda=1$ potential,
but the wave packet of the lighter particle spreads much more quickly as it 
passes through the caldera region (indeed, the spreading is fast enough, that 
for the $^{1}$H with $\langle E^{\star}\rangle_{t=0}=1$ kcal mol$^{-1}$ packet, 
regardless of the value of $\lambda$, a small but significant amount of amplitude 
quickly passes out the upper TS and into the NIP region without ever entering the caldera).  
By the time it reaches the lower saddle region, the wave packet is spread out enough that a 
small portion of it reflects off the potential walls adjacent to the lower energy saddle or, 
particularly for the $\langle E^{\star}\rangle_{t=0}=1$ kcal mol$^{-1}$ packet, 
lower energy components reflect back off the lower TS itself.  This 
reflected 
amplitude spends some time bouncing about in the caldera region before passing out 
via barrier passage or tunneling.  The lower right panel of Figure \ref{fig:z_wp_5} 
shows a renormalized snapshot of the remaining amplitude.  

The direct caldera passage time of the $^{1}$H quantum wave packet
is approximately 55 fs for $\langle E^{\star}\rangle_{t=0}=1$ kcal mol$^{-1}$ and $\sim30$ fs for
$\langle E^{\star}\rangle_{t=0}=6$ kcal mol$^{-1}$.  In the corresponding classical simulations, the estimated 
direct passage time of 58 fs (1 time unit) compares well with the $^1$H wave packet result.

Nonetheless, at all values of $\langle E^{\star}\rangle_{t=0}$, a 
majority of the packet passes over the lower energy TS and is absorbed 
just as in the $^{12}$C system, and the direct mechanism, the unhindered 
passage between the dynamically matching the upper and lower TSs, is 
exhibited in both systems with reflections within the caldera being
either absent or insignificant.

\subsubsection{$\lambda=0.6$}
In the $\lambda=0.6$ PES the dynamical matching  between the diametrically opposed upper and lower 
TSs is reduced.  The wave packet
can no longer pass directly between the TSs following the momentum in the reaction coordinate direction. 
Instead, the wave packet will collide with the bottom potential wall.  
The collision with the 
potential wall has important consequences on the ensuing dynamics,
whose details depend on the energy and the mass of the particle.  
This quantum mechanical behavior is fully consistent 
with the classical trajectory results for $\lambda = 0.6$
reported in the previous Section.

\begin{figure}[htbp!]
\centerline{\includegraphics[height=90mm]{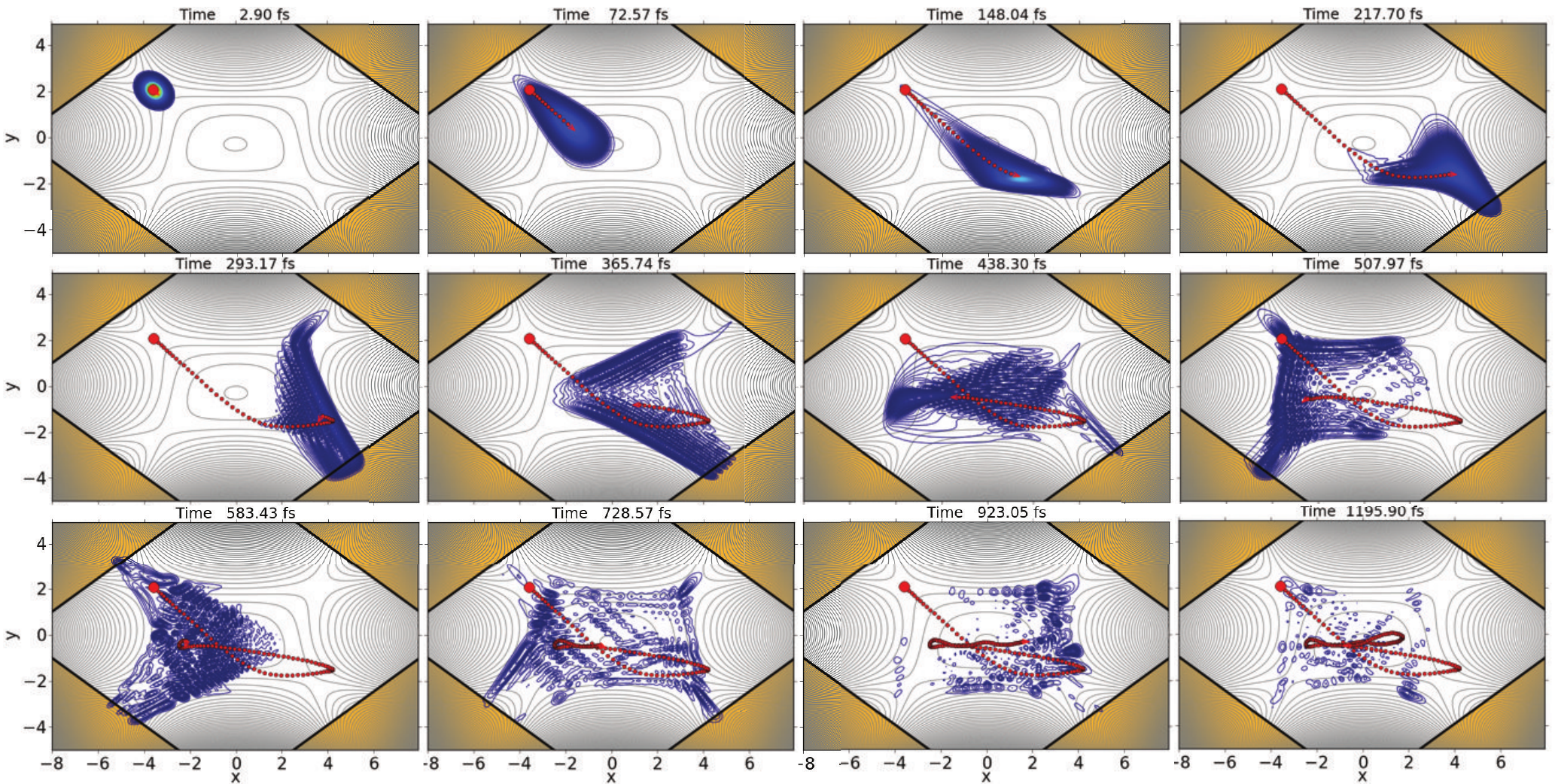}}
\caption{Similar to Figure \ref{fig:z_wp_1}, but for a carbon particle on the 
$\lambda=0.6$ PES with  $\langle E \rangle_{t=0}=1$ kcal mol$^{-1}$.  
While most of the wave packet still passes directly from the upper TS to the lower TS, 
due to the scaling of the potential in the $x$ direction a 
significant portion of the wavepacket collides with the repulsive wall of the caldera and undergoes 
a more complex motion.}
\label{fig:z_wp_2}
\end{figure}

Snapshots of the $\langle E^{\star}\rangle_{t=0}=1$ kcal mol$^{-1}$ $^{12}$C example are shown
in Figure \ref{fig:z_wp_2}.
For the $^{1}$H packets and the 1 kcal mol$^{-1}$ $^{12}$C packet, 
the collision has the effect of both reflecting the packet toward the 
right potential wall and of spreading the packet.  After this first reflection, 
a portion of the spread-out packet passes out the diametrically opposed lower TS. 
The remaining portion of the wavepacket bifurcates 
during a collision with the right PES wall, and some of 
the packet passes into the region of the right upper TS, where some of the higher 
energy components pass out into the NIP region, and lower energy components are reflected back into the caldera.  
The parts of the packet that remain in the caldera are now quite delocalized in the $y$ 
direction between the right upper and lower TSs and have momentum directed toward the left.  
The subsequent dynamics of the packet, seen in Figure \ref{fig:z_wp_2}, consists of a back 
and forth reflection of a packet wave front in the $x$ direction every $\sim 300$ fs over the 
next picosecond of propagation.  As the packet reaches its turning points, portions of the 
amplitudes pass over the lower and upper TS regions and most of the packet is 
absorbed after 1 ps.

\begin{figure}[htbp!]
\centerline{\includegraphics[height=90mm]{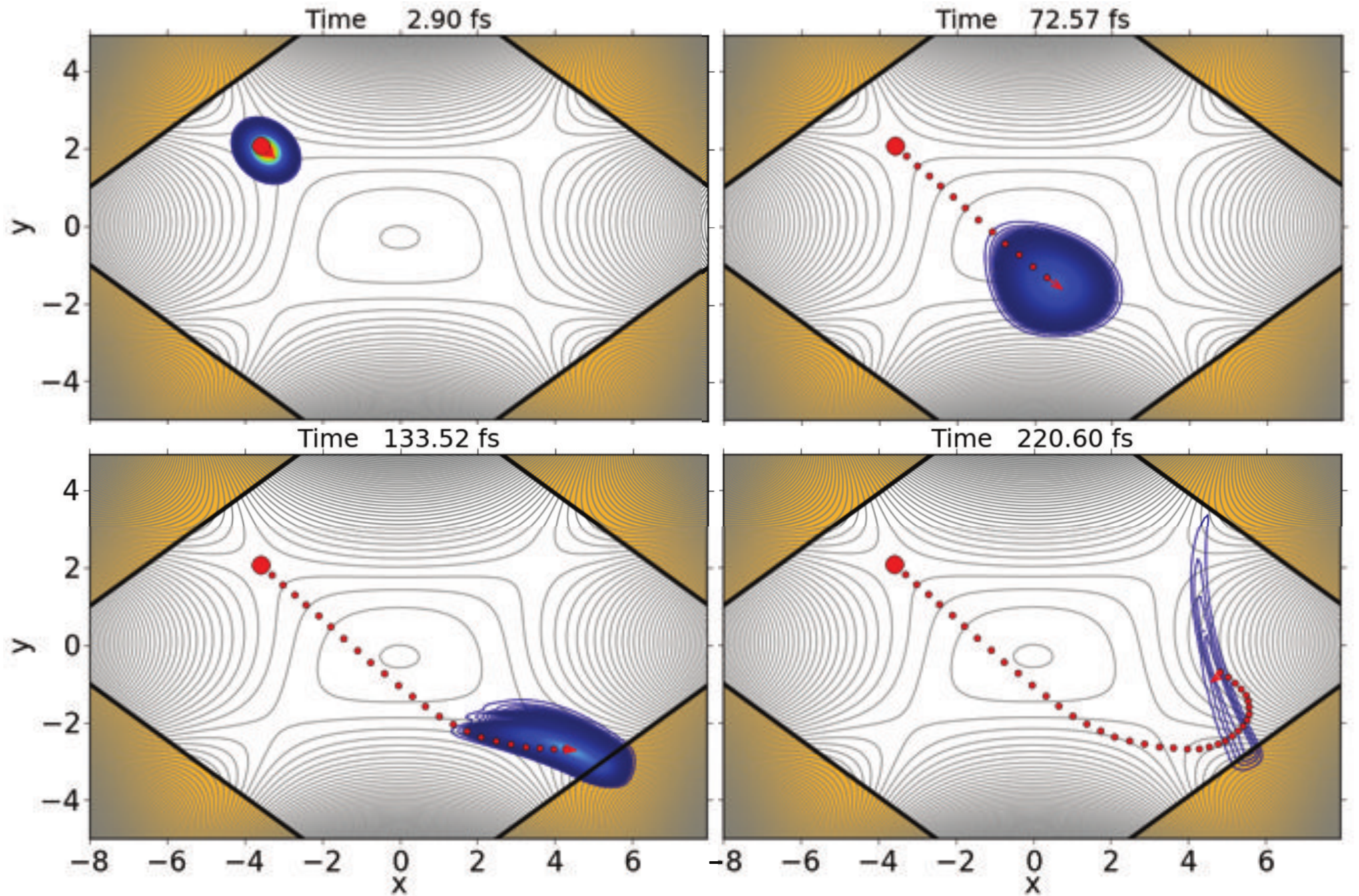}}
\caption{Similar to Figure \ref{fig:z_wp_2}, but for a carbon particle on 
the $\lambda=0.6$ PES with  $\langle E \rangle_{t=0}=6$ kcal mol$^{-1}$, i.e.,
a higher energy wave packet with more momentum in the ``reactive" direction.  
This higher energy wave packet can pass over the PES features that reflected the
packet in Figure \ref{fig:z_wp_2}, and so most of the packet passes out of the lower TS.}
\label{fig:z_wp_3}
\end{figure}

In Figure \ref{fig:z_wp_3} we show the $\langle E^{\star}\rangle_{t=0}=6$ kcal mol$^{-1}$ packet.
For the higher energy $^{12}$C wave packet, the observed mechanism changes.  
Now the $^{12}$C has sufficient lateral  momentum that,
upon collision with this PES wall, the packet spreads toward the diametrically opposing lower TS 
about the bottom wall.  Most of the amplitude then passes over the lower energy opposing TS.  
For the higher energy wave packet, the diametrically opposing upper and lower TSs have again become 
effectively dynamically matching and the direct mechanism between them is again dominant.  
A small portion in of the packet, seen in the last panel of Figure \ref{fig:z_wp_3}, is 
still reflected and undergoes motion similar to that of the lower energy case.

\subsubsection{$\lambda=0.4$}
\begin{figure}[htbp!]
\centerline{\includegraphics[height=120mm]{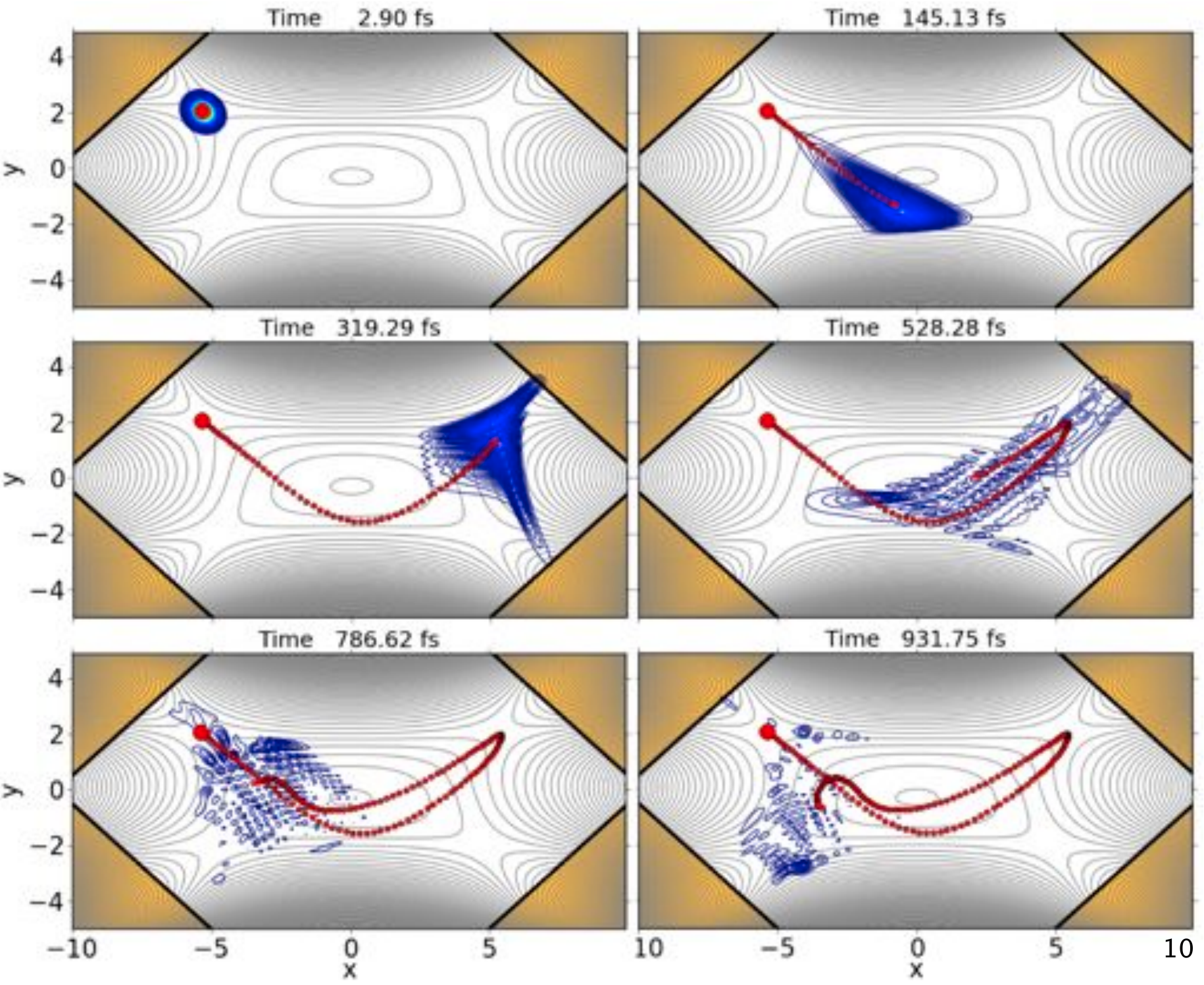}}
\caption{Similar to Figure \ref{fig:z_wp_1}, but for a carbon particle on the $\lambda=0.4$ 
PES with  $\langle E \rangle_{t=0}=1$ kcal mol$^{-1}$.  
The elongated shape of the caldera region causes the wave packet to be reflected 
off the bottom of the wall in its initial motion and toward the opposing upper TS, 
through which a significant portion of the packet is lost.  
The subsequent dynamics sees the remaining portion of the wave packet 
spreading out in the caldera region and portions passing out through the four TS dividing surfaces.}
\label{fig:z_wp_4}
\end{figure}

\begin{figure}[htbp!]
\centerline{\includegraphics[height=80mm]{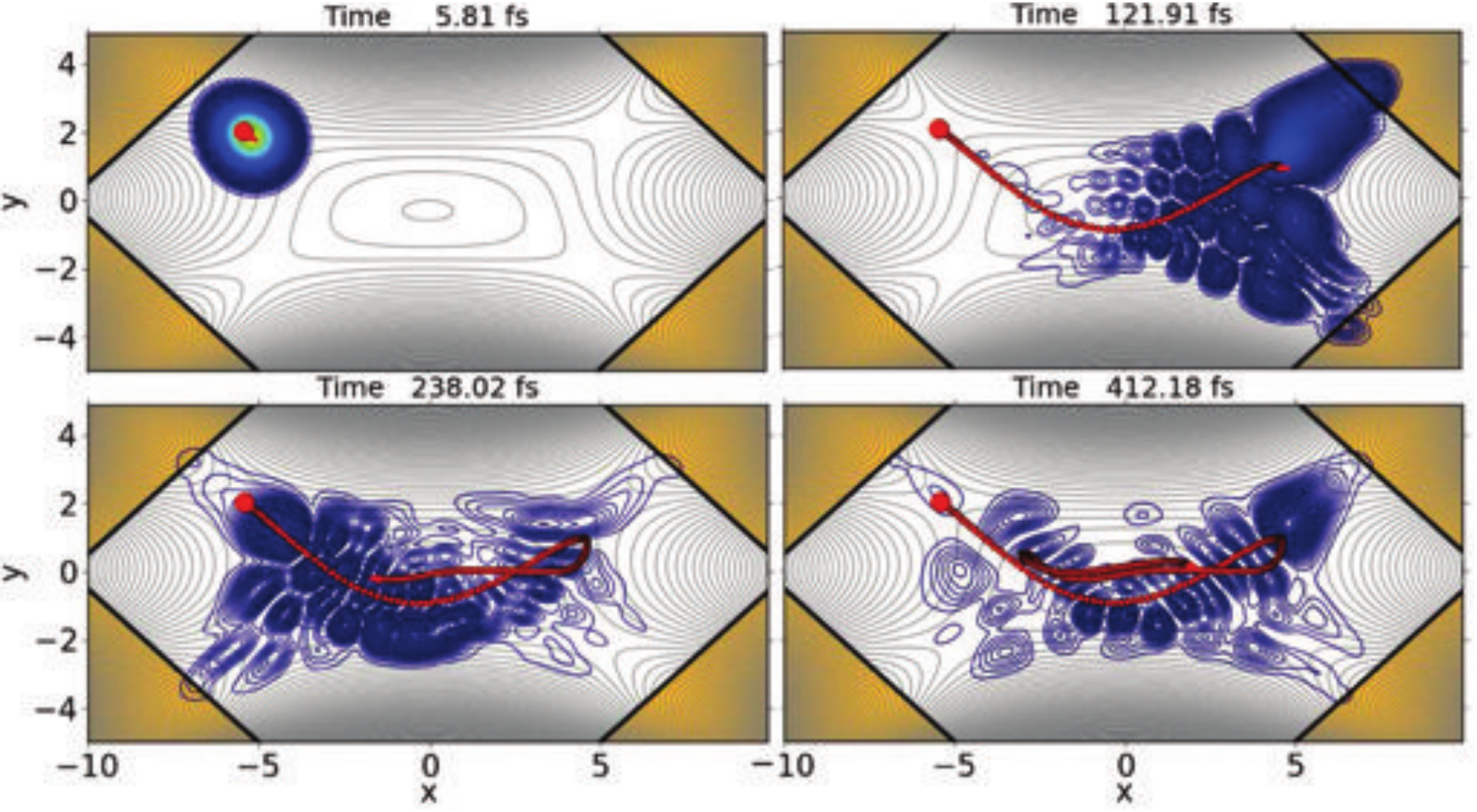}}
\caption{Similar to Figure \ref{fig:z_wp_1}, but for a Hydrogen particle on the $\lambda=0.4$ 
PES with  $\langle E \rangle_{t=0}=1$ kcal mol$^{-1}$.}
\label{fig:z_wp_9}
\end{figure}

The dynamical matching effect is further reduced for wave packets on the stretched potential with
$\lambda=0.4$.  
Here the initial momentum of the gaussian wave packet is again directed towards 
the bottom PES wall with the lower energy RH TS even further displaced from the collision point 
than in the $\lambda=0.6$ case.  For the lower energy packets 
with $\langle E^{\star}\rangle_{t=0}=1$ kcal mol$^{-1}$,
both the $^{12}$C and $^1$H wave packets (Figures \ref{fig:z_wp_4} and \ref{fig:z_wp_9}, respectively),  
the collision with the PES bottom wall spreads the wave packet creating a wave front proceeding to the right.  
Some of the  front subsequently passes through the diametrically\--opposing lower energy RH TS 
or is reflected off the rightmost PES wall.  However, most of the 
packet front is directed toward the RH higher energy TS region where a 
significant portion of the wave packet passes through to the NIP and 
some portion is reflected back into the caldera.  The portions of the packet 
remaining in the caldera at this point are reflected back to the left portion of the 
caldera where more amplitude is lost through the left TSs.  Some coherence of 
the packet moving back and forth in the $x$ direction is evident in the remnants of 
the $^{12}$C packet at longer times and a second reflection of the wave packet off the 
bottom PES wall is observed. Even in the case of the $^1$H packet, some structure in the 
remaining wave function is evident as time progresses, 
see the bottom two panels of Figure \ref{fig:z_wp_9}.  

\begin{figure}[htbp!]
\centerline{\includegraphics[height=60mm]{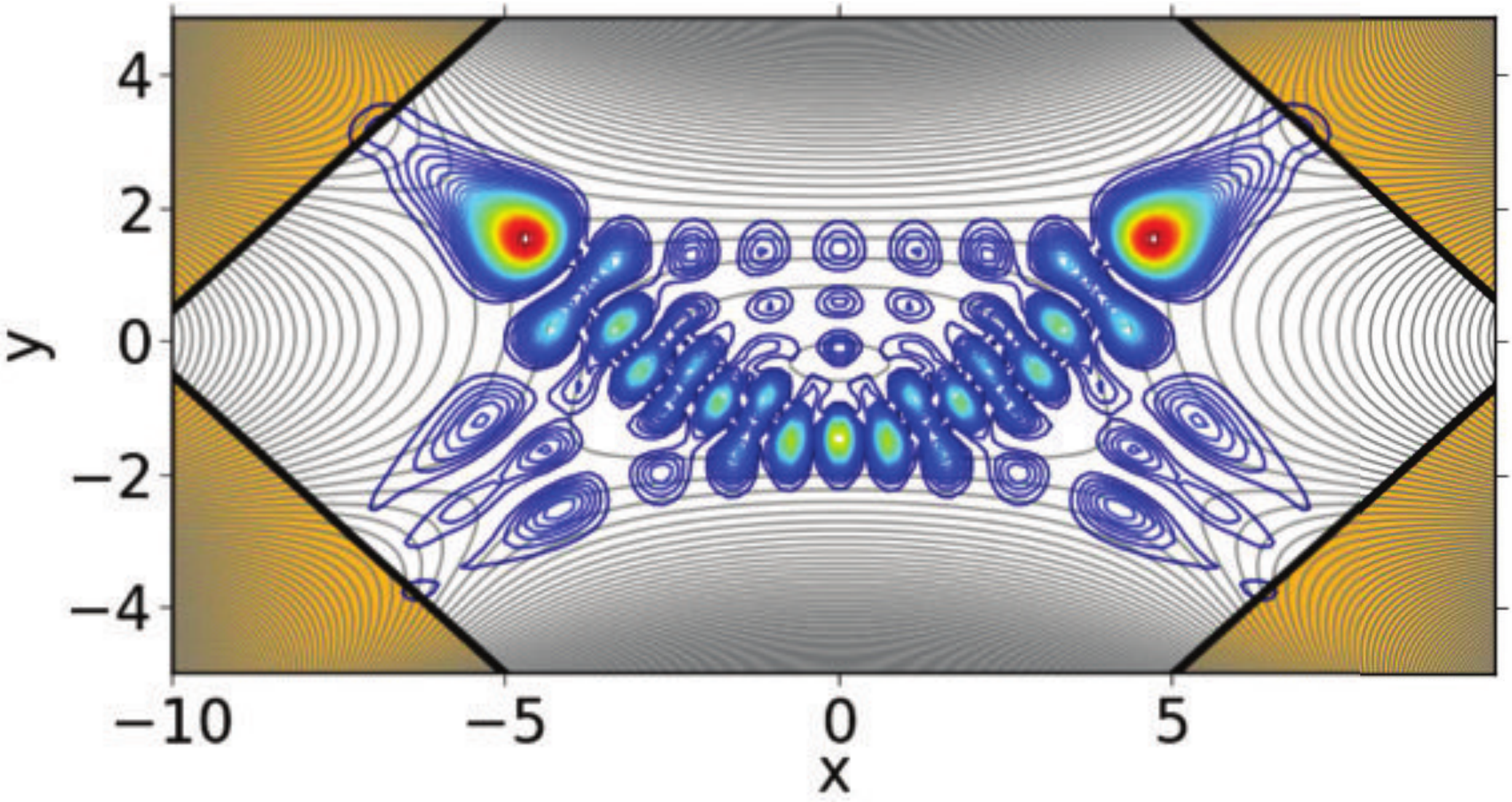}}
\caption{A complex eigenstate of the $^1$H system on the complex $\lambda=0.4$ PES 
with a energy $\sim1.4$ kcal mol$^{-1}$ above of the upper transition state 
and a lifetime of $\sim370$ fs.  This eigenstate, and those with a similar structure, 
were a contributing component of the $\langle E^{\star} \rangle=1$ kcal mol$^{-1}$ wave packet.}
\label{fig:z_wp_10}
\end{figure}

The remaining packet seems to follow an `arc' between the two higher energy TS regions, 
bouncing off the bottom PES wall.  In order to isolate 
such a structure, a time\--independent analysis of the $^{1}$H system on the 
$\lambda=0.4$ was performed by diagonalizing the complex Hamiltonian to obtain 
eigenstates with energies and lifetimes determined by the real and imaginary 
components of the corresponding eigenvalues, respectively \cite{DieterMeyer}.  
The contributions of these complex normalized eigenstates to the initial wave packet were computed.  
While most of the wave packet was composed of states with very short lifetimes, 
several important contributing states with longer lifetimes were present.  
One of these states is shown in Figure \ref{fig:z_wp_10}, and 
is very similar to the structure of the wave packet snapshots at late 
propagation times shown in Figure \ref{fig:z_wp_9}.  
(Other long\--lifetime contributing states for this 
system have similar structure.)
The structure of the state suggests that it is associated with a classical periodic
orbit moving between regions of configuration space in the vicinity 
of the two higher energy TSs with a reflection off 
the bottom PES wall between.  This structure is consistent with 
the observed quantum dynamics, 
where the dominant reaction mechanism seems to involve a bounce off the bottom caldera wall 
and passage over the upper TS, and also the classical trajectory results obtained
for $\lambda = 0.4$.

\begin{figure}[htbp!]
\centerline{\includegraphics[height=120mm]{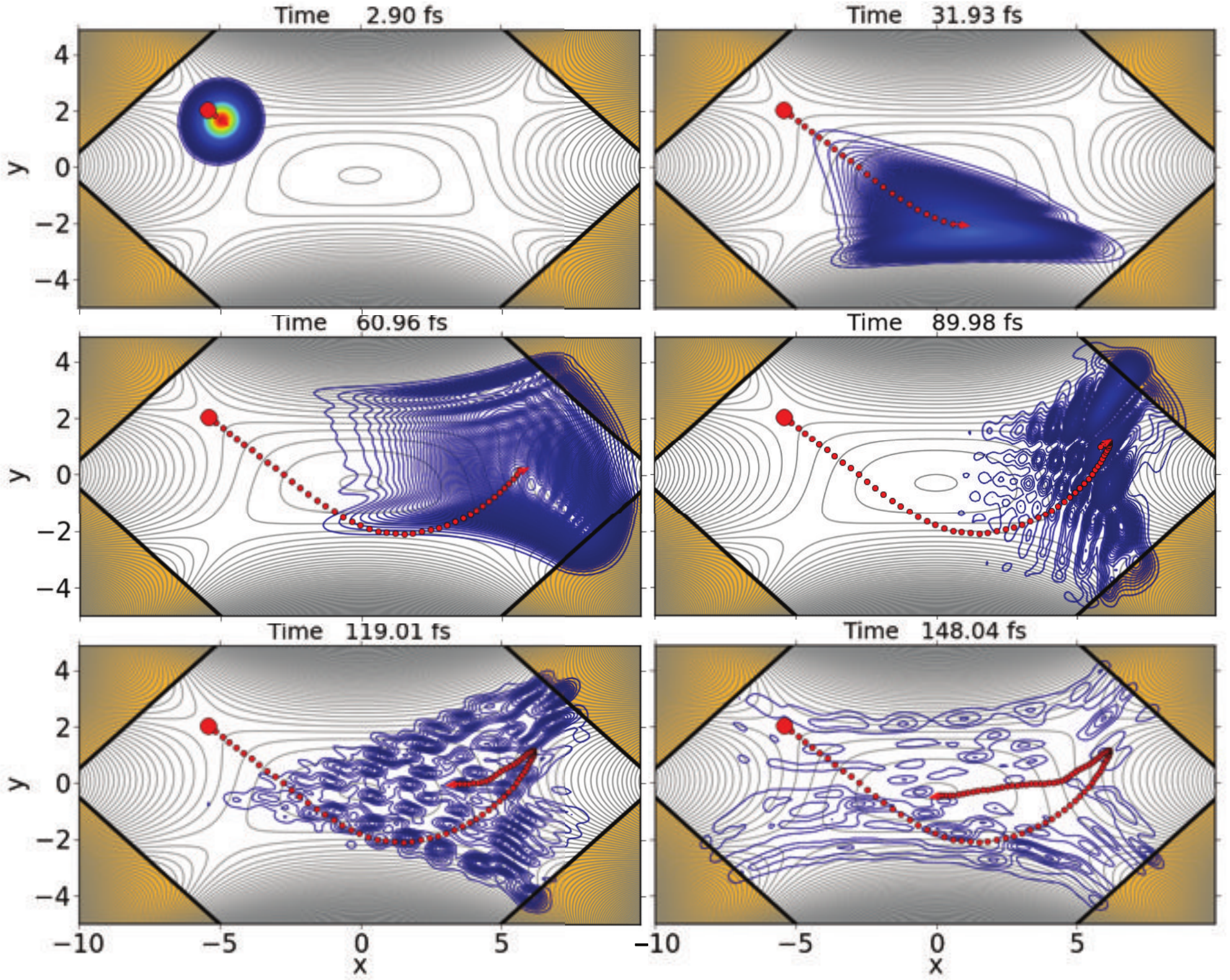}}
\caption{Similar to Figure \ref{fig:z_wp_4}, but for the hydrogen particle with 
$\langle E \rangle_{t=0}=6$ kcal mol$^{-1}$.  The dynamics are quite similar to 
those observed for the carbon particle, save they occur on a shorted time scale and 
the spreading of the wave packet is much faster.}
\label{fig:z_wp_6}
\end{figure}

When the energy of the initial wave packet is raised, a similar 
effect to that discussed in the $\lambda=0.6$ potential occurs.   
The reflection of the wave packet off the bottom potential wall does not 
deflect the wave packet's course as effectively as at lower energy, 
and most of the reflected wave front,
especially for the $^{12}$C system, passes out both the diametrically\--opposing lower 
energy RH TS and the upper RH TS.  Snapshots of the $\langle E^{\star}\rangle_{t=0}=6$ kcal mol$^{-1}$ 
wave packet of the $^{1}$H system are shown in Figure \ref{fig:z_wp_6} and can be 
compared to the corresponding lower energy packet of Figure \ref{fig:z_wp_4}.  
The residence time of the higher energy wave packet in the caldera region is relatively short, and
the structure of the packet at long propagation times is no longer observed
in the high energy packet.  Indeed, decomposing the high energy wave packet 
into the eigenstates of the complex Hamiltonian reveals that contributions of 
longer lifetimes states are drastically reduced relative to the low energy wave packet.  

\subsection{Wave packet comparison and summary}
\begin{figure}[htbp!]
\centerline{\includegraphics[height=120mm]{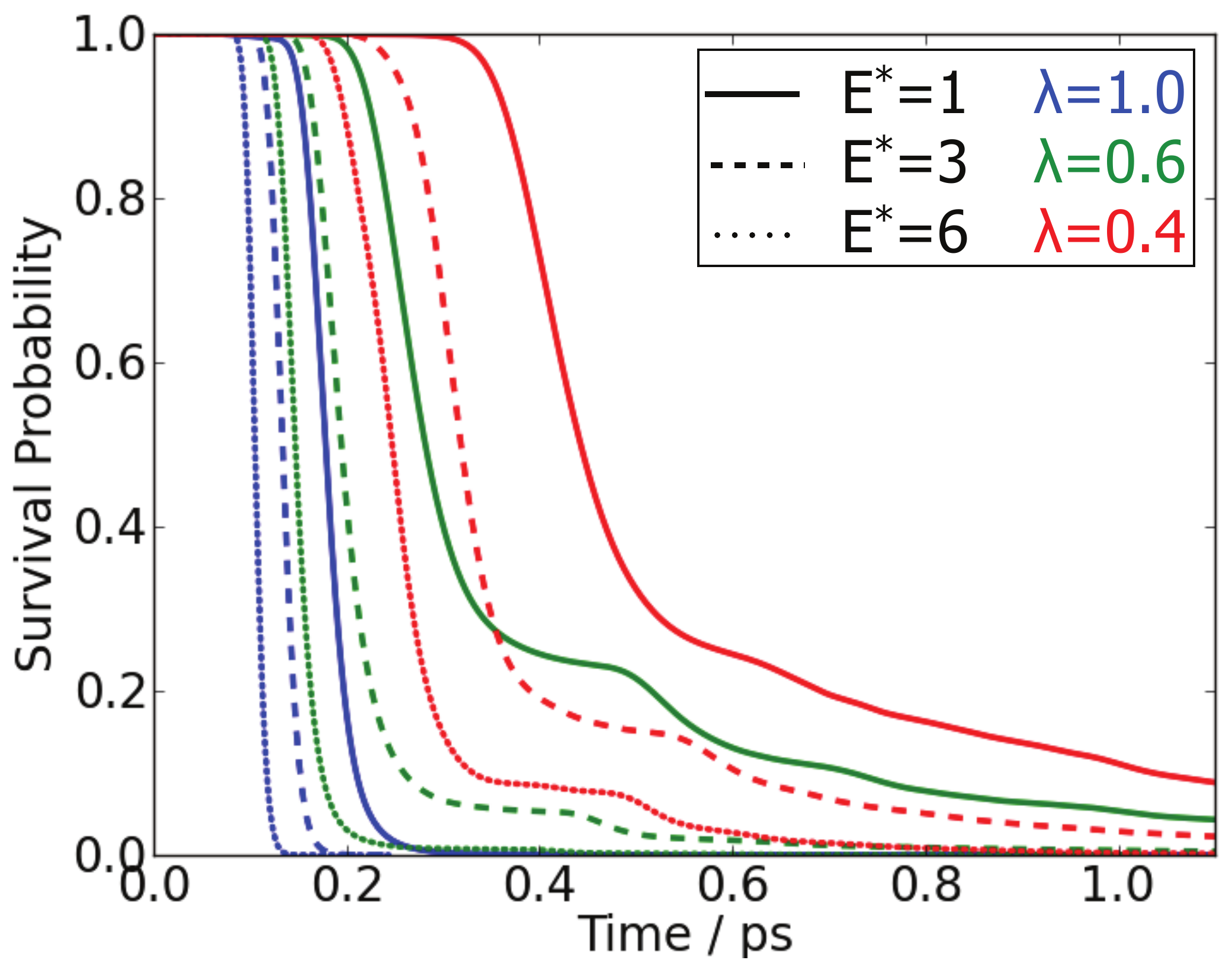}}
\caption{The survival probability, defined as the remaining wave packet probability 
that has not been absorbed by the NIP at a time $t$, $\langle \Psi \vert \Psi \rangle_t$ 
for the carbon\--particle wave packets on the various caldera potentials.}
\label{fig:z_wp_7}
\end{figure}

\begin{figure}[htbp!]
\centerline{\includegraphics[height=120mm]{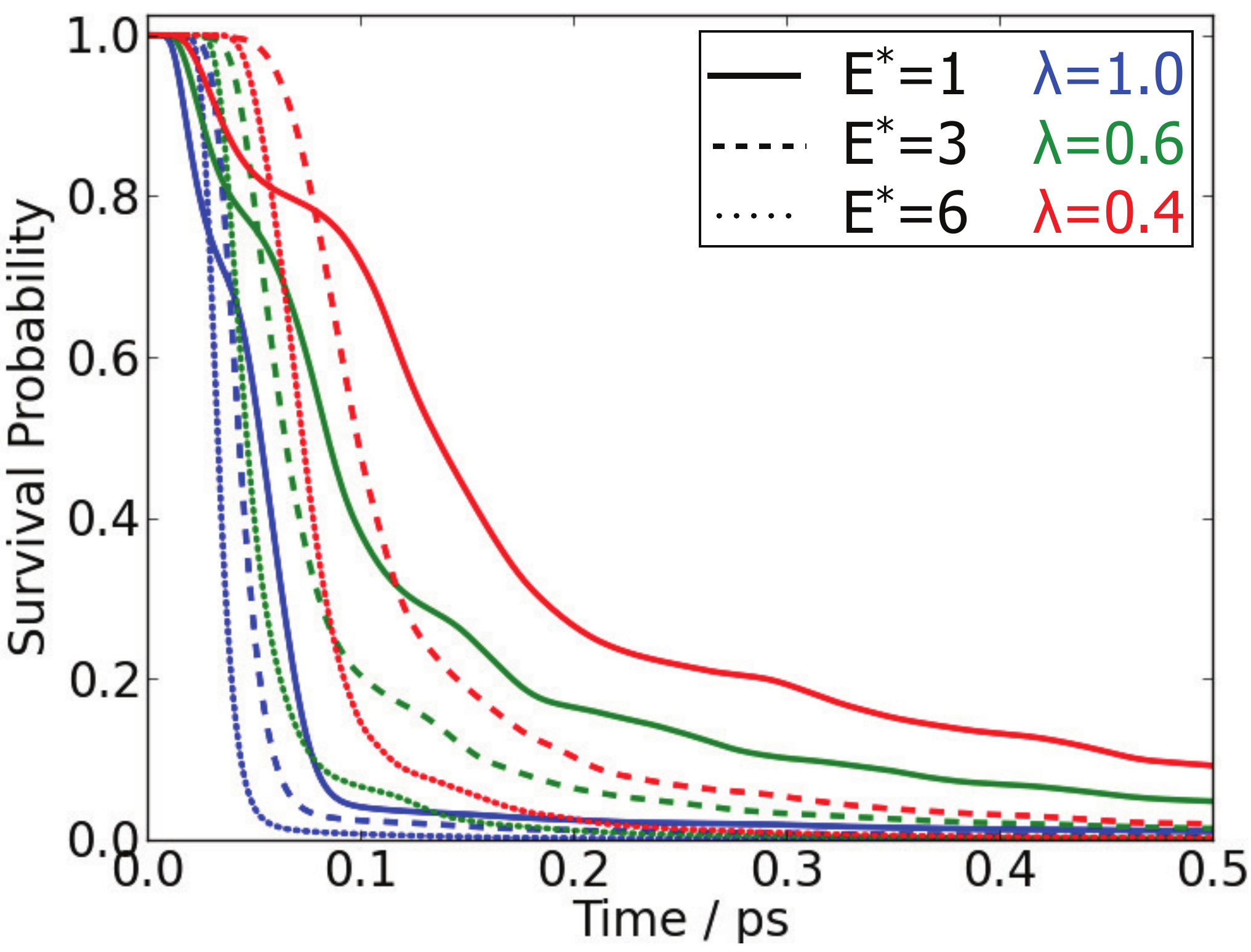}}
\caption{Similar to Figure \ref{fig:z_wp_7} for the hydrogen\--particle wave packets.  
The initial drop in the low energy wave packets ($\sim20\%$ at times $<100$ fs) is 
mainly attributed to the spreading of the wave packet backward in the reactive direction, 
and this amplitude never enters the caldera region. }
\label{fig:z_wp_8}
\end{figure}

To complement the wavepacket results just presented, it
is useful to look at the survival probability of the wave packet in the 
caldera region over the propagation time.  The survival probability 
is defined as the fraction of the wave packet that remains in the caldera region, 
which we approximate as the portion unabsorbed by the NIP.  
Since the wave packet is initially normalized to unity, we approximate 
the survival probability as the normalization of the wave packet at a 
given time $t$. The survival probability of the wave packet in the caldera region 
is shown in Figures \ref{fig:z_wp_7} and \ref{fig:z_wp_8} for the $^{12}$C and $^1$H systems, respectively.  
The curves corresponding to the $\langle E^{\star}\rangle_{t=0}=1$ kcal mol$^{-1}$ 
wave packets of $^1$H have a quick initial drop in survival probability that 
corresponds to the spreading of the initial packet out the upper LH entry TS and
represents the portion of the wave packet that does not enter the caldera region.  
It will not be discussed further as it is not important to the caldera dynamics.  
The fast and uniform decay of the survival probability for all the packets after 
a short period on the $\lambda=1$ PES are a clear result of the direct passage mechanism.  
The main differences in these curves are mainly a result of
the different passage times for wave packets over 
the caldera region, i.e. the packet velocity, between the diametrically opposing TSs.  
At longer times, the effects of the reflected portions of the $^1$H wave packets can 
be seen as the initial decay between $\sim50\--100$ fs, depending on $\langle E^{\star}\rangle_{t=0}$ 
as a small slowly decaying portion of the wave packet remains.

The survival probabilities for the $\lambda=0.6$ and $0.4$ potentials are more structured.  
After a passage period where the wave packet collides with the bottom PES wall, 
the first significant drop in the probability occurs when a major portion of the wave 
packet passes through the right upper and lower TSs.  The remaining portion is 
then reflected back to the left TSs and after a passing period, when the main 
portion of the wave packet is passing back to the left over the bowl of the caldera, 
the next drop in the survival probability occurs when the packet front passes over the left TSs.  

The survival probabilities all show that where the direct mechanism is more 
important, that is, where the reaction coordinate at the upper entry TS is aligned along 
the direction pointing towards the diametrically\--opposing lower exit TS and for 
higher masses and higher energies, the decay of the wave packet is much 
faster and more complete at shorter time scales.  When the wave packet must 
reflect off a caldera wall, the magnitude of the 
survival probability is appreciable to longer times
for several reasons.  
First, and most obvious, the wave packet must travel a greater distance between TS regions.  
Second, the collision with the wall of the PES tends to significantly broaden 
the wave packet so that portions of the amplitude not only proceed to the lower RH TS, for 
those packets that entered via the upper LH TS,
but to the right wall and the upper RH TS where it is possible for the reflection, 
and broadening, to occur again.  Thus, the initial collision with the wall 
can prolong the survival of the packet in the caldera region.  

\newpage

\section{Summary and conclusions}
\label{sec:summary}

In this paper we have explored both classical and quantum dynamics of a model potential
exhibiting a caldera: that is, a shallow potential well with two pairs of symmetry 
related index one saddles associated with the entrance/exit channels.

Classical trajectory simulations at several different energies 
confirm the existence of the `dynamical matching' phenomenon 
originally proposed by Carpenter, where the momentum direction
associated with an incoming trajectory initiated at a high energy TS
determines to a considerable extent the outcome of the 
reaction (passage through the opposite exit channel).  By studying
a `stretched' version of the caldera model, we have uncovered a generalized
dynamical matching: bundles of trajectories can reflect off a hard potential wall
so as to end up exiting predominantly via the TS opposite the 
reflection point.  
In this respect, the stretched caldera provides an example of a `molecular billiard'.

The effects on the caldera reaction dynamics due to higher dimensional coupling and environmental factors were
explored by introducing energy dissipation into the equations of motion of the caldera system, similar to our previous
work on a model potential with a VRI point \cite{Collins13}.  The results indicated that environmental and coupling effects
may serve to divide the caldera reaction mechanism into a direct component, which occurs quickly and is highly dependent
upon the caldera dynamics, and a statistical component where a population of reagents loses energy in the caldera intermediate
region, and therefore a population of a  long\--lived caldera intermediate forms that reacts in accord with a statistical model.

In addition to classical trajectory studies, we 
have examined the dynamics of quantum wavepackets on the caldera potential (stretched and unstretched).
Our computations reveal a quantum mechanical analogue of the `dynamical matching'
phenomenon, where the initial expectation value of the momentum 
direction for the wavepacket determines the exit channel through which most
of the probability density passes to product.

\acknowledgments
PC and SW acknowledge the support of the Office of Naval Research 
(Grant No.~N00014-01-1-0769). 
PC, BKC and SW acknowledge the support 
of the UK Engineering and Physical Sciences Research Council (Grant No.\ EP/K000489/1).
The work of ZCK and GSE is supported by the US National Science Foundation under 
Grant No.\ CHE-1223754.

\def\cprime{$'$}

\newpage

\begin{table}[htpb]
\begin{center}
\caption{Stationary points of the potential from Equation \eqref{eq:pot1} or the $\lambda=1$ potential in Equation \eqref{eq:pot2}, where `RH' stands for `righthand' and `LH' stands for `lefthand'.}
\label{tab:equi_pts} 
\begin{tabular}{| c || c c || c |} \hline
 Critical point  & $x$ & $y$  & $E$ \\      \hline \hline
Central minimum  & 0.000 & -0.297 & -0.448 \\ 
Upper LH saddle & -2.149& 2.0778 & 27.0123 \\ 
Upper RH saddle & 2.149 & 2.0778 & 27.0123 \\ 
Lower LH saddle & -1.923 & -2.003 & 14.767 \\ 
Lower RH saddle & 1.923 & -2.003 & 14.767 \\ 
\hline
\end{tabular}
\end{center}
\end{table}

\begin{table}[H]
\begin{center}
 \caption{The initial parameters and energy expectation values (relative to the upper saddle point energy) 
of the wave packets.  All wave packets were initialized in the $v=0$ local eigenstate of the bath coordinate.}
\begin{tabular}{| c c | c c |}
\hline
Atom & $\langle$E$\rangle_{t=0}$  &   $p_r$ \small{($\lambda = 1, 0.6, 0.4$)}  & $\delta_r$ \\ 
 (mass) & (kcal mol$^{-1}$) & (amu bohr fs$^{-1}$) &  (bohr)  \\ \hline \hline
C &   1.0   &  0.173, 0.173, 0.172 & 0.25 \\
($m=12.00$ amu)  &  3.0 & 0.319, 0.319, 0.318 & 0.25 \\
 & 6.0 & 0.457, 0.457, 0.457 & 0.25 \\ \hline
H &   1.0   & 0.0150, 0.0211, 0.0238  & 0.35 \\
($m=1.01$ amu)  &  3.0 & 0.0791, 0.0804, 0.0812 & 0.35 \\
 & 6.0 & 0.124, 0.125, 0.125 & 0.35 \\ \hline
 \end{tabular}
\end{center}
\end{table}

\begin{table}[H]
\begin{center}
  \caption{The potential and coordinate grid parameters used in the wave packet calculations.  
  The magnitude of the optical potential is given in kcal mol$^{-1}$ and all distances are in bohr.  
  The potentials used are identical to those described Sec.\ \ref{sec:pes}, but are 
  modified by a scaling factor $\epsilon_{{\textnormal{scale}}}$ in units of hartree. 
  The specification of the boundary in the last column represents the four possible lines that can be 
constructed from the given slope and intercept magnitudes.}
\begin{tabular}{ |  c  |  c  ||  c  | c   c   | c |   c    c  || c  |   c  |}
\hline
 $\lambda$   & $\epsilon_{{\textnormal{scale}}}$ &  $N_x$ & $x_{{\textnormal{min}}}$ & $ x_{{\textnormal{max}}}$ 
 & $N_y$ & $y_{{\textnormal{min}}}$  & $ y_{{\textnormal{max}}}$ & $U_0$ & $u\left(x,y\right)$ Boundary \\ \hline \hline
 0.4 & $0.20/627.5095 $ & 200 & -10.0 & 10.0 & 100 & -5.0  & 5.0 & 36.7 & $y=\left(\pm0.900\right)x\pm9.5$ \\
  0.6 & $3.1872\times10^{-4}$ & 200 & -8.0 & 8.0 & 150 & -5.0  & 5.0 & 36.7 & $y=\left(\pm0.727\right)x\pm6.82$  \\
   1.0 & $3.1872\times10^{-4}$ & 150 & -5.0 & 5.0 & 150 & -5.0  & 5.0 &36.7  & $y=\left(\pm1.00\right)x\pm6.75$ \\ \hline
  \end{tabular}
\end{center}
\end{table}


\newpage

\newpage

\newpage

\newpage

\newpage

\newpage

\newpage

\newpage

\newpage

\newpage

\newpage

\newpage

\newpage

\newpage

\newpage

\newpage

\newpage

\end{document}